\documentclass[11pt]{article}
\usepackage[utf8]{inputenc}
\usepackage[T1]{fontenc}
\usepackage[utf8]{inputenc}
\usepackage{authblk} 

\title{\Large{Exploring the Evolution of Altruistic Punishment with a PDE Model of Cultural Multilevel Selection}}
\author[1,2]{Daniel B. Cooney}
\affil[1]{Department of Mathematics, University of Illinois Urbana-Champaign, Urbana, IL, USA}
\date{\today}

\usepackage{latexsym}
\usepackage{amssymb,amsmath}
\usepackage{amsthm}
\usepackage{mathtools}
\usepackage{graphicx}
\usepackage{color}
\usepackage{blkarray}
\usepackage{multirow}
\usepackage{hyperref}
\usepackage{float}
\usepackage{subfig}
\usepackage{relsize}
\usepackage{array,makecell}

\usepackage{cleveref}

\usepackage[numbers,sort&compress]{natbib}

\hypersetup{
  colorlinks   = true, 
  urlcolor     = blue, 
  linkcolor    = blue, 
  citecolor   = red 
}

\usepackage{breqn}

\usepackage{xcolor}
\usepackage{cancel}

\usepackage{hyperref}
\usepackage{empheq}

\usepackage[utf8]{inputenc}
\usepackage{nicefrac}

\usepackage{setspace}

\usepackage{enumerate}
\usepackage{comment}
\usepackage{makecell}

\usepackage[titletoc,toc]{appendix}

\usepackage{tocbasic}
\DeclareTOCStyleEntry[
  beforeskip=.2em plus 1pt,
  pagenumberformat=\textbf
]{tocline}{section}

\usepackage[font=small,labelfont=bf]{caption}

\newcommand{\ds}{\displaystyle}
\newcommand{\mc}{\mathcal}

\DeclareMathOperator{\var}{Var}

\newcommand{\bbm}{\begin{bmatrix}}
\newcommand{\bpm}{\begin{pmatrix}}
\newcommand{\ebm}{\end{bmatrix}}
\newcommand{\epm}{\end{pmatrix}}

\newcommand{\revision}[1]{{\bf{#1}}}

 \newcommand{\dsdel}[2]{\displaystyle\frac{\partial #1}{\partial #2}}

\newcommand{\dsddt}[1]{\displaystyle\frac{d #1}{dt}}

\renewcommand{\abstractname}{Abstract}
\numberwithin{equation}{section}
\numberwithin{figure}{section}
\renewcommand{\thesection}{\arabic{section}}

\newcommand{\changelocaltocdepth}[1]{%
  \addtocontents{toc}{\protect\setcounter{tocdepth}{#1}}%
  \setcounter{tocdepth}{#1}%
}
\begin{document}

\maketitle

\newtheorem{definition}{Definition}[section]
\newtheorem{theorem}{Theorem}[section]
\newtheorem{lemma}[theorem]{Lemma}
\newtheorem{corollary}[theorem]{Corollary}
\newtheorem{claim}[theorem]{Claim}
\newtheorem{fact}[theorem]{Fact}
\newtheorem{proposition}[theorem]{Proposition}
\newtheorem{remark}[theorem]{Remark}
\newtheorem{example}[theorem]{Example}
\newtheorem{observation}[theorem]{Observation}


\renewcommand{\thesection}{\arabic{section}}
\setcounter{section}{0}

\begin{abstract}
Two mechanisms that have been used to study the evolution of cooperative behavior are altruistic punishment, in which cooperative individuals pay additional costs to punish defection, and multilevel selection, in which competition between groups can help to counteract individual-level incentives to cheat. Boyd, Gintis, Bowles, and Richerson have used simulation models of cultural evolution to suggest that altruistic punishment and pairwise group-level competition can work in concert to promote cooperation, even when neither mechanism can do so on its own. In this paper, we formulate a PDE model for multilevel selection motivated by the approach of Boyd and coauthors, modeling individual-level birth-death competition with a replicator equation based on individual payoffs and describing group-level competition with pairwise conflicts based on differences in the average payoffs of the competing groups. Building off of existing PDE models for multilevel selection with frequency-independent group-level competition, we use analytical and numerical techniques to understand how the forms of individual and average payoffs can impact the long-time ability to sustain altruistic punishment in group-structured populations. We find several interesting differences between the behavior of our new PDE model with pairwise group-level competition and existing multilevel PDE models, including the observation that our new model can feature a non-monotonic dependence of the long-time collective payoff on the strength of altruistic punishment. Going forward, our PDE framework can serve as a way to connect and compare disparate approaches for understanding multilevel selection across the literature in evolutionary biology and anthropology.     
\end{abstract}

\singlespacing

%
{\hypersetup{linkbordercolor=black, linkcolor = black}
\begin{spacing}{8}
\renewcommand{\baselinestretch}{0.8}\normalsize
\tableofcontents
\addtocontents{toc}{\protect\setcounter{tocdepth}{2}}
\end{spacing}
\singlespacing

\section{Introduction} \label{sec:intro}

\subsection{Altruistic Punishment and Cultural Multilevel Selection}
 
The emergence of cooperative behavior has played an important role in evolutionary biology in settings ranging from the major transitions in the evolution of complex cellular life to the establishment of cooperative animal societies \citep{szathmary1995major}. A common feature in many biological and social systems is a fundamental tension between the individual costs of cooperative behavior and the social benefits conferred by cooperation, presenting a dilemma in which there is conflict between individual and collective evolutionary interests. Many mechanisms have been proposed to explain how collective benefits can be achieved in spite of the individual advantages of cheating behaviors, with the emergence and survival of cooperation attributed to features of population structure or social forces including like-with-like assortment \citep{grafen1979hawk,bergstrom2003algebra,wilson1997group}, rewarding cooperation or punishing defection through direct or indirect reciprocity \citep{trivers1971evolution,axelrod1981evolution,ohtsuki2006leading,nowak1998evolution}, and multilevel selection featuring group-level competition favoring cooperative groups over groups of defectors \citep{wilson1975theory,traulsen2006evolution,traulsen2008analytical}.   Punishment and reciprocity are often described as mechanisms that can help to establish cooperation among non-relatives and to describe the emergence of cooperative social norms in human societies.  

A particularly interesting form of reciprocity is altruistic punishment, in which cooperative individuals can pay additional costs to confer punishments upon defectors. While incurring additional costs to punish defectors may seem to make cooperative behavior even more disadvantageous, experiments in public goods and common-pool resource games have shown that humans can facilitate increased cooperative behavior in the presence of the option to pursue costly punishment \citep{ostrom1992covenants,fehr2000cooperation,fehr2002altruistic,janssen2010lab,boyd2010coordinated}. Altruistic punishment can also give rise to so-called ``second-order free-rider problems'', in which individuals who cooperate in an underlying game can choose to cheat in the task of costly punishment of defectors, but the ability of costly punishment to decrease the individual incentive to defect can allow a population with sufficiently many altruistic punishers to withstand potential invasion from defectors \citep{sethi1996evolution,fowler2005altruistic} or to fix in a finite population when initially rare \citep{bowles2004evolution}. Cross-cultural studies have suggested that altruistic punishment is a common feature in human societies and have shown that costly punishment can be correlated with cooperative behavior in game-theoretic experiments \citep{henrich2006costly}. Beyond the case of human cooperation, costly punishment behaviors have also been observed in a variety of non-human animal species \citep{clutton1995punishment}.
Another mechanism that is often attributed to the evolution of human cooperation is cultural group selection, in which group-level competition can help to promote collectively beneficial behaviors \citep{henrich2004cultural,wilson1994reintroducing,richerson2016cultural,smith2020cultural,soltis1995can}. Mathematical models of cultural group selection have been used to study the establishment and maintenance of social norms of indirect reciprocity \citep{bowles2004evolution,chalub2006evolution,santos2007multi,scheuring2009evolution,scheuring2010coevolution} {as well as the emergence of parochial altruism and in-group cooperation \cite{choi2007coevolution,garcia2011evolution,efferson2024super}}, and the role of cultural group selection has {also} been explored as a potential mechanism for establishing the sustainable management of common-pool resources \citep{waring2017coevolution,waring2018evidence,tam2021measuring,hillis2018applying,wilson2013generalizing,wilson2023multilevel}. Cultural group selection has shown to be helpful in promoting beneficial outcomes when within-group mechanisms of reciprocity or social norms feature bistability of all-defector and all-cooperator equilibria, allowing group-level competition to select the within-group equilibria that promotes collective success of the group \citep{boyd1990group}. Recent theoretical and experimental work has also suggested that cultural group selection and direct reciprocity can work in concert to promote cooperative behavior in cases for which neither of the two mechanisms could do so on its own \citep{efferson2024super}, and experiments have suggested that between-group competition can facilitate increased levels of altruistic punishment within groups \citep{rebers2012altruistic}.  
 
Boyd and coauthors introduced a model of cultural group selection to explore the evolution of altruistic punishment, studying how the presence of altruistic punishers can help to promote greater levels of cooperation via multilevel selection  \cite{boyd2003evolution}.  The authors considered a stochastic model of a group-structured population, assuming that social interactions followed a donation game and allowed for the possibility for cooperative individuals to pursue costly punishment of defectors. Considering individual-level selection following a birth-death process and group-level competition taking place through pairwise conflicts between groups, Boyd and coauthors showed that altruistic punishment could work in concert with group-level competition to promote additional cooperative behavior even for scenarios in which altruistic punishment was not sufficient to sustain cooperation by individual-level selection alone. Further applications of this stochastic framework for cultural group selection have been used to study the evolution of parochial altruism featuring cooperation with in-group peers \citep{garcia2011evolution}, the role of norm internalization on the promotion of cooperation \citep{odouard2023polarize}, and the role of social preferences on the evolution of cooperation \citep{janssen2014effect}.

Simulations of the stochastic model by Boyd and coauthors produced a variety of qualitative conclusions about the promotion of cooperative behavior under the combined forces of altruistic punishment and cultural group selection \citep{boyd2003evolution}. In particular, the authors showed that the presence of altruistic punishment facilitated a greater achievement of cooperative behavior than was achieved in multilevel competition in a group-structured population composed only of defectors and pure cooperators who do not engage in costly punishment. Furthermore, the authors showed in simulations that the level of cooperative behavior achieved under cultural multilevel selection increased with the strength of punishment imposed on defectors and with the rate of between-group conflict events, while cooperative behavior decreased with the cost incurred to punish defection. The authors also found that altruistic punishment was more conducive to the achievement of cooperation via multilevel selection when the costs of punishing were only incurred through interactions with defectors, while altruistic punishment was more difficult to achieve for scenarios in which individuals paid a fixed cost to punish potential defectors regardless of the current strategic composition of the group \citep{boyd2003evolution}. These predictions made by Boyd and coauthors provide useful expectations for further exploration of mathematical models combining the social forces of altruistic punishment and between-group competition, and serve as a point for comparison that we will use for analytical and numerical results obtained in the deterministic approach we use in this paper to model multilevel selection with altruistic punishment. 

\subsection{PDE Models for the Evolution of Altruistic Punishment via Multilevel Selection with Pairwise Group-Level Conflict}

While the stochastic simulations of models for cultural multilevel selection have revealed many qualitative insights into the evolution of altruistic punishment, it can also be helpful to {examine these multilevel dynamics of cultural evolution with a mathematical framework that is tractable with analytical techniques. One such framework has introduced by Luo and coauthors \citep{luo2014unifying,luo2017scaling,van2014simple}, which starts with a stochastic process in group-structured populations that models competition within and among groups using individual-level and group-level replication events.} In the limit in which the number and size of groups tend to infinity, it is possible to derive a partial differential equation (PDE) description of this two-level birth-death process, with an advection term describing the individual-level advantage of defection and a nonlocal term describing the group-level benefits of cooperation. {This PDE approach has been further generalized to consider evolutionary competition based upon individual-level and group-level replication events based on the payoffs of two-player, two-strategy games \citep{cooney2019replicator,cooney2020analysis}, as well as to explore competition that incorporates arbitrary forms of frequency dependence within groups \citep{cooney2022long,cooney2022assortment,cooney2022pde}}. These PDE models often provide analytically tractable ways to understand the tradeoffs between individual-level and group-level incentives, highlighting the conditions under which between-group competition can allow for the establishment of long-time cooperation in a population.   %
One particular advantage of this PDE framework for studying multilevel selection is that there is substantial flexibility for modifying payoffs and interaction structure to incorporate within-group mechanisms and explore how these individual-level mechanisms can work in concert with between-group competition to promote cooperative behaviors via multilevel selection. Recent work has used these PDE models to explore how mechanisms including within-group network structure, like-with-like assortment, other-regarding preferences, and both direct and indirect reciprocity can interact with group-level competition to help promote cooperative behavior \citep{cooney2022assortment,cooney2023evolutionary}. It is shown that these mechanisms can work synergistically with between-group competition, allowing the two mechanisms to promote long-time cooperation in parameter regimes for which neither mechanism could allow for the evolution of cooperation when operating alone. As the stochastic model explored by  \citep{boyd2003evolution} {have} also shown beneficial interaction between altruistic punishment and cultural group selection, it is natural to ask whether such synergistic effects can be seen as well in PDE models for multilevel selection. 

In  this paper, we take inspiration from the stochastic model by Boyd and coauthors to formulate a PDE describing the evolution of altruistic punishment via multilevel selection with pairwise between-group competition. While existing work on PDE models for multilevel selection based on the approach of Luo and coauthors have considered group-level replication events occurring at a rate depending only on the strategic competition of the replicating group \citep{luo2014unifying,luo2017scaling,cooney2019replicator,cooney2022long}, {this model of pairwise group-level conflict} between groups and allows us to study frequency-dependent between-group competition. {In particular, we consider a variety of group-level victory probabilities that incorporate different ways to describe how the strategic compositions of competing groups can be used to determine the winning group in a pairwise conflict, including the form of victory probability introduced by Boyd and couathors \cite{boyd2003evolution} as well as group-level analogues of the Fermi update rule \citep{traulsen2007pairwise}, the local update rule for individual-level selection \citep{traulsen2005coevolutionary}, and the Tullock contest function \cite{tullock2008efficient} that have been used previously to decribe pairwise individual-level selection dynamics. To explore how altruistic punishment and multilevel competition can interact to support the evolution of cooperative behavior, we consider both multilevel selection scenarios featuring groups composed of two possible strategies (either defectors and cooperators or defectors and altruistic punisher as well as scenarios for multilevel selection with groups featuring all three strategies.}

{For the case of two-strategy dynamics, we apply existing analytical results from Cooney and Mori \cite{cooney2022long} to describe the long-time behavior of the multilevel dynamics in the special case of additively separable group-level victory probabilities, and we study the more general class of pairwise group-level victory probabilities using a mix of numerical simulation and preliminary analytical predictions that have been derived in a companion paper by Alexiou and Cooney \cite{alexiou2024steady}. We see that the analytical results for two-strategy competition with the group-level victory probability of Boyd and coauthors  \cite{boyd2003evolution} feature results similar to the stochastic model in terms of the dependence of the collective outcome achieved on the costs and strengths of punishment. Furthermore,} both the analytical and numerical results for pairwise competition based on group-level payoff display some surprising dynamical behaviors, such as the possibility that increasing the cost of punishing defectors can increase the level of altruistic punishment at steady state (albeit at a lower average payoff) as well as a non-monotonic dependence of long-time average payoff on the strength of punishment of defection (with the increase of punishment strength sometimes producing decreasing collective payoff at intermediate levels of punishment before facilitating greater collective payoff when punishment is sufficiently strong). {In addition, we perform numerical simulations of the  multilevel dynamics for groups composed of defectors, cooperators, and altruistic punishers, showing that the collective outcome achieved by the population correponds well with the analytical predictions for the collective outcome for two-strategy multilevel dynamics for the cases of group-level victory probability based on the difference in cooperative individuals (as assumed by Boyd and coauthros \cite{boyd2003evolution}) or on a normalized difference in the average payoff of competing groups.}  %

\subsection{Outline of the Paper}

{The remaining sections of the paper are organized as follows.} In Section \ref{sec:gametheoryindividual}, we formulate our model for altruistic punishment by defining the payoffs received by cooperators, defectors, and altruistic punishers, and we also review results for the individual-level replicator dynamics in the presence of altruistic punishment. In Section \ref{sec:PDEmodels}, we present PDE models for multilevel selection featuring within-group altruistic punishment and a pairwise model of between-group competition inspired by the work of Boyd and coauthors \citep{boyd2003evolution}. {We then recall in Section \ref{sec:existingresults} existing analytical results for PDE models of multilevel selection in two-strategy games based on either the case of frequency-independent group-level competition from Cooney and Mori \cite{cooney2022long} or the case of pairwise group-level competition from Alexiou and Cooney \cite{alexiou2024steady}}. In Section \ref{sec:analytical}, we present analytical results for two-strategy multilevel competition for the case of additively separable pairwise group-level victory probabilities, {and we study the more general class of models pairwise group-level competition defector-punisher groups in Section \ref{sec:nonlineargroup}}.  %
In Section \ref{sec:trimorphicnumerics}, {we consider the case of three-strategy dynamics by exploring} numerical simulations of multilevel selection with additively separable group-level victory probabilities in groups featuring cooperators, defectors, and altruistic punishers, providing comparisons with the analytical results from Section \ref{sec:analytical} {for the two-strategy case}. Finally, in Section \ref{sec:discussion}, we provide a discussion of our results {and perspectives on future work related to PDE models of multilevel selection with frequency-dependent group-level competition and applying such models to explore the evolution of cooperation via cultural evolution}. 

The appendix also features some additional analysis and derivations of our PDE models and numerical schemes. {In Section \ref{sec:PDEderivation}, we show how to adapt the approach introduced by Luo and coauthors \citep{luo2014unifying,van2014simple} to present heuristic derivations of our main PDE models starting from a stochastic model of multilevel selection with pairwise between-group competition, illustrating the derivation for the cases of multilevel competition featuring groups with either two or three strategies.} In Section \ref{sec:densityderivation}, we provide the full derivations for the forms of the steady state densities achieved via multilevel selection for scenarios in which pairwise between-group competition can be reduced to a two-level replicator equation. We also provide comparisons of our models with within-group altruistic punishment and pairwise between-group competition with other recent work on existing PDE models of multilevel selection, exploring the dynamics of a two-level replicator equation model to see how altruistic punishment can work in concert with frequency-independent group-level competition to promote cooperative behavior (Section \ref{sec:LMaltruisticpunishment}) and showing the effect of pairwise between-group competition when paired with the within-group mechanism of indirect reciprocity (Section \ref{sec:indirectreciprocity}). We discuss the schemes for numerically solving our PDE models in Section \ref{sec:numericalderivation}, presenting a derivation of an upwind finite-volume scheme to describe the effects of pairwise group-level conflicts and adapting existing approaches to describe finite volume models for trimorphic multilevel dynamics.

\section{Baseline Game-Theoretic Model and the Evolution of Altruistic Punishment via Individual-Level Selection}
\label{sec:gametheoryindividual}

In this section, we formulate our baseline model for cooperative behavior with altruistic punishment. Following the approach used by  Boyd and coauthors \cite{boyd2003evolution}, we consider a game-theoretic interaction consisting of a donation game, and then allow individuals who cooperate in the donation game to have the option to participate in altruistic punishment by paying a cost to confer a punishment upon defectors. In Section \ref{sec:trimorphicpayoffs}, we formulate the payoffs received by cooperators, defectors, and altruistic punishers under a generalized version of the model of Boyd and coauthors, and then we formulate replicator equations describing individual-level selection and calculate the average payoff a group in the presence of all three strategies. In Section \ref{sec:twostrategypayoffs}, we restrict attention to the calculation of individual and collective payoff in groups that feature only defectors and altruistic punishers, preparing us to study in subsequent sections the dynamics of multilevel selection for groups featuring two strategies. 

\subsection{Formulation of Three-Strategy Game with Cooperators, Defectors, and Altruistic Punishers}
\label{sec:trimorphicpayoffs}
We will consider groups composed of individuals that can play one of three strategies: cooperation ($C$), defection ($D$), and altruistic punishment ($P$). In each interaction between pairs of individuals, cooperators pay a cost $c$ to confer a benefit $b$ to their opponent, while defectors pay no cost and confer no benefits to interaction partners. Altruistic punishers also pay a cost $c$ to confer a benefit $b$ in each interaction, but also pay additional costs to confer a punishment $p$ when they interact with a defector. We will consider two ways in which altruistic punishers can pay a cost to confer a punishment on defectors: punishers can pay a fixed cost $k$ for each interaction with a defector and punishers can pay a single fixed cost $q$ to confer punishment on defectors (regardless of whether they interact with any defectors). The per-interaction punishment cost $k$ represents a cost that individuals pay to directly punish an individual who has defected against them in the donation game, while the fixed cost $q$ models the payment of a cost to maintain the ability to punish defectors. {Boyd and coauthorsconsider separate simulations featuring only fixed costs or per-interaction costs of punishment  \cite{boyd2003evolution} , but other models of altruistic punishment have considered the combined effects of both forms of costs \citep{gavrilets2017collective}. In particular,  Gavrilets and Richerson use a fixed cost to describe an investment in monitoring cooperative behavior and identifying defectors in the game-theoretic interaction, while a per-interaction cost is used to model the cost of imposing a peer punishment when an altruistic punisher interacts with and identifies a defector.} 

We now consider a group of individuals with a composition featuring a fraction $x$ of pure cooperators, a fraction $y$ of defectors, and a fraction $z = 1 - x -y$ of altruistic punishers. If individuals play the game against all other members of a large group with this strategic composition, we can use the assumptions for costs, benefits, and punishments described above to see that the average payoffs obtained received by defectors, cooperators, and altruistic punishers in such a group are given by
\begin{subequations} \label{eq:trimorphicpayoffs}
\begin{align}
    \pi_C(x,y) &= b \left( 1 - y\right) - c \\
    \pi_D(x,y) &= b \left( 1 - y\right) - p \left( 1 - x - y\right)\\
    \pi_P(x,y) &= b \left( 1 - y \right) - c - q - k y
\end{align}
\end{subequations}

To describe the evolutionary dynamics due to individual-level selection within groups, we may use a system of ODEs known as the replicator dynamics to study how the frequencies of strategies change within the group due to differences in payoffs between individuals. For a group featuring cooperators, defectors, and altruistic punishers, the replicator dynamics for individual-level selection tells us that the fractions $x$, $y$, $z$ of cooperators, altruistic punishers, and defectors change according to the system of ODEs
\begin{subequations}
\begin{align}
\dsddt{x} &= x \left[ \pi_C(x,y) - \left( x \pi_C(x,y) + y \pi_D(x,y) + z \pi_P(x,y) \right) \right] \\
\dsddt{y} &= y \left[ \pi_D(x,y) - \left( x \pi_C(x,y) + y \pi_D(x,y) + z \pi_P(x,y) \right) \right] \\
\dsddt{z} &= z \left[ \pi_P(x,y) - \left( x \pi_C(x,y) + y \pi_D(x,y) + z \pi_{P}(x,y) \right)\right].
\end{align}
\end{subequations}
The biological assumption or interpretation of the replicator equation is that strategies increase in frequency in a group when the payoff of individuals using that strategy exceeds the average payoff received by all members of the group. 

Using the constraint $x + y + z = 1$ that the fractions of strategies add up to $1$, we can use that fact that the fraction of defectors is given by $z = 1 - x -y$ to reduce our replicator equation for the three-strategy dynamics down to the following system of two ODEs
\begin{subequations} \label{eq:replicatortrimorphicxy}
\begin{align}
\dsddt{x} &= x \left[ (1-x) \left( \pi_C(x,y) - \pi_P(x,y) \right) - y \left( \pi_D(x,y) - \pi_P(x,y) \right) \right] \\
\dsddt{y} &= y \left[ (1-y) \left( \pi_D(x,y) - \pi_P(x,y) \right) - x \left(\pi_C(x,y)  - \pi_P(x,y) \right) \right].
\end{align}
\end{subequations}

Using the payoffs defined in Equation \eqref{eq:trimorphicpayoffs}, we can then characterize how the stability of equilibria for the replicator dynamics of Equation \eqref{eq:replicatortrimorphicxy} depend on the strength of punishment $p$ and the fixed and per-interaction costs $q$ and $k$ for punishing defectors. In Proposition \ref{prop:withintrimorphic}, we characterize the equilibrium behavior for the replicator dynamics for the cases of both per-interaction and fixed costs of punishment. In the case of fixed punishment costs, the behavior seen here is analogous to the behavior of the model of costly punishment by \cite[Proposition 4]{sethi1996evolution} for the case of an underlying game-theoretic interaction corresponding to a donation game rather than a common-pool resource game. 

\begin{proposition} \label{prop:withintrimorphic}
For the case in which altruistic punishment features only per-interaction punishment costs ($p > 0$, $q = 0$), we have the following equilibrium behavior:
\begin{itemize}
    \item The all-defector state $(0,1)$ is an equilbrium and is asymptotically stable for all payoff parameters
    \item The punisher-cooperator edge of the simplex $\Delta^2_{PC} := \{ (x,0) : x \in [0,1]\}$ is an interval of equilibria for all payoff parameters (as $\pi_C(x,0) = \pi_D(x,0)$ when $q = 0$, yielding neutral competition between cooperators and altruistic punishers in the absence of defectors).
    
    A point $(x,0) \in \Delta^2_{PC}$ is neutrally stable if $x \leq \frac{p-c}{p} \leq 1$ (in which case $\pi_D(x,0) \leq x \pi_C(x,0) + (1-x) \pi_P(x,0)$ and defectors do not receive a payoff exceeding the average payoff at the equilibrium point). Such neutrally stable equilibria can only exist if $p > c$ (and the punishment for defecting exceeds the cost of cooperating in the donation game).
    
    A point $(x,0) \in \Delta^2_{PC}$ is unstable if $x > \frac{p-c}{c}$ (in which case  defectors outcompete the average individual at the equilibrium composition). All points $(x,0) \in \Delta^2_{PC}$ satisfy this condition when $p < c$ (and the punishment for defection does not outweigh the cost of cooperating).

    \item {There is an equilibrium point $\left(0, 1 - \frac{c}{p} \right)$ on the defector-punisher edge of the simplex if $p > c$, and this equilibrium is an unstable node whenever it exists. }
\end{itemize}

For the case in which altruistic punishment features only fixed punishment costs (corresponding to $q > 0$ and $k = 0$), the replicator dynamics of Equation \eqref{eq:replicatortrimorphicxy} have the following equilibrium behavior. 

\begin{itemize}
    \item The all-defector state $(0,1)$ is an asymptotically stable equilibrium for all payoff parameters.
    \item The all-cooperator state $(1,0)$ is an unstable equilibrium for all payoff parameters.
    \item The all-punisher state $(0,0)$ is a saddle equilibrium when $p > c + q$ (the punishment for defection exceeds the costs a punisher incurs to both punish defectors and confer benefits to their opponents in their game-theoretic interactions). In this case, the all-punisher state is stable with respect to perturbations along the defector-punisher edge of the simplex, but will be unstable to invasion by cooperators. The all-punisher equilibrium will be an unstable node when $p < c + q$. 
    \item There is an equilibrium point {$(x,y) = \left( 0, 1 - \frac{c + q}{p}\right)$} on the interior of the defector-punisher edge of the simplex if $p > c + q$, {and this equilibrium is an unstable node whenever it exists.} 
\end{itemize}
\end{proposition}

We can illustrate the equilibrium behavior of the replicator dynamics for the cases of per-interaction and fixed punishment by plotting the vector fields for Equation \eqref{eq:replicatortrimorphicxy}. In Figure \ref{fig:withintrimorphic}, we consider vector fields for both the case of per-interaction punishment costs $k > 0$ (top panels) and fixed punishment costs $q > 0$ (bottom panels), exploring the case of a weaker punishment $p = 0.5$ (left panels) and a stronger punishment $p = 2.5$. For the case of weaker punishment, we see that the within-group dynamics favor convergence to the all-defector state for all initial conditions in which any fraction of defectors are initially present. For the case of {stronger} punishments {with} $p = 2.5$, we see that there is multistability between the all-defector equilibrium and a line of equilibria on the cooperator-punisher edge of the simplex in the case of per-interaction punishment rates (Figure \ref{fig:withintrimorphic}, bottom-right). We see from Figure \ref{fig:withintrimorphic}(bottom-right) that, for fixed punishment costs $q > 0$ and strong punishment $p = 2.5$, the all-defector {equilibrium} still attracts all initial conditions from the interior of the simplex, but that the all-punisher equilibrium does attract initial conditions on the defector-punisher edge of the simplex provided that there is initially a sufficiently large cohort of altruistic punishers. 

\begin{figure}[!htb]
    \centering
    \includegraphics[width = 0.48\textwidth]{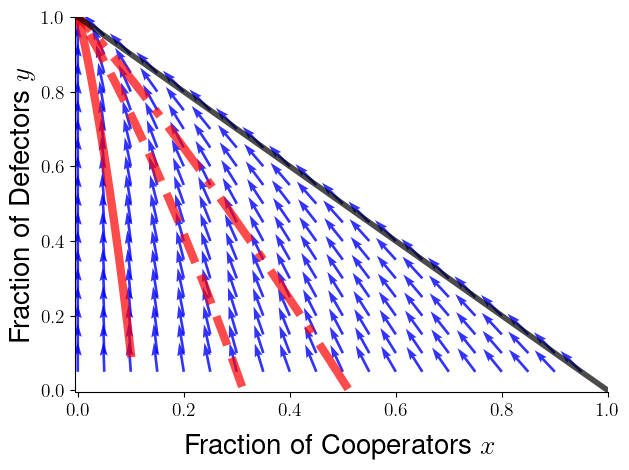}
    \includegraphics[width = 0.48\textwidth]{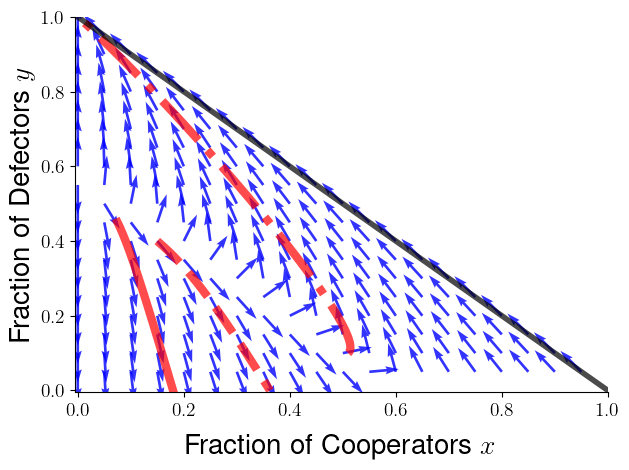}
       \includegraphics[width = 0.48\textwidth]{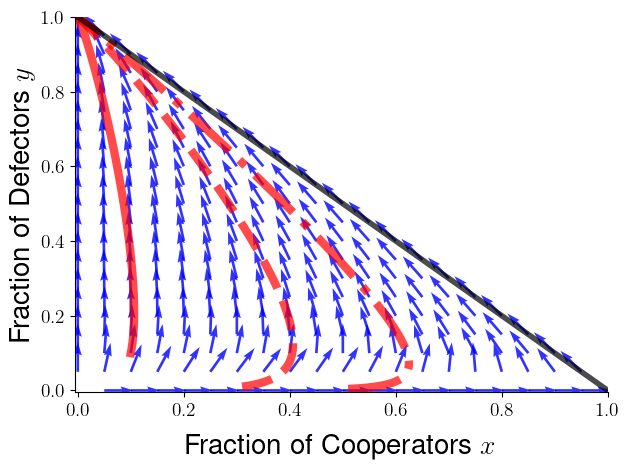}
    \includegraphics[width = 0.48\textwidth]{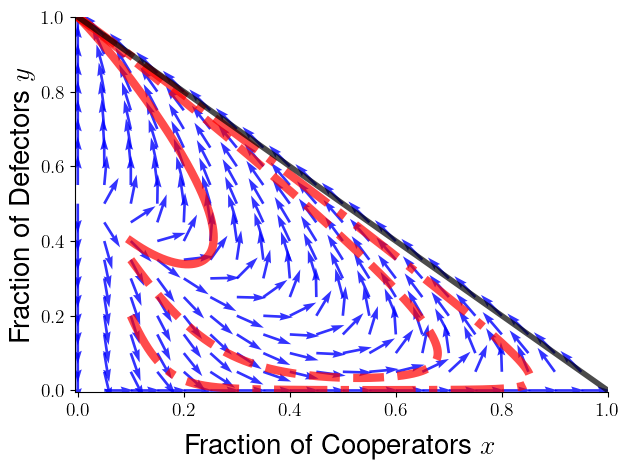}
    \caption{Example vector fields for the replicator dynamics for individual-level selection in model of altruistic punishment with either per-interaction or fixed cost of punishment. Vector fields for per-interaction punishment case are presented for $k = 0.2$ and $q = 0.0$ for punishment strengths $p = 0.5$ (top-left) and $p = 2.5$ (top-right). Vector fields for fixed punishment costs are presented for $q = 0.2$ and $k = 0$ for punishment costs $p = 0.5$ (bottom-left) and $p = 2.5$ (bottom-right). For each scenario, we present trajectories to the replicator dynamics for three initial conditions, plotted as red solid, dashed, and dash-dotted curves. Payoff parameters were fixed at $b = 2$ and $c=1$, and trajectories from numerical solutions were solved over {$8,000$ time-steps with step-size of $\Delta t = 0.01$.}}
    \label{fig:withintrimorphic}
\end{figure}

\subsubsection{Average Payoff of Group Members for Three-Strategy Game}

In addition to studying the dynamics of individual-level replication competition, we can also ask how the composition of a group impacts the average payoff of group members. For groups that can feature cooperators, altruistic punishers, and defectors, we see that the average payoff for a group with composition $(x,y)$ is defined as
\begin{equation}
G(x,y) = x \pi_C(x,y) + y \pi_D(x,y) + (1-x-y) \pi_P(x,y),
\end{equation}
and we can use the expressions for payoffs in Equation \eqref{eq:trimorphicpayoffs} to write this average payoff as
\begin{equation}
G(x,y) = b - c - q + q x - \left( b - c - q+ p + k \right) y + \left( p + k \right) x y + \left( p + k \right) y^2.
\end{equation}
We note that the average payoffs for groups combined only of cooperators, only of defectors, and only of altruistic punishers are $G(1,0) = b - c$, $G(0,1) = 0$ and $G(0,0) = b - c - q$. Therefore we see that the all-cooperator and all-punisher groups have equal average payoff in the case of per-interaction costs of punishment (when $k > 0$ and $q = 0$), but that the all-cooperator group will have a higher average payoff than the all-punisher group when there is a fixed cost of punishing defectors ($q > 0$ and $k = 0$) regardless of the strategic composition of the group. The all-cooperator group will always feature higher average payoff than the all-defector group whenever the benefit of cooperation exceeds the cost of cooperation ($b > c$), while the all-punisher group has a higher average payoff than the all-defector group when the benefit of cooperation exceeds the sum of the cost of cooperation and any fixed cost of punishing defectors ($b > c + q$).

\subsection{Two-Strategy Individual-Level Competition Between Defectors and Altruistic Punishers}
\label{sec:twostrategypayoffs}

For the case where we restrict the strategy space to defection and altruistic punishment, we set $x = 0$, and can consider the state of the group as just the fraction $z$ of altruistic punishers and the fraction $1-z$ of defectors. In this case, we can plug $x = 0$ and $y = 1-z$ into our expressions from Equation \eqref{eq:trimorphicpayoffs} to see that the payoffs for altruistic punishers and defectors are given by
\begin{subequations} \label{eq:payoffsDPedge}
\begin{align}
    \pi_P(z) &= \left(b + k \right) z - \left( c + q + k\right) \\
    \pi_D(z) &= \left( b - p \right) z.
\end{align}
\end{subequations}

We can also describe replicator dynamics for individual-level selection in groups featuring only altruistic punishers and defectors. In this case, the fraction of altruistic punishers changes according to the ODE
\begin{equation}
\dsddt{z} = z (1-z) \left[ \pi_P(z) - \pi_D(z)\right].
\end{equation}
We note that the all-punisher equilibrium becomes locally stable when we choose punishment parameters $k$, $p$, and $q$ such that
\begin{equation}
\pi_P(1) > \pi_D(1),
\end{equation}
so a group of altruistic punishers can withstand invasion from a small number of defectors when the payoff parameters satisfy
\begin{equation}
b - c - q > b - p \Longrightarrow p > c + q.
\end{equation}
In the case of fixed punishment costs, this means that the all-punisher equilibrium will be stable when the punishment received by a defector exceeds the sum of the cost $c$ required for a punisher to cooperate in the donation game and the cost $q$ of punishing defectors. In the case of per-interaction punishment costs (with $k > 0$ and $q = 0$), this condition reduces to $p > c$, so punishers can withstand invasion by defectors once the punishment for defection exceeds the cost to cooperate in the game.

\subsubsection{Average Payoff of Group Members}

For evolutionary competition in groups featuring only altruistic punishers and defectors, we can define the average payoff of group members in a $z$-punisher group as
\begin{equation}
G(z) = z \pi_P(z) + (1-z) \pi_D(z).
\end{equation}
We can then use the payoffs given in Equation \eqref{eq:payoffsDPedge} to write this average payoff as
\begin{equation}
G(z) = \left( b - \left[ c + q + p + k \right] \right) z + (p+k) z^2. 
\end{equation}
Here we see that the payoffs of the all-punisher and all-defector groups are given by $G(1) = b - c - q$ and $G(0) = 0$, so $G(1) > G(0)$ if the benefit generated by cooperation exceeds the sum of the cost of cooperation and any fixed cost for punishing defectors. 

\begin{remark}
We note that the individual-level advantage of defectors $\pi_D(1) - \pi_D(1) = c + q - p$ is an all-punisher group and the average payoffs $G(1)$ and $G(0)$ of the all-punisher and all-defector groups do not depend on per-interaction costs $k$. This will have implications for the role played by per-interaction punishment costs on achieving altruistic punishment via multilevel selection in a two-level replicator equation model. We will revisit this  point in Section \ref{sec:LMaltruisticpunishment} of the appendix when we explore the role of altruistic punishment in two-level replicator equation models for multilevel selection. 
\end{remark}

\section{PDE Models For Multilevel Selection with Pairwise Between-Group Competition} \label{sec:PDEmodels}

{We will now} formulate PDE models for multilevel selection when group-level competition takes place through pairwise conflicts between groups.  %
We first present a formulation of a PDE model for multilevel selection in groups that may feature only defectors and altruistic punishers in Section \ref{sec:PDEmodelDP}, and then we present a similar model for describing trimorphic multilevel competition for groups that may contain cooperators, defectors, and altruistic punishers in Section \ref{sec:PDEmodeltrimorphic}.

\subsection{PDE Model for Multilevel Dynamics Featuring Defectors and Altruistic Punishers}
\label{sec:PDEmodelDP}

To formulate models for multilevel selection, we modify the approach introduced by Luo and coauthors to describe the impacts of strategic competition using a two-level birth-death process driven by within-group and between-group competition \citep{luo2014unifying,van2014simple,luo2017scaling}.  For a group-structured population with $m$ groups and $n$ members for group, we calculate average payoffs $\pi_P(\frac{i}{n})$ and $\pi_D(\frac{i}{n})$ received by altruistic punishers and defectors in a group composed of $i$ altruistic punishers $n-i$ defectors. We assume that individual altruistic punishers reproduce at rate $1 + w_I \pi_P^n(\frac{i}{n})$ and individual defectors reproduce at rates $1 + w_I \pi_D(\frac{i}{n})$, where 1 is a background birth rate independent of payoff and $w_I$ describes the sensitivity of individual-level replication rates 
to the payoffs received by individuals. We model between-group competition by assuming that a group engages in pairwise between-group conflict at a rate $\Lambda$, and we assume that a group is matched up for a pairwise conflict with an opponent group that is randomly sampled from the population of groups. We assume that an $i$-punisher group defeats a $j$-punisher group with probability $\rho\left( \frac{i}{n},\frac{j}{n} \right)$. The group that wins the pairwise conflict produces a copy of itself, and this offspring group replaces the group that was defeated in pairwise conflict. 

In the limit of infinite group and infinitely many members per group ($m,n \to \infty$), we can describe the composition of the group-structured population by the probability density $f(t,z)$ of having groups with fraction $z$ altruistic punishers and fraction $1-z$ defectors at time $t$. We show in Section \ref{sec:PDEderivation} that, in this limit of infinite group size and infinite number of groups, the individual-level and group-level competition will allow the strategic composition $f(t,z)$ of the group-structured population to be described by the following partial differential equation 
\begin{equation} \label{eq:multilevelPDErhodiff}
    \dsdel{f(t,z)}{t} = - \dsdel{}{x} \left[z (1-z) \left( \pi_P(z) - \pi_D(z) \right) f(t,z) \right] + \lambda f(t,z) \left[  \int_0^1 \left\{ \rho(z,u) - \rho(u,z) \right\} f(t,u) du   \right].
\end{equation}
Here, the parameter $\lambda := \frac{\Lambda}{w_I}$ describes the relative strength of pairwise between-group competition. 

Because $\rho(z,u)$ describes the probability the an $z$-punisher group defeats a $u$-punisher group in a pairwise conflict, we have that the probability that a $u$-punisher group wins such a conflict is $\rho(u,z) = 1 - \rho(z,u)$. Using this observation, we may rewrite our PDE in the following form
\begin{equation} \label{eq:multilevelPDEtworho}
    \dsdel{f(t,z)}{t} = - \dsdel{}{z} \left[z (1-z) \left( \pi_P(z) - \pi_D(z) \right) f(t,z) \right] + \lambda f(t,z) \left[ 2 \int_0^1 \rho(z,u) f(t,u) du - 1  \right]
\end{equation}

\begin{remark} \label{lem:replicatorreduced}
{As a point of comparison with the current model of multilevel selection with pairwise group-level selection, we recall the following PDE model for multilevel selection in evolutionary games that has been studied in previous work} 
\begin{equation} \label{eq:twolevelreplicator}
\dsdel{f(t,z)}{t} = -\dsdel{}{z} \left[ z(1-z) \left(\pi_P(z) - \pi_D(z) \right) f(t,z) \right] + \lambda f(t,z) \left[G(z) - \int_0^1 G(u) f(t,u) du \right],
\end{equation}
{which we refer to as a two-level replicator equation \citep{cooney2022long}. This PDE model was derived under identical assumptions on the rule for individual-level replication and the assumption that groups produce copies of themselves at a rate $\Lambda G(x)$ that depend only on the composition $x$ of the replicating group, with the offspring group replacing a randomly chosen group from the population \citep{cooney2019replicator,cooney2022long}.}

Unlike the model featuring pairwise between-group competition, the term $\lambda G(z) f(t,z)$ {in Equation \eqref{eq:twolevelreplicator}} describing group-level replication in the two-level replicator equation is linear in the density $f(t,x)$. As a result, it is possible to study the dynamics of a two-level replicator equation by considering solutions to an associated linear PDE
\begin{equation} \label{eq:twolevellinear}
\dsdel{g(t,z)}{t} = - \dsdel{}{z} \left[ z(1-z) \left(\pi_P(z) - \pi_D(z) \right) g(t,z) \right] + \lambda G(z) g(t,z),
\end{equation}
which only considers the effect of within-group birth-death dynamics and group-level birth events, ignoring the possibility of group-level death events. A solution $f(t,z)$ to the two-level replicator equation can be obtained by normalizing a solution to the associated linear equation, as the function
\begin{equation}
f(t,z) = \frac{g(t,z)}{\int_0^1 g(t,z) dz}
\end{equation}
will be a solution to the full nonlinear multilevel dynamics of Equation \eqref{eq:twolevelreplicator} if $g(t,z)$ is a solution to the associated linear PDE of Equation \eqref{eq:twolevellinear}. 

Unlike the two-level replicator equation, the PDE model from Equation \eqref{eq:multilevelPDEtworho} for multilevel selection with pairwise between-group competition cannot be reduced to an associated linear equation describing a growing population of groups. Instead, the group-level reproduction term is itself nonlinear, which means that incorporating group-level frequency dependence imposes additional mathematical challenges in describing the long-time behavior for multilevel selection with pairwise group-level competition. 
\end{remark}

Using the probability of group-level victory 
\begin{equation}
 \rho(z,u) = \frac{1}{2} + \frac{1}{2}\left(z-u \right)    
\end{equation}
considered by Boyd and coauthors in the special case of groups composed entirely of defectors and altruistic punishers \cite{boyd2003evolution} , we can see that the dynamics of Equation \eqref{eq:multilevelPDEtworho} reduce to the following PDE
\begin{equation}
 \dsdel{f(t,z)}{t} = - \dsdel{}{z} \left[z (1-z) \left( \pi_P(z) - \pi_D(z) \right) f(t,z) \right] + \lambda f(t,z) \left[ z - \int_0^1 u f(t,u) du  \right],
\end{equation}
with between-group competition resembling that of the Luo-Mattingly model for multilevel selection when group-level reproduction is proportional to the fraction of cooperators in the group \citep{luo2014unifying,luo2017scaling,van2014simple}. {We can also consider between-group competition based on differences in average payoff between groups, which we can model through the group-level victory probability given by}\begin{equation} \label{eq:grouplocalmaxmin}
\rho(x,u) = \frac{1}{2} \left[ 1 + \frac{G(z) - G(u)}{G^* - G_*},\right], \end{equation} 
where $G^* = \max_{z \in [0,1]}G(z)$ and $G_* = \min_{z \in [0,1]} G(z)$ {describe the maximum and minimum possible average payoffs of groups}. For this choice of group-level probability of group-level victory, the multilevel dynamics of Equation \eqref{eq:multilevelPDEtworho} can be written in the form
\begin{equation}
 \dsdel{f(t,z)}{t} = - \dsdel{}{z} \left[z (1-z) \left( \pi_P(z) - \pi_D(z) \right) f(t,z) \right] + \left(\frac{\lambda}{G^* - G_*}\right) f(t,z) \left[ G(z) - \int_0^1 G(u) f(t,u) du  \right],
\end{equation}
so the dynamics take the form of a generalized two-level replicator equation in this case \citep{cooney2019replicator,cooney2022long}.

More broadly, we can consider a class of additively separable group-level victory probabilities of the form
\begin{equation}
\rho(z,u) = \frac{1}{2} + \frac{1}{2} \left[ \mc{G}(z) - \mc{G}(u) \right]
\end{equation}
where $\mc{G}(z) \in C^1([0,1])$ and $\max_{z \in [0,1]} |\mc{G}(z)| \leq 1$. By plugging this form of $\rho(x,u)$ into Equation \eqref{eq:multilevelPDEtworho}, we see that our PDE model for multilevel dynamics with pairwise group-level competition can be written in the following form
\begin{equation} \label{eq:PDEadditivelyseparable}
 \dsdel{f(t,z)}{t} = - \dsdel{}{z} \left[z (1-z) \left( \pi_P(z) - \pi_D(z) \right) f(t,z) \right] + \lambda f(t,x) \left[\mc{G}(z) - \int_0^1 \mc{G}(u) f(t,u) du \right].
\end{equation}
This takes the form of a two-level replicator equation, and we will present the long-time behavior for solutions of this PDE in Section \ref{sec:existingresults}.

We may also consider families of group-level victory probabilities that are not additively separable. One example of such a function takes the form
\begin{equation} \label{eq:groupFermi}
\rho(z,u) = \frac{1}{2} \left[ 1 +  \tanh\left( s \left[ G(z) - G(u)\right]\right) \right],
\end{equation}
where $s$ is a non-negative parameter governing the sensitivity of group-level victory probability to the difference in average payoffs of the two competing groups. This group-level victory probability is inspired by the Fermi update rule introduced to describe noisy social learning for individual-level selection in evolutionary games \citep{traulsen2005coevolutionary}. Another group-level victory probability we may consider is 
\begin{equation} \label{eq:grouppairwiselocal}
\rho(z,u) = \frac{1}{2} \left[ 1 + \frac{G(z) - G(u)}{|G(z)| + |G(u)|} \right],
\end{equation}
which is the group-level generalization of an alternative formulation of the local / pairwise update rule that is normalized by the payoffs of the two groups engaged in a pairwise conflict \citep{morgan2003pairwise,schluter2016robustness,tavoni2012survival}. The distinction between these two forms of the group-level local update rule presented in Equations \eqref{eq:grouplocalmaxmin} and \eqref{eq:grouppairwiselocal} is that the form normalizing by the maximum possible group-payoff difference $G^* - G_*$ is additively separable, while the normalization by the absolute value of payoffs $|G(z)| + |G(u)|$ prevents us from reducing our PDE model to the form of Equation \eqref{eq:PDEadditivelyseparable}.

Finally,  we consider a group-level victory probability of the form
\begin{equation} \label{eq:groupvictoryTullock}
\rho(z,u) =  \frac{\left( G(z) - G_* \right)^{1/a}}{\left( G(z) - G_* \right)^{1/a} + \left( G(u) - G_* \right)^{1/a}},
\end{equation}
where $a > 0$ measures the relative sensitivity of differences in average payoff of groups to the probability of group-level victory,  which is a function form motivated by existing work in the literature on biological and cultural evolution.  Note that we consider the probability of group-level victory depending on the difference $G(\cdot) - G_{*}$ between average group payoff $G(\cdot)$ and the the minimum possible group payoff $G_{*}$ so that we can map negative average payoff $G(\cdot)$ into our model for the probability of winning a pairwise group-level competition. 
\begin{remark}
This form of group-level competition is motivated by the contest competition function introduced in the economics literature by Tullock \cite{tullock2008efficient}, and has been previously applied to model intergroup competition in animal populations using a variety of frameworks for multilevel selection \citep{reeve2007emergence,baik2008contests,crowley2010variable,gavrilets2015collective,rusch2020logic,tverskoi2021dynamics}. With the history of using contest functions of this form to describe group-level competition using other modeling frameworks, it is natural to ask how this form of victory probability for group-level conflict could impact the long-time behavior in a PDE model of selection at two levels. 
\end{remark}

\subsection{Three-Strategy Models of Multilevel Selection with Pairwise Between-Group Competition}
\label{sec:PDEmodeltrimorphic}

We can describe the strategic composition of a group-structured population featuring cooperators, defectors, and altruistic punishers using a probability density $f(t,x,y)${, where the set of possible strategic compositions of groups with fraction $x$ cooperators, fraction $y$ defectors, and fraction $z = 1 -x - y$ is described the points $(x,y)$ on the two-dimensional simplex }
{
\begin{equation}
\Delta^2 := \left\{(x,y) : x \geq 0, y \geq 0, x + y \leq 1 \right\}.
\end{equation}
}
By considering a probability of victory of a group $\rho(x,y;u,v)$ with composition $(x,y)$ over a group with composition $(u,v)$, we can describe multilevel selection in three-strategy groups using the PDE
\begin{equation} \label{eq:PDEtrimorphicrho}
\begin{aligned}
\dsdel{f(t,x,y)}{t} &= - \dsdel{}{x} \left[ x \left\{ (1-x)  \left(\pi_C(x,y) - \pi_P(x,y) \right) - y \left( \pi_D(x,y) - \pi_P(x,y) \right) \right\} f(t,x,y) \right] \\
&-   \dsdel{}{y} \left[ y \left\{ (1-y)  \left(\pi_D(x,y) - \pi_P(x,y) \right) - x \left( \pi_C(x,y) - \pi_P(x,y) \right) \right\} f(t,x,y) \right]  \\
&+ \lambda f(t,x,y) \left[ \int_0^1 \int_{0}^{1-u} \left\{ \rho(x,y;u,v) - \rho(u,v;x,y) \right\} f(t,u,v) dv du \right].
\end{aligned}
\end{equation}
A heuristic derivation of this PDE is provided in Section \ref{sec:PDEderivation} of the appendix, which is based on a similar derivation for a three-strategy PDE model for multilevel selection in a population of protocells \citep{cooney2022pde}. 

As with the case of two strategies, we can choose different forms of the group-level victory probability $\rho(x,y;u,v)$ in order to model different assumptions about competition between groups. The victory probability of the form
\begin{equation}
\rho(x,y;u,v) = { \frac{1}{2} + \frac{\left(1 - y \right) - \left( 1 - v \right)}{2}= \frac{1}{2} + \frac{v - y}{2}}
\end{equation}
captures the assumption used by Boyd and coauthors in their simulation models for multilevel selection and the evolution of altruistic punishment, making the assumption that the victory probability of groups is determined by the fraction of individuals displaying cooperative behavior in the game (i.e. the fraction of non-defectors $1-y$ given by the sum of the fraction of pure cooperators $x$ and the fraction altruistic punishers $1 - x - y$). Other group-level victory probabilities we could consider include the generalization of the normalized pairwise group-level update rule

\begin{equation}
\rho(x,y;u,v) = \frac{1}{2} + \frac{1}{2} \left(\frac{G(x,y) - G(u,v)}{G^*(\Delta^2) - G_*(\Delta^2)}\right) 
\end{equation}
where $G^*(\Delta^2) = \max_{(x,y) \in \Delta^2}$ and $G_*(\Delta^2) = \min_{(x,y) \in \Delta^2}$, as well as the three-strategy analogue of the group-level Fermi update rule
\begin{equation}
\rho(x,y;u,v) = \frac{1}{2} + \frac{1}{2} \tanh\left( s  \left[G(x,y) - G(u,v) \right] \right).
\end{equation}
These group-level victory probabilities allow us describe relative chances of group-level victory as a function of the difference in average payoffs between the two groups engaged in a pairwise conflict.

{\section{Summary of Existing Results on Long-Time Behavior for PDE Models of Multilevel Selection} \label{sec:existingresults}

{Before studying the dynamics of multilevel selection with within-group altruistic punishment, we will first recall existing general results for PDE models of multilevel selection. We start by summarizing the results of Cooney and Mori \cite{cooney2022long} on the long-time behavior of two-level replicator equations that can be applied to study pairwise group-level competition in the special case of additively separable group-level victory probabilities (Section \ref{sec:longtimeadditive}, and then we proceed to summarize preliminary analytical predictions derived by Alexiou and Cooney \cite{alexiou2024steady} for pairwise group-level competition for a more general class of group-level victory probabilities (Section \ref{sec:pairwiseresults}).}

\subsection{Results for Two-Level Replicator Equations Corresponding to Additive Group-Level Victory Probabilities}
\label{sec:longtimeadditive}
}

In this section, we describe existing results for a class of PDE models of multilevel selection taking the form of a two-level replicator equation, with group-level competition that can be described using a net collective reproduction rate $\mc{G}(\cdot)$ that depends only on the composition of the reproducing group. As we see in Equation \eqref{eq:PDEadditivelyseparable}, our models for pairwise between-group competition with additively separable group-level victory probabilities $\rho(z,u)$ can be rewritten as a special case of PDE models in this class of two-level replicator equations. We consider a nonlinear, hyperbolic PDE of the form 

\begin{equation} \label{eq:twotypeLM}
 \dsdel{f(t,z)}{t} = \dsdel{}{x} \left[z (1-z) {\Pi(z)} f(t,z) \right] + \lambda f(t,z) \left[ \mc{G}(z) - \int_0^1 \mc{G}(w) f(t,w) dw \right],
\end{equation}
where $\Pi(z)$ describes the relative advantage of the altruistic punisher strategy and $\mc{G}(z)$ models the relative group-level replication rate for groups featuring a fraction $z$ of altruistic punishers. Here, $\mc{G}(z)$ is a continuously differentiable function, which can correspond to $\mc{G}(z) = z$ for the case of {group-level victory probability determined by the difference in} fraction of cooperative individuals and corresponds to $\mc{G}(z) = \frac{G(z)}{G^* - G_*}$ for the case in which probability of group victory depends on the average payoff of group members. $\Pi(z)$ is a continuously differentiable function that describes the individual-level advantage for defectors over altruistic punishers, and is given by $\Pi(z) := \pi_D(z) - \pi_P(z)$ for the examples discussed in Section \ref{sec:PDEmodels}.

The long-time behavior of solutions to Equation \eqref{eq:twotypeLM} has been characterized in previous work {for a broad class of} continuously differentiable $\mc{G}(z)$ and $\Pi(z)$ \citep{cooney2022long}, so we can apply existing results to study how the models we consider for altruistic punishment can work in concert with multilevel selection to achieve cooperation within social groups. For the purposes of our current motivating problem of studying altruistic punishment within groups, we can focus on previously studied cases in which either $\Pi(z) > 0$ for $z \in [0,1]$ (referred to as a generalized Prisoners' Dilemma / PD scenario) or the case in which there is a point $z_{eq} \in (0,1)$ such that $\Pi(z_{eq}) = 0$, $\Pi(z) > 0$ for $z \in [0,z_{eq})$, and $\Pi(z) < 0$ for $z \in (z_{eq},1]$ (referred to as a generalized Stag-Hunt / SH scenario). 

For the case in which defectors have an individual-level advantage over altruistic punishers at any group composition (i.e. when $\Pi(z) > 0$ for all $z \in [0,1]$), there exists an family of steady state densities for the multilevel dynamics of Equation \eqref{eq:twotypeLM} for any fixed replication rates $\Pi(z)$ and $\mc{G}(z)$ and strength of between-group competition $\lambda$. It has been shown that the family of steady states can be characterized by the tail behavior of the density near the all-punisher equilibrium $z  = 1$. Namely, we can parametrize the family of steady state densities by the H{\"o}lder exponent of their cumulative distribution functions $\theta$ near $z = 1$, which we define below. 

\begin{definition} \label{def:holderexponent}\label{def:Holder} A probability distribution with density $f(t,x)$ has a H{\"o}lder exponent $\theta_t$ with associated H{\"o}lder constant $C_{\theta_t}$ near $x=1$ if the density has the following limiting behavior %
\begin{equation} \label{eq:Holderlimit}
\lim_{x \to 0} \frac{\int_{1-x}^1 f(t,y) dy}{x^{\Theta}} = \left\{
     \begin{array}{lr}
       0 & : \Theta < \theta_t \\
       C_{\theta_t} & : \Theta =  \theta_t \\
       \infty & : \Theta > \theta_t
     \end{array}
   \right. .
\end{equation}
 \end{definition}
An example family of probability densities with H{\"o}lder exponent $\theta$ near $z=1$ is given by $f_{\theta}(z) = \theta \left( 1 - z \right)^{\theta - 1}$, of which the uniform density $f_{1}(z) = 1$ is a special case. It can be shown that the H{\"o}lder exponent near $z=1$ is preserved by the multilevel dynamics of Equation \eqref{eq:twotypeLM} \citep{cooney2020analysis,cooney2022long}, and so it is helpful to consider the H{\"o}lder exponent near $z=1$ of the initial strategic distribution of groups to determine the long-time behavior of the multilevel dynamics.

We can use the H{\"o}lder exponent near $z=1$ to parametrize our infinite family of steady state densities by mass densities $f^{\lambda}_{\theta}(z)$ of the form 
\begin{equation} \label{eq:flambdatheta}
\begin{aligned}
f^{\lambda}_{\theta}(z) &= \frac{1}{Z_f} z^{\pi(0)^{-1} \left( \lambda \left[ \mc{G}(1) - \mc{G}(0) \right] - \theta \Pi(1) \right)} \left( 1 - z \right)^{\theta - 1} \frac{\Pi(1)}{\Pi(z)} \exp\left( - \lambda \int_{z}^1 \frac{C(s)}{\Pi(s)} ds \right)
\end{aligned}
\end{equation}
where the term $-\lambda C(s)$ is given by
\begin{equation}
\begin{aligned}
- \lambda C(s) &=  \lambda \left( \frac{\mc{G}(s) - \mc{G}(0)}{s} \right) + \left( \frac{\lambda \left[ \mc{G}(1) - \mc{G}(0) \right] - \theta \Pi(1)}{\Pi(0)} \right) \left( \frac{\Pi(s) - \Pi(0)}{s} \right) \\
&+ \lambda \left( \frac{\mc{G}(s) - \mc{G}(1)}{1-s} \right) - \theta \left( \frac{\Pi(s) - \Pi(1)}{1-s} \right)
\end{aligned}
\end{equation}
and $Z_f$ is a normalizing constant given by
\begin{equation}
Z_f = \int_0^1 \left[ u^{\pi(0)^{-1} \left( \lambda \left[ \mc{G}(1) - \mc{G}(0) \right] - \theta \Pi(1) \right)} \left( 1 - u \right)^{\theta - 1} \frac{\Pi(1)}{\Pi(u)} \exp\left( - \lambda \int_{u}^1 \frac{C(s)}{\Pi(s)} ds \right) \right] du.
\end{equation}

Given an initial density $f(0,x) := f_0(x)$ with H{\"o}lder exponent $\theta > 0$ near $x=1$, we can characterize whether multilevel selection in a generalized PD scenario will support coexistence of cooperative individuals and defectors in terms of the following critical threshold on the relative strength of between-group selection:

\begin{equation} \label{eq:lambdastargeneral}
\lambda^* := \frac{\theta \Pi(1)}{\mc{G}(1) - \mc{G}(0)}
\end{equation}
Using the characterization of the individual advantage of defection by $\Pi(z) = \pi_D(z) - \pi_P(z)$, we may rewrite this threshold quantity in the following form:
\begin{equation} \label{eq:lambdastartugofwar}
\lambda^* := \frac{ \left( \pi_D(1) - \pi_P(1) \right) \theta}{\mc{G}(1) - \mc{G}(0)}
\end{equation}

In Theorem \ref{thm:longtimePD}, we recall existing results for the long-time behavior of solutions to our PDE for multilevel selection for the case in which the within-group and between-group dynamics resemble a PD game in the sense that defectors have an individual-level replication advantage over cooperators ($\Pi(z) > 0$ for all $z \in [0,1]$) and the all-cooperator has greater group-level success than an all-defector group ($\mc{G}(1) > \mc{G}(0)$). We show that for an initial population $f(0,z)$ with H{\"o}lder exponent $\theta > 0$ and associated nonzero, finite H{\"o}lder  constant $C_{\theta}$ near $x=1$, then the population will concentrate upon a delta-function at the all-defector equilibrium $\delta(x)$ if $\lambda \leq \lambda^*$ and that the population will converge to the steady-state density $f^{\lambda}_{\theta}(x)$ with H{\"o}lder exponent $\theta$ if $\lambda > \lambda^*$. 

\begin{remark}
In Theorem \ref{thm:longtimePD}, we describe the long-time behavior of solutions to Equation \eqref{eq:twotypeLM} using the notion of weak convergence. We say that a probability density $f(t,x)$ converges weakly to the density $f_{\infty}(x)$ as $t \to \infty$ if, for each test-function $v(x) \in C\left([0,1] \right)$, we have that $\lim_{t \to \infty} v(x) f(t,x) dx = \int_0^1 v(x) f_{\infty}(x)$. In this case, we write that $f(t,x) \rightharpoonup f_{\infty}(x)$ as $t \to \infty$ to denote this weak convergence. This notion of weak convergence was initially motivated by the fact that the results of long-time behavior in Theorem \ref{thm:longtimePD} were originally demonstrated for measure-valued solutions to the two-level replicator equation \citep[Theorem 4]{cooney2022long}, but it has also been shown that is possible to prove pointwise convergence to steady-state densities for the two-level replicator equation starting from a continuously differentiable initial density $f(0,x)$ \citep{cooney2023evolutionary}. 
\end{remark}

\begin{theorem}[Long-Time Behavior of PDE Model of Multilevel Selection for Generalized Prisoners' Dilemma Case: Combination of {\citep[Theorem 1]{cooney2022long}} and {\citep[Theorem 4]{cooney2022long}}] \label{thm:longtimePD} 

Suppose that $\mathcal{G}(z), \pi(z) \in C^1([0,1])$, $\mathcal{G}(1) > \mathcal{G}(0)$, and $\Pi(z) > 0$, and consider an initial density $f(0,z) = f_0(z)$ with H{\"o}lder exponent $\theta$ near $x=1$ and finite H{\"o}lder constant $C_{\theta} > 0$. If $\lambda > \lambda^*(\theta)$, then the solution $f(t,x)$ to Equation \eqref{eq:twotypeLM} satisfies $f(t,x) \rightharpoonup f^{\lambda}_{\theta}(x)$ as $t \to \infty$. If instead $\lambda \leq \lambda^*$, then $f(t,x) \rightharpoonup \delta(x)$ as $t \to \infty$. 
\end{theorem}

For the case in which $\Pi(z) < 0$ on an interval of the form $[z_,1]$, the all-cooperator equilibrium will be locally stable under the within-group replicator dynamics. This is the case for multilevel replicator equations for Stag-Hunt {games (which features bistability of the all-defector and all-cooperator equilibrium)  as well as Harmony / Prisoners' Delight games (for which cooperation is the dominant strategy)}, and will also be the dynamical scenario seen in our model of altruistic punishment when the strength of punishment $k$ is sufficiently large. For the case in which within-group replicator dynamics feature local stability of the all-punisher equilibrium, Theorem \ref{thm:longtimePDel} allows us to conclude that the population will concentrate upon a delta-function at the all-punisher equilibrium provided that there is any between-group competition (corresponding to $\lambda > 0$) and that there is any initial proportion of groups with compositions located within the basin of attraction for the all-punisher equilibrium under the within-group dynamics. 

\begin{theorem}[Long-Time Behavior of Multilevel Dynamics in Generalized Stag-Hunt Case] \label{thm:longtimePDel}
Suppose that $\mathcal{G}(z) \in C^1([0,1])$ and $\Pi(z) \in C^2([0,1])$, $\mathcal{G}(1) > \mathcal{G}(0)$, and that there is a $z_{eq}$ such that $\Pi(z) < 0$ for $z \in (z_{eq},1]$ and $\Pi(z) > 0$ for $z \in [0,z_{eq})$. If $\lambda > 0$ and the initial density $f(0,z) = f_0(z)$ satisfies $\int_{z_{-}}^1 f_0(z) dz > 0$ for some $z_{-} > z_{eq}$, then the solution $f(t,z)$ to the multilevel dynamics of Equation \eqref{eq:twotypeLM} satisfies $f(t,z) \rightharpoonup \delta(1-x)$ as $t \to \infty$. 
\end{theorem}

\begin{remark}
This result provides an analogue in our PDE of multilevel selection to the result of Boyd and Richerson \cite{boyd1990group}, which indicates that group-level selection can promote collectively beneficial outcomes when within-group evolutionary dynamics feature alternative stable states. 
\end{remark}

For both the cases of multilevel Prisoners' Dilemma or Prisoners' Delight scenarios covered by Theorem \ref{thm:longtimePD} and \ref{thm:longtimePDel}, it is also possible to calculate the average payoff achieved at the long-time steady state population achieved when starting with an initial distribution with H{\"o}lder exponent $\theta > 0$ \citep{cooney2022long}. For the case of $\Pi(z) > 0$ and $\mc{G}(1) > \mc{G}(0)$ corresponding to the multilevel PD dynamics,  it has been shown that the average group-level replication rate across the density $f(t,z)$ that is achieved in the long-time limit of the population, which is given by
\begin{equation} \label{eq:longtimeGseparable}
\langle \mc{G}(\cdot) \rangle_{f^{\lambda}_{\theta}} :=  \lim_{t \to \infty} \int_0^1 \mc{G}(z) f(t,z) dz = \left\{
    \begin{array}{cr}
      \mc{G}(0) & : \lambda < \lambda^*\\
     \mc{G}(1) - \frac{\theta}{\lambda} \Pi(1) & : \lambda \geq \lambda^*
    \end{array}
  \right.
\end{equation}
\citep{cooney2022long}, which can also be expressed in terms of the threshold selection strength $\lambda^*$ as
\begin{equation} \label{eq:longtimemcG}
\langle \mc{G}(\cdot) \rangle_{f^{\lambda}_{\theta}} := \lim_{t \to \infty} \int_0^1 \mc{G}(z) f(t,z) dz = \left\{
    \begin{array}{cr}
      \mc{G}(0) & : \lambda < \lambda^*\\
      \left( \frac{\lambda^*}{\lambda} \right) \mc{G}(0) + \left( 1  - \frac{\lambda^*}{\lambda} \right) \mc{G}(1)  & : \lambda \geq \lambda^*.
    \end{array}
  \right.
\end{equation}

For the case of multilevel dynamics corresponding to SH or Prisoners' Delight games, we have that
\begin{equation}
\langle \mc{G}(\cdot) \rangle_{f^{\lambda}_{\theta}} := \lim_{t \to \infty} \int_0^1 \mc{G}(z) f(t,z) dz = \mc{G}(1)
\end{equation}
for any $\lambda > 0$, so the all-punisher outcome is always achieved in the long-time limit. 

{
\subsection{Existing Results and Suggestions for a More General Class of Pairwise Group-Level Victory Probabilities}
\label{sec:pairwiseresults}

Now we summarize recent results and conjectures for the PDE model of multilevel selection for more general models of pairwise group-level victory probabilities in the form provided by Equation \eqref{eq:multilevelPDEtworho}, which have been explored in a recent paper by Alexiou and Cooney \cite{alexiou2024steady}. While a full analysis of long-time behavior is not yet available for this model, it has been possible to obtain necessary conditions for properties steady state solutions for generalized Prisoners' Dilemma scenarios and and prove convergence of the population to a delta-function at the all-punisher equilibrium for the case of a generalized Stag-Hunt scenario. While these preliminary results only provide a partial description of the dynamical and steady-state behavior for our model of multilevel selection with pairwise group-level competition, they will allow us to gain intuition on how altruistic punishment can interact with pairwise group-level competition and to set baseline expectations for the behavior we see in numerical explorations of our model in Section \ref{sec:nonlineargroup}. %
}

{In Proposition \ref{prop:pairwisesteady}, we provide necessary conditions for the population to have a steady-state density $f(x)$ with H{\"o}lder exponent $\theta$ near $z = 1$ for the case of within-group and group-level dynamics corresponding to a generalized Prisoners' Dilemma scenario. In particular, we find the expression that the average group-level victory probability $\int_0^1 \rho(x,1) f(x) dx$ that such a steady-state population must have when paired in group-level competition with an all-punisher group. This measurement for the collective success of the steady-state population also allows us to obtain a conjectured threshold quantity $\lambda^*_{PW}$ for the survival of all-punisher groups in the long-time population under a generalized Prisoner's Dilemma scenario, with defectors taking over the population when $\lambda < \lambda^*_{PW}$ and the population converging to a steady-state density supporting cooperation for $\lambda > \lambda^*_{PW}$. This conjecture has been supported by numerical simulations of Equation \eqref{eq:multilevelPDEtworho} for a range of group-level victory probabilities in the case payoffs arising from a two-player, two-strategy symmetric game \cite[Section 6]{alexiou2024steady}. }
{
\begin{proposition}[Necessary Condition for Existence of Steady-State Density Supporting Positive Level of Altruistic Punishment (Originally {\cite[Proposition A.1.]{alexiou2024steady}})] \label{prop:pairwisesteady}
Suppose that the group-level victory probability satisfies $\rho(z,u) \in C^1\left([0,1]^2\right)$ and that the individual-level advantage to defect satisfies $\Pi(z) \in C^1\left( [0,1] \right)$ and $\Pi(z) > 0$ for all $z \in [0,1]$. If Equation \eqref{eq:multilevelPDEtworho} has a steady-state density solution $f(x)$ with H{\"o}lder exponent near $x=1$, then the average group-level victory probability of the steady-state population when paired with the all-punisher group is given by
\begin{equation}
\int_0^1 \rho(x,1) f(x) dx = \frac{1}{2} - \frac{\theta \Pi(1)}{2 \lambda}.
\end{equation}
Furthermore, this group-level victory probability $\int_0^1 \rho(x,1) f(x) dx$ will exceed the victory probability $\rho(0,1)$ of an all-defector group when paired with an all-punisher group provided that the strength of group-level competition satisfies
\begin{equation} \label{eq:lambdastarpairwise}
\lambda > \lambda^*_{PW} := \frac{\theta \Pi(1)}{\rho(1,0)- \rho(0,1)}
\end{equation}
\end{proposition}
}

{A more complete characterization of the long-time behavior under pairwise group-level competition for the case of a generalized Stag-Hunt scenario. For the case in which when the all-punisher equilibrium is locally stable under the within-group dynamics, we note in Proposition \ref{prop:pairwiseSHdelta} that the population will converge upon a delta-function at the all-punisher equilibrium when there is any group-level competition (corresponding to $\lambda > 0$) and if there are any groups in the initial population featuring a fraction of altruistic punishers that falls within the basin of attraction of the all-punisher group $z=1$ under the within-group dynamics. 
}
{
\begin{proposition}[Convergence of Population to Delta-Function at All-Punisher Group in Generalized Stag-Hunt Scenario {\cite[Proposition 5.2]{alexiou2024steady}}]
\label{prop:pairwiseSHdelta}
Consider a within-group relative reproduction rate $\Pi(z) \in C^2\left( [0,1] \right)$ and an equilibrium point $x_{eq}$ of the within-group dynamics satisfying $\Pi(z_{eq}) < 0$ for $z \in (z_{eq},1]$, and consider a group-level victory probability $\rho(z,u) \in C^1\left([0,1]^2\right)$ such that there is a fraction of altruistic punishers $z_{min} > z_{eq}$ such that $\rho(z,u)$ satisfies the following two properties:
\begin{itemize}
    \item $\rho(z,u) > \rho(w,u)$ for any $z > z_{min} > w$ and for all $u \in [0,1]$
    \item $\rho(z,u) > \frac{1}{2} > \rho(u,z)$ for any $z,u$ satisfying $z > z_{min} > u$
\end{itemize}
If $\lambda > 0$ and the initial density $f(0,z) = f_0(z)$ satisfies $\int_{z_{eq}}^1 > 0$, then the solution $f(t,z)$ to the multilevel dynamics of Equation \eqref{eq:multilevelPDEtworho} will satisfy $f(t,z) \rightharpoonup \delta(1-x)$ as $t \to \infty$. 
\end{proposition}
}
{
This result allows us to see that the population will be able to achieve an all-punisher outcome under the dynamics of Equation \eqref{eq:multilevelPDEtworho} provided that the strength of altruistic punishment is sufficiently strong to stability the all-punisher group under the within-group replicator dynamics. In addition, we note that, if the population converges to a delta-function at the all-punisher equilibrium, we can use the fact that $\rho(z,z) = \frac{1}{2}$ for any $z \in [0,1]$ to see that the long-time collective success of the population in pairwise competition with the all-punisher group is given by
\begin{equation}
\lim_{t \to \infty} \rho(z,1) f(t,z) dz = \int_0^1 \rho(z,1) \delta(1-z) dz = \rho(z,z) = \frac{1}{2}.
\end{equation}

We can then combine the preliminary results we have obtained from Propositions \ref{prop:pairwisesteady} and \ref{prop:pairwiseSHdelta} for pairwise group-level competition in generalized Prisoners' Dilemma and Stag-Hunt scenarios to obtained a conjectured formula for the that the long-time limit of the average success of the population in pairwise group-level competition. In particular, we conjecture that, if the population $f(0,z)$ starts with H{\"o}lder exponent $\theta$ near $z=1$, then the long-time collective success has the following piecewise characterization

\renewcommand{\arraystretch}{2}
\begin{equation} \label{eq:pairwiserholongtime}
\lim_{t \to \infty} \int_0^1 \rho(z,1) f(t,z) dz =
\left\{ 
\begin{array}{cl}
\rho(0,1) &: \Pi(z) > 0 \: \: \mathrm{for} \: \: z \in [0,1], \lambda < \lambda_{PW}^*  \\
\ds\frac{1}{2} - \ds\frac{\theta \Pi(1)}{2 \lambda} &: \Pi(z) \: \: \mathrm{for} \: \: z \in [0,1], \lambda < \lambda_{PW}^*  \\ 
\ds\frac{1}{2} &: \Pi(1) < 0
\end{array}
\right.
\end{equation}
\renewcommand{\arraystretch}{1}
}

{
\section{Analysis of Multilevel Competition Between Defectors and Altruistic Punishers with Additive Group-Level Victory Probabilities}
\label{sec:analytical}
}

In this section, we consider multilevel competition in groups featuring only defectors and altruistic punishers, {restricting our analysis to families of group-level victory probabilities that are amenable to the analytical techniques described in Section \ref{sec:existingresults}. We first characterize the long-time behavior with group-level victory probabilities based on the difference in the fraction of altruistic punishers in the competing groups (Section \ref{sec:PDtwotypecooperativefraction}) and the case of group-level victory based on the difference in average payoff of competing groups normalized by the maximum possible range of group payoffs (Section \ref{sec:grouplocalupdate}). We then proceed to characterize the long-time collective outcomes of the population for these two group-level competition scenarios (Section \ref{eq:longtimegrouppayoff}) and consider how these group-level victory probabilities impact multilevel selection in the case of groups composed only of cooperators of defectors (Section \ref{sec:payoffCDcompare}).}

\vspace{5mm}

\subsection{Group-Level Competition Based on the Fraction of Altruistic Punishers}
\label{sec:PDtwotypecooperativefraction}

We now consider the dynamics of multilevel selection when pairwise between-group competition is based on the group-level victory probability
\begin{equation}
\rho(z,u) = \frac{1}{2} + \frac{1}{2} \left[ z - u \right],
\end{equation}
which is the case motivated by the form of victory probability considered by Boyd and coauthors \cite{boyd2003evolution}. As discussed in Section \ref{sec:PDEmodelDP}, the mutlilevel dynamics can be reduced in this case to the PDE model of Equation \eqref{eq:PDEadditivelyseparable} with net group-level replication rate $\mc{G}(z) =z$. 

In Proposition \ref{prop:steadycooperators}, we summarize the results for long-time behavior for this choice of group-level victory probability. This result follows from applying Theorems \ref{thm:longtimePD} and \ref{thm:longtimePDel} for the case of net reproduction rate $\mc{G}(z) = z$ and for the choice of individual-level advantage $\Pi(z)$ of defectors over altruistic punishers. 

\begin{proposition} \label{prop:steadycooperators}
For the choice of net replication rates $\mc{G}(z) = z$ and $\Pi(z) = \pi_D(z) - \pi_P(z)$, we have that the family of steady states from Equation \eqref{eq:flambdatheta} are given by
\begin{equation} \label{eq:flambdathetaBGBR}
\begin{aligned}
f^{\lambda}_{\theta}(z) &= \frac{1}{Z_f} z^{\frac{1}{c+k+q} \left[ \lambda - \theta \left(c+q -p\right) \right] - 1} \left( 1 - z \right)^{\theta - 1} \left(c + k + q - (p+k) z \right)^{\left( \frac{- \lambda + \theta (k+p)}{c+q+k} \right) - 1} \\
Z_f &= \int_0^1 \left[  z^{\frac{1}{c+k+q} \left[ \lambda - \theta \left(c+q -p\right) \right] - 1} \left( 1 - z \right)^{\theta - 1} \left(c + k + q - (p+k) z \right)^{\left( \frac{- \lambda + \theta (k+p)}{c+q+k} \right) - 1} \right] dz.
\end{aligned}
\end{equation}
The threshold strength of group-level selection for these steady states to be integrable is 
\begin{equation} \label{eq:cooplambdastar}
\lambda^* = \theta \left( c + q - p \right),
\end{equation}
and we can apply Theorems \ref{thm:longtimePD} and \ref{thm:longtimePDel} to see that solutions to our PDE have the following long-time behavior starting from an initial density $f(0,z) = f_0(z)$ whose distribution has H{\"o}lder exponent $\theta > 0$. 
\begin{itemize}
    \item If $p < c + q$ and $\lambda < \lambda^*$, the solution $f(t,z) \rightharpoonup \delta(z)$ as $t \to \infty$.
    \item If $p < c + q$ and $\lambda > \lambda^*$, then $f(t,z) \to f^{\lambda}_{\theta}(z)$ as $t \to \infty$. 
    \item If $p > c + q$, then $f(t,z) \rightharpoonup \delta(z-1)$ as $t \to \infty$ provided that $\lambda > 0$. 
\end{itemize}
\end{proposition}

{In Figure \ref{fig:BRBGsteady}, we plot the steady-state densities $f^{\lambda}_{\theta}(z)$ described by Equation \eqref{eq:flambdathetaBGBR} for the cases of per-interaction alone ($k > 0, q = 00$) and fixed punishment costs alone ($q > 0$, $k =0$), different punishment strengths, and several H{\"o}lder exponents $\theta$ near $z=1$. For each of the cases considered in Figure \ref{fig:BRBGsteady}, we consider the steady state densities for a fixed initial condition and strength of between-group selection, shwoing that increasing the strength of punishment $p$ results in steady states featuring groups with greater proportions of altruistic punishers. We also see that the level of altruistic punishment achieved is greater when we compare the case of per-interaction punishment cost $k = 0.5$ (Figure \ref{fig:BRBGsteady}, top-left and top-right) to the case of fixed punishment cost $q = 0.5$ (Figure \ref{fig:BRBGsteady}, bottom-left and bottom-right), while greater levels of altruistic punishment are achieved for H{\"o}lder exponent $\theta = 1$ near $x = 1$ (Figure \ref{fig:BRBGsteady}, top-left and bottom-left) relative to the case of H{\"o}lder exponent $\theta = 2$ near $x=1$ (Figure \ref{fig:BRBGsteady}, top-right and bottom-right).   }

\begin{figure}[!htp]
    \centering
    \includegraphics[width=0.48\linewidth]{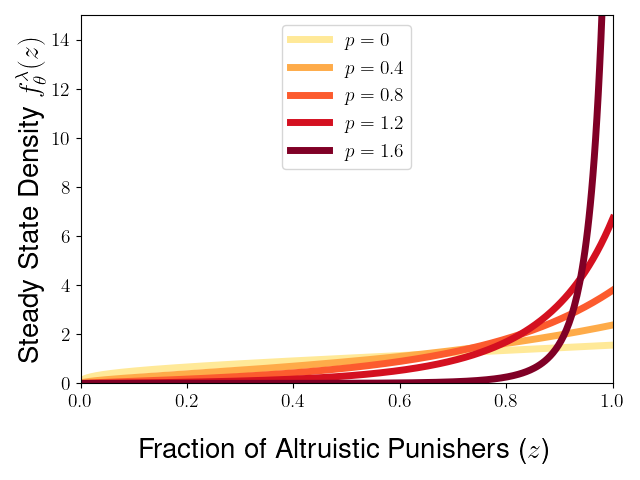}
    \includegraphics[width=0.48\linewidth]{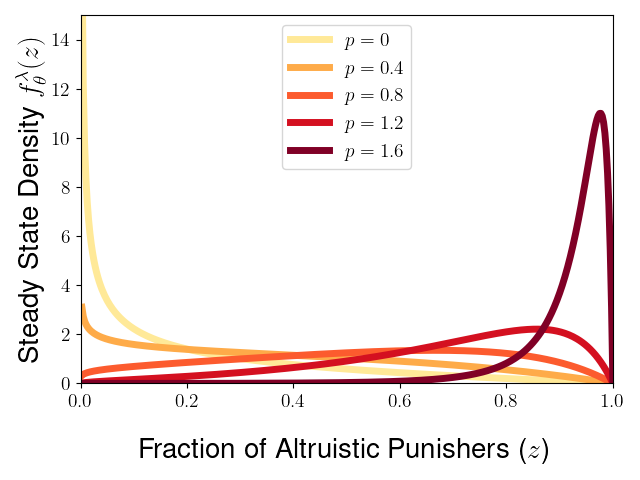}
    \includegraphics[width=0.48\linewidth]{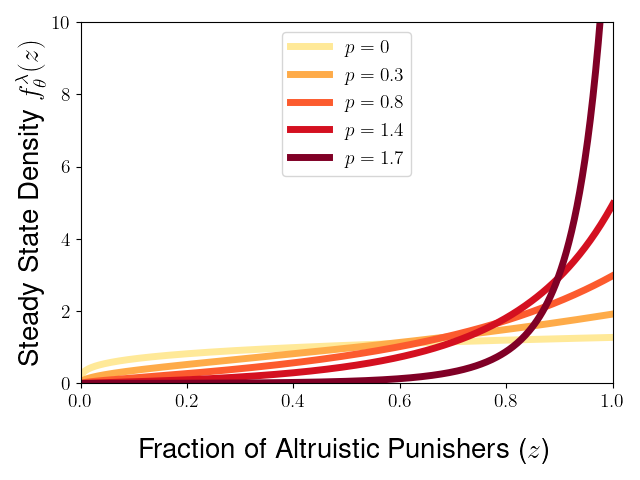}
    \includegraphics[width=0.48\linewidth]{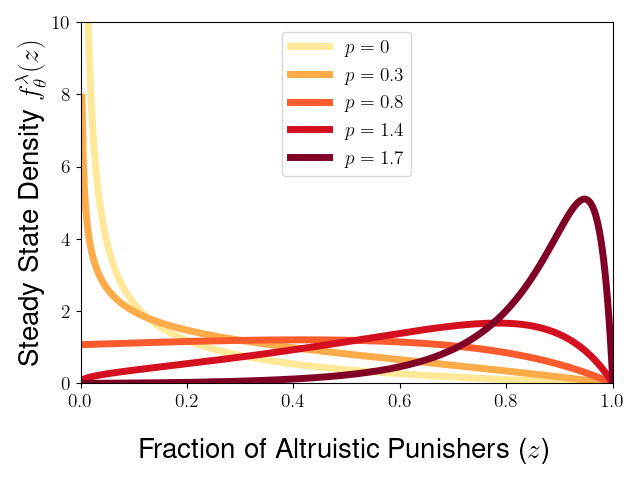}
    \caption{{Example steady state densities for PDE the model of multilevel selection with pairwise competition following the group-level victory probability based on the difference in the fraction of altruistic punishers. We plot steady densities calculated from Proposition \ref{prop:steadycooperators} for various strengths $p$ of punishment. We consider the cases of per-interaction punishment costs with $k = 0.5$ and $q = 0$ for initial conditions with H{\"o}lder exponent $\theta = 1$ (top-left) and $\theta = 2$ (top-right) near $z=1$, as well as the case of fixed punishment costs with $k = 0$ and $q = 0.5$ for initial conditions with H{\"o}lder exponents $\theta = 1$ (bottom-left) and $\theta = 2$ (bottom-right) near $z=1$. The game-theoretic parameters are fixed to $b = 2$ and $c = 1$ in each panel. } }
    \label{fig:BRBGsteady}
\end{figure}

{We can also use Theorem \ref{thm:longtimePD} to determine the average fraction of altruistic punishers achieved at steady-state for this model. Using the fact that this group-level victory probability corresponds to the net group-level replication rate} $\mc{G}(z) = z$, we can use Equation \eqref{eq:longtimemcG} to see that
\begin{equation} \label{eq:avgcoopDPedge}
  \lim_{t \to \infty} \int_0^1 z f(t,z) dz = \lim_{t \to \infty} \int_0^1 \mc{G}(z) f(t,z) dz = \left\{
    \begin{array}{cr}
     0  & : \lambda < \lambda^*\\
       \\ {1 - \left(\ds\frac{\theta}{\lambda} \right) \left( c + q - p \right)} & : \lambda \geq \lambda^*.
    \end{array}
  \right.
\end{equation}

{
\begin{remark}
The form of $\lambda^*$ and $\langle z \rangle_{f^{\lambda}_{\theta}}$ suggests that having fixed costs for punishing defectors is less conducive to achieving cooperation via multilevel selection than having per-interaction punishment costs alone. This observation agrees with simulation results in the stochastic model by Boyd and coauthors \cite{boyd2003evolution}, who found that it was difficult to sustain altruistic punishment via cultural group selection in large groups for the case of fixed costs of punishment. In our PDE model, we see the key difference between the effects of fixed and per-interaction punishment costs by noting that the threshold selection strength $\lambda^*$ is proportional to the payoff difference
\begin{equation}
\pi_D(1) - \pi_P(1) = c + q - p
\end{equation}
between defectors and punishers in a group otherwise composed entirely of punishers. Because the effect of per-interaction costs of punishment vanish in the limit of the all-punisher composition as $z \to 1$ but the effect of fixed punishment costs do not, we see that a group of punishers are at less of an individual-level disadvantage against invasion by defectors in the per-interaction cost case. The decrease in individual-level disadvantage then allows for a greater ability for the collective success of all-punisher groups to sustain cooperative behavior via multilevel selection, yielding better collective outcomes in the absence of fixed punishment costs. 
\end{remark}
}

\subsection{Group-Level Competition Depending on Average Payoff of Group Members} \label{sec:grouplocalupdate}

In this section, we consider the case of group-level {victory probability} \[ {\rho(z,u) = \frac{1}{2} + \frac{1}{2} \left[ \frac{G(z) - G(u)}{G^* - G_*}\right]}, \]
{which can be described in terms of a net group-level reproduction rate} of the form $\mc{G}(z) = \frac{G(z)}{G^* - G_{*}}$. {In order to analyze the dynamics of multilevel selection for this victory probability, we must} start by calculating the maximum and minimum possible average payoffs achieved by groups with a fraction $z$ of altruistic punishers. {We can show that, for our model of altruistic punishment within groups, the values of $G^*$ and $G_*$ are given by}
\begin{subequations}
\begin{equation}
{G^* = b - c - q}
\end{equation}
{and}
\begin{equation}
G_{*} = \left\{
    \begin{array}{cr}
      0 & : b \geq c + p + k + q\\
      - \ds\frac{\left( b - (c+p+k+q)\right)^2}{4 (p+k)} & : b < c + p + k + q
    \end{array}
  \right. ,
\end{equation}
\end{subequations}
{and we refer the reader to Lemma \ref{lem:minmaxpayoff} for a full derivation of these quantities}.In particular, the minimum value $G_*$ of $G(z)$ can decrease due to a marginal increase in the strength punishment $p$ or the costs of punishment $k$ or $q$. {We note that} the maximum payoff $G^* = b - c -q$ is constant in the punishment $p$, and therefore we the normalization factor $G^* - G_{*}$ can increase with the strength of punishment $p$ for a range of parameters in this case. This can potentially cause an increase in the strength of punishment of defectors to actually decrease the group-level victory probability of groups featuring high numbers of altruistic punishers. 

\begin{remark}
We note that the condition required for $G(z)$ to be minimized by an intermediate fraction of altruistic punishers $y \in (0,1)$ can only be achieved when the strength the punishment $p$ or the the costs of punishment $k$ or $q$ are sufficiently large. Intuitively, the possibility that average group payoff can achieve an outcome worse than the all-defector payoff arises because having a small number of altruistic punishers can introduce both costs and punishments that diminish average payoff, but without a sufficient critical mass of altruistic punishers to achieve the baseline benefits of cooperation in the underlying donation game. {The possibility that $G_{*} < G(0)$ in the case of altruistic punishment is different from the behavior seen in prior work when studying multilevel selection in the presence of other within-group mechanisms to that modify payoffs from an underlying donation game in the case of a two-level replicator equation \citep{cooney2022assortment, cooney2023evolutionary}. For this reason, we provide a comparison in Section \ref{sec:indirectreciprocity} between our results with altruistic punishment to a model of multilevel selection with pairwise group-level competition featuring within-group interactions featuring indirect reciprocity based on an image scoring rule \citep{nowak1998evolution}. }
\end{remark}

In Proposition \ref{prop:steadypayoff}, we use the form of net group-level replication rates $\mc{G}(\cdot) = \frac{\mc{G}(z)}{G^*-G_*}$ for the group-level local update rule, applying Theorems \ref{thm:longtimePD} and \ref{thm:longtimePDel} to provide the following characterization of the long-time behavior for multilevel selection for this group-level victory probability.  The result of Proposition \ref{prop:steadypayoff} features generally analogous qualitative behavior for the long-time dynamics as seen in Proposition \ref{prop:steadycooperators} for the case of a net group-level replication rate $\mc{G}(z) = z$, with appropriate modifications made to reflect the difference in net group-level reproduction rate for the two models. For completeness, we present the derivation of the family of steady states $f^{\lambda}_{\theta}(z)$ for this case of $\mc{G}(z)$ in Section \ref{sec:densityderivation} of the appendix. 

\begin{proposition} \label{prop:steadypayoff}
For the case of net group-level replication rate $\mc{G}(z) = \frac{G(z)}{G^*-G_*}$ depending on normalized differences in average payoff of group members, there is a family of density density steady states of the form
\begin{equation} \label{eq:grouplocaldensity}
\begin{aligned}
f^{\lambda}_{\theta}(z) &= \frac{1}{Z_f} z^{\frac{1}{c+k+q} \left[\frac{\lambda (b-c-q)}{G^* - G_{*}} - \theta \left(c+q -p\right) \right] - 1} \left( 1 - z \right)^{\theta - 1} \left(c + k + q - (p+k) z \right)^{-\left( \frac{\lambda (b+k) - \theta (k + p) (G^* - G_*)}{(G^* - G_*)(c+k+q)} \right)-1} \\
Z_f &= \int_0^1 z^{\frac{1}{c+k+q} \left[\frac{\lambda (b-c-q)}{G^* - G_{*}} - \theta \left(c+q -p\right) \right] - 1} \left( 1 - z \right)^{\theta - 1} \left(c + k + q - (p+k) z \right)^{-\left( \frac{\lambda (b+k) - \theta (k + p) (G^* - G_*)}{(G^* - G_*)(c+k+q)} \right)-1} dz.
\end{aligned}
\end{equation}

The threshold strength of group-level selection for these steady states to be integrable is 
\begin{equation}
\label{eq:lambdastargrouplocal}
\lambda^* = \frac{\theta \Pi(1)}{\mc{G}(1) - \mc{G}(0)} = \frac{\theta (c+q-p)}{\left( \frac{G(1) - G(0)}{G^* - G_*} \right)} = \frac{\theta (G^* - G_*) (c+q-p)}{b-c-q}.
\end{equation}
and we can apply Theorems \ref{thm:longtimePD} and \ref{thm:longtimePDel} to see that solutions to our PDE have the following long-time behavior starting from an initial density $f(0,z) = f_0(z)$ whose distribution has H{\"o}lder exponent $\theta > 0$. 
\begin{itemize}
    \item If $p < c + q$ and $\lambda < \lambda^*$, the solution $f(t,z) \rightharpoonup \delta(z)$ as $t \to \infty$.
    \item If $p < c + q$ and $\lambda > \lambda^*$, then $f(t,z) \to f^{\lambda}_{\theta}(z)$ as $t \to \infty$. 
    \item If $p > c + q$, then $f(t,z) \rightharpoonup \delta(z-1)$ as $t \to \infty$ provided that $\lambda > 0$. 
\end{itemize}

\end{proposition}

In Figure \ref{fig:group_local_density_k_model}, we use the expression from Equation \eqref{eq:grouplocaldensity} to plot steady-state densities achieved under the multilevel dynamics with the local group-level comparison rule for two different values of the per-interaction punishment cost $k$. For both the case of $k = 0.1$ (Figure \ref{fig:group_local_density_k_model}, left) and $k = 4$ (Figure \ref{fig:group_local_density_k_model}, right), we see that increasing the strength of punishment $p$ leads to steady state densities featuring groups with greater proportions of altruistic punishers. However, a surprising behavior we see is that a greater proportion of altruistic punishers is achieved at steady state for the case of $k = 4$ than in the case in which $k = 0.1$, meaning that more cooperative behavior is achieved when altruistic punishers encounter a greater cost for punishing defectors.

\begin{figure}[!htp]
    \centering
    \includegraphics[width = 0.48\textwidth]{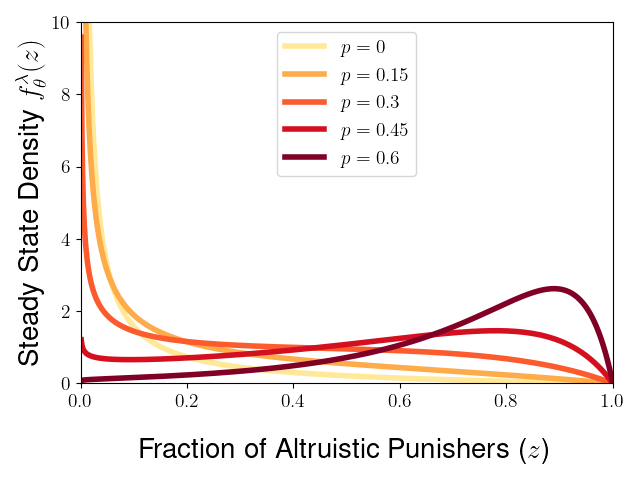}
     \includegraphics[width = 0.48\textwidth]{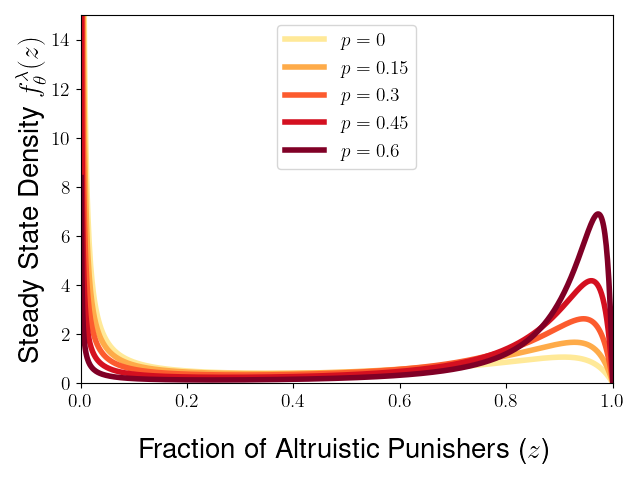}
    \caption{Comparison of steady state densities achieved for multilevel dynamics with local group update rule with various punishment strengths $p$ and for per-interaction punishment costs of $k = 0.1$ (left) and $k = 4$ (right). Other payoff parameters are $b = 2$, $c = 1$, $p = 0.5$, and the steady state densities are calculated for between-group selection strength $\lambda = 2$, and the H{\"o}lder exponent $\theta = 2$ near $z = 1$.}
    \label{fig:group_local_density_k_model}
\end{figure}

We can also quantify the collective success of groups under this group-level victory probability by calculating the average payoff of groups in the steady-state population achieved at steady state for this choice of group-level victory probability. We note that, for $\lambda > \lambda^*$, the average group-level reproduction rate $\mc{G}(z) = \frac{G(z)}{G^*-G_*}$ across the steady state density $f^{\lambda}_{\theta}(z)$ is given by
\begin{equation}
\begin{aligned}
\langle \mc{G}(\cdot) \rangle_{f^{\lambda}_{\theta}} &= \left\langle  \frac{G(\cdot)}{G^* - G_{*}}\right\rangle_{f^{\lambda}_{\theta}}  %
 = \frac{1}{G^* - G_*} \int_0^1 G(z) f^{\lambda}_{\theta}(z) dz = \frac{1}{G^* - G_*} \langle G(\cdot) \rangle_{f^{\lambda}_{\theta}}.
\end{aligned}
\end{equation}
We can then combine this expression with Equation \eqref{eq:longtimemcG} {and use the fact that $G(1) = b -c - q$ and $G(0) = 0$} to see that, for $\lambda > \lambda^*$, the average payoff of group members at steady state is given by
\begin{equation} \label{eq:averageGgrouplocal}
\langle G(\cdot) \rangle_{f^{\lambda}_{\theta}} = \left( \frac{\lambda^*}{\lambda} \right) G(0) + \left( 1 - \frac{\lambda^*}{\lambda} \right) G(1) \: { = \left( b - c - q \right) \left( 1 - \frac{\lambda^*}{\lambda} \right)}.
\end{equation}

{
\subsection{Evaluating Collective Outcomes for Additively Separable Group-Level Victory Probabilities}
\label{eq:longtimegrouppayoff}
}

{Now that we have characterized the long-time behavior for the two additively separable group-level victory probabilities under consideration, we can look to characterize how the collective outcomes achieved depend on the parameters for our model of altruistic punishment. In Table \ref{tab:replicatortable}, we summarize the behavior seen for the group-level victory probabilities based on difference in the fraction of altruistic punishers ($\rho(z,u) = \frac{1}{2} + \frac{x-y}{2}$) and on the normalized difference in average payoffs of the competing groups $\left(\rho(z,u) = \frac{1}{2} + \frac{1}{2} \left[\frac{G(z) - G(u)}{G^* - G_*} \right]\right)$. From the table, we see that the threshold selection strength $\lambda^*$ and average fraction of altruistic punishers $\langle z \rangle_{f^{\lambda}_{\theta}}$ can be readily interpreted in terms of their dependence on the parameters for the cost and strength of altruistic punishment. However, due to the piecewise characterization for the minimum possible average group payoff $G_*$, it is difficult to determine a complete analytical characterization how the threshold selection strength $\lambda^*$ and average steady state payoff $\langle G(\cdot) \rangle_{f^{\lambda}_{\theta}}$ depend on the punishment parameters. Instead, we will numerically explore these quantifies for several example punishment scenarios, which allows us to determine that these key quantities can have a non-monotonic dependence on the strength of punishment $p$.}

\renewcommand{\arraystretch}{2}
\begin{table}[!ht]
\caption{{Summary of key quantities for multilevel selection for pairwise group-level competition for victory probabilities based on either the difference in fraction of altruistic punishers between the competing groups or on the difference in average payoff of competing groups normalized by the maximum possible difference in average payoffs. For each group-level victory probability considered, we present the threshold punishment strength required to stabilize altruistic punishment within groups, the threshold selection strength $\lambda^*$ required to achieve altruistic punishment by multilevel selection when defection dominates within groups, as well as a key quantity derived from the average net group-level reproduction rate $\mc{G(\cdot)}_{f^{\lambda}_{\theta}}$ for the cases of per-interaction punishment costs alone ($q > 0$, $p = 0$) and of fixed punishment costs alone ($p > 0$, $q = 0$).} }
\begin{center}
\small
\begin{tabular}{|c|c|c|c|c|}
\hline
\makecell{Group \\ Competition \\ Scenario} & \makecell{Punishment \\ Cost Type}  & \makecell{All-$P$ \\ Stable \\ in Group}  & \makecell{Threshold \\ Group \\ Competition $\lambda^*$}  & \makecell{Collective Success \\ for Density \\ Steady-State \\ } \\
    \hline
  \multirow{2}{*}{$\begin{aligned}\rho(z,u) &= \ds\frac{1}{2} + \ds\frac{z-u}{2} \\ \mc{G}(z) &= z \end{aligned}$} & 
Interaction & $p > c$ & $\theta (c-p)$  & $\langle z \rangle_{f^{\lambda}_{\theta}} = 1 - \left( \frac{\theta}{\lambda}\right) \left( c - p \right)$  \\
  \cline{2-5}
   & Fixed &  $p > c + q$ & $\theta (c+q-p)$  &  $\langle z \rangle_{f^{\lambda}_{\theta}} = 1 -\left(\frac{\theta}{\lambda} \right) \left( c + q  - p \right)$   \\
   \cline{2-5}
   \hline
   \multirow{2}{*}{$\begin{aligned}\rho(z,u) &= \ds\frac{1}{2} + \ds\frac{G(z) - G(u)}{G^* - G_*} \\ \mc{G}(z) &= \ds\frac{G(z)}{G^* - G_*} \end{aligned}$} & 
   Interaction  & $p > c$  & $\frac{\theta (G^* - G_*) (c-p)}{b-c}$ & $\langle G(\cdot) \rangle_{f^{\lambda}_{\theta}} = (b-c) \left( 1 - \frac{\lambda^*}{\lambda}\right)$   \\
  \cline{2-5}
   & Fixed & $p > c + q$ & $\frac{\theta (G^* - G_*) (c+q-p)}{b-c}$  & $\langle G(\cdot) \rangle_{f^{\lambda}_{\theta}} = (b-c - q) \left( 1 - \frac{\lambda^*}{\lambda} \right)$  \\
   \hline 
\end{tabular}
\end{center}
\label{tab:replicatortable}
\end{table}
\renewcommand{\arraystretch}{1}

To further explore the impact of {the strength and costs of punishment} on the long-time outcome of the multilevel dynamics with the local group update rule, we can examine the threshold selection strength $\lambda^*$ required to sustain altruistic punishment at steady state and the average payoff $\langle G(\cdot) \rangle_{f^{\lambda}_{\theta}}$ achieved at density steady states for a fixed strength of between-group selection $\lambda$. In Figure \ref{fig:lambda_thresh_avg_G_k_model}, we use the expressions from {Equations \eqref{eq:lambdastargrouplocal} and \eqref{eq:averageGgrouplocal}} to plot $\lambda^*$ and $\langle G(\cdot) \rangle_{f^{\lambda}_{\theta}}$ as a function of punishment strength $p$ for various punishment costs $k$. For all values of $k$ considered, we see from Figure \ref{fig:lambda_thresh_avg_G_k_model}(left) that the threshold selection strength $\lambda^*$ decreases with increasing punishment cost $p$, reaching a threshold of $\lambda^* = 0$ when $p \to c$ and the within-group dynamics feature bistability of the all-defector and all-punisher equilibria. From Figure \ref{fig:lambda_thresh_avg_G_k_model}(right), we further see that average payoff at steady state $\langle G(\cdot) \rangle_{f^{\lambda}_{\theta}}$ increases with the strength of punishment $p$ until reaching the average payoff of the all-punisher group when $p \to c$. 

\begin{figure}[!htp]
    \centering
    \includegraphics[width = 0.48 \textwidth]{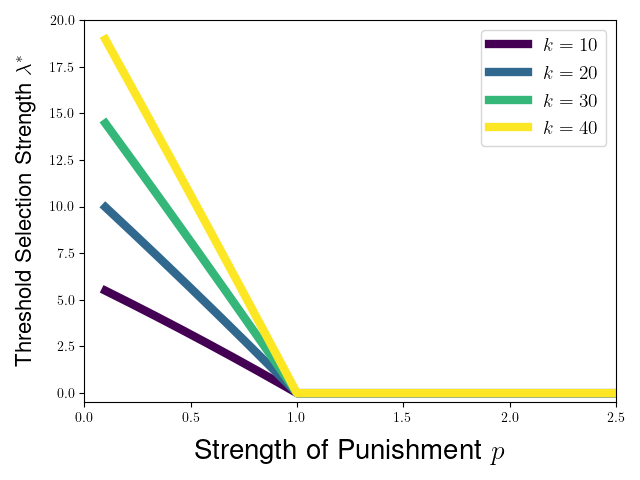}
    \includegraphics[width = 0.48 \textwidth]{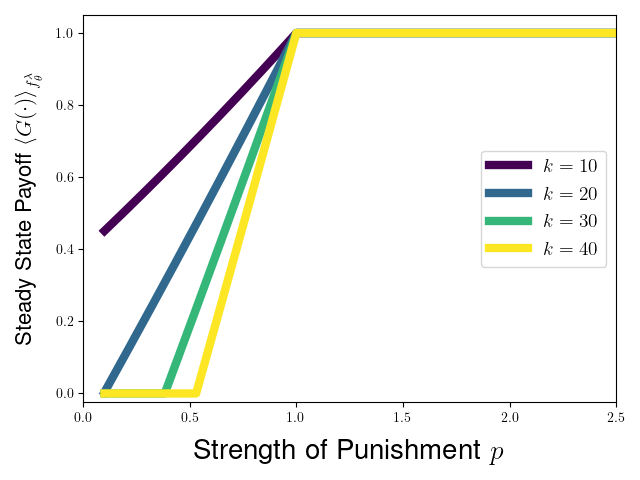}
    \caption{Threshold selection strength $\lambda^*$ (left) and average payoff $\langle G(\cdot) \rangle_{f^{\lambda}_{\theta}}$ (right) for altruistic punishment model with various levels of the per-interaction punishment cost $k$, plotted as a function of the strength of punishment $p$. Other payoff parameters are fixed at $b = 2$, $c = 1$, and $q = 0$, and we calculate $\lambda^*$ and $\langle G(\cdot) \rangle_{f^{\lambda}_{\theta}}$ under the assumption of a fixed group-level selection strength $\lambda = 10$ and H{\"o}lder exponent $\theta = 2$ near $x=1$.}
    \label{fig:lambda_thresh_avg_G_k_model}
\end{figure}

Notably, we can also see from Figure \ref{fig:lambda_thresh_avg_G_k_model} that the threshold selection strength appears to increase with the cost of punishment $k$, while the average payoff at steady state decreases with increasing punishment cost. Although we saw in Figure \ref{fig:group_local_density_k_model}(right) and example of an increased density of groups featuring many altruistic punishers for $k = 4$ relative to the case of $k = 0.1$, it turns out that that steady state density for $k = 4$ actually features a lower average payoff than the average payoff achieved at the steady state for $k = 0.1$. While it may be natural to associate increasing proportions of cooperative behavior with better collective outcomes and higher average payoffs, it turns out that is possible to worsen the collective utility by achieving greater proportions of altruistic punishment. This behavior may be attributable to the fact that our model of altruistic punishment allows the possibility for the collective costs incurred by altruistic punishment to outweigh the collective benefit $b$ gained from the punishers' cooperation in the underlying donation game (due to both the cost $k$ paid by the punisher and the punishment $p$ they impose on defectors). Such a decrease in collective payoff coupled with an increase in cooperative behavior is not possible under multilevel selection {for an underlying donation game} in the presence of {within-group} mechanisms {like} assortment and reciprocity studied in prior work \citep{cooney2022assortment,cooney2023evolutionary}.

{Finally, we consider the case of altruistic punishment when $k = 0$ and there are only fixed costs to punish defectors}.In Figure \ref{fig:lambda_thresh_avg_G_q_model}, we plot the threshold selection strength $\lambda^*$ and the average payoff of group members at steady state $\langle G(\cdot) \rangle_{f^{\lambda}_{\theta}}$ as a function of the punishment for defection {$p$} for various values of the fixed cost of punishment $q$. For certain values of the cost of punishment, we find that both the threshold selection strength and the average payoff have a non-monotonic dependence of the punishment value $k$. In particular, we see that at low levels of the fixed punishment cost $q$, the increase in strength of punishment $p$ always facilitates more cooperative behavior and a greater average payoff at steady state. However, when the fixed punishment $q$ is sufficiently strong, the threshold selection strength can increase and the average payoff can decrease with increasing punishment strength $p$ for certain ranges of punishment strength. In fact, there are cases in which the threshold selection strength (for $q = 0.86$ in Figure \ref{fig:lambda_thresh_avg_G_q_model}, left) and the average payoff (for $q = 0.925$ in Figure \ref{fig:lambda_thresh_avg_G_q_model}, right) reflect a worse collective outcome than for the case of a complete absence of punishment with $p = 0$. We see in Section \ref{sec:LMaltruisticpunishment} of the appendix that such a non-monotonic dependence of threshold selection strength and long-time average payoff on the punishment strength $p$ turns out not to be possible for altruistic punishment in a two-level replicator equation, so we see a genuinely different qualitative behavior by considering the possibility of multilevel selection with pairwise between-group selection.

\begin{figure}[ht]
    \centering
    \includegraphics[width = 0.48 \textwidth]{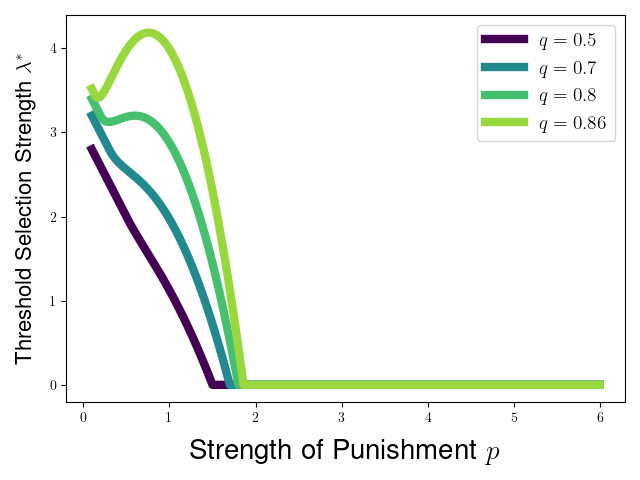}
    \includegraphics[width = 0.48 \textwidth]{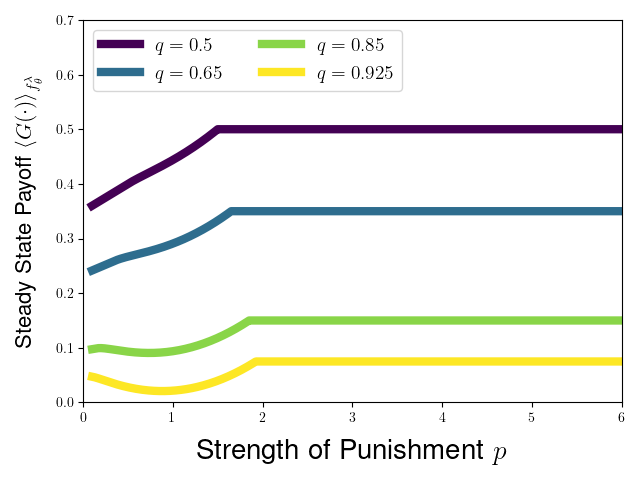}
    \caption{Threshold selection strength $\lambda^*$ and average payoff $\langle G(\cdot) \rangle_{f^{\lambda}_{\theta}}$ for altruistic punishment model with various levels of the fixed cost of punishment $q$, plotted as a function of the strength of punishment $p$. Other payoff parameters are fixed at $b = 2$, $c = 1$, and $k = 0$, and we calculate $\lambda^*$ and $\langle G(\cdot) \rangle_{f^{\lambda}_{\theta}}$ under the assumption of a fixed group-level selection strength $\lambda = 10$ and H{\"o}lder exponent $\theta = 2$ near $x=1$.}
    \label{fig:lambda_thresh_avg_G_q_model}
\end{figure}

\subsection{Comparison with Multilevel Selection Featuring Only Cooperators and Defectors} \label{sec:payoffCDcompare}

{
We can also consider the dynamics of multilevel selection occurring in groups composed only of cooperators and defectors. This will allow us to provide a comparison between the analytical results we can obtain for the multilevel dynamics on the defector-punisher and defector-cooperator edges of the simplex with the behavior of numerical solutions we will obtain in Section \ref{sec:trimorphicnumerics} for multilevel competition on the full defector-cooperator-punisher simplex.

In groups featuring $x$ fraction cooperators and $1-x$ fraction defectors, we see that the payoffs received by cooperators $\pi_C(x)$ and defectors $\pi_D(x)$ are given by
\begin{subequations}
\begin{align}
\pi_C(x) &= b x - c \\
\pi_D(x) &= b x,
\end{align}
\end{subequations}
so the defector's individual-level payoff advantage is
\begin{equation}
\Pi_{DC}(x) = \pi_D(x) - \pi_C(x) =  c
\end{equation}
and the average payoff of group members is given by
\begin{equation}
G(x) = x \pi_C(x) + (1-x) \pi_D(x) = (b-c)x.
\end{equation}
We therefore see that the maximum and minimum possible average group payoffs are given by $G^* = \max_{x \in [0,1]} (b-c)x = b-c$ and $G^* = \min_{x \in [0,1]} (b-c)x = 0$, and we note that the group-level probability based on globally normalized differences in average payoffs takes the form
\begin{equation}
\rho(x,u) = \frac{1}{2} + \frac{1}{2} \left[ \frac{G(x) - G(y)}{G^* - G_*} \right] = \frac{1}{2} + \frac{1}{2} \left[ \frac{\left[(b-c)x\right] - \left[ (b-c) y \right]}{(b-c) - (0)} \right] = \frac{1}{2} + \frac{x-y}{2}.
\end{equation}
Therefore we see that the two group-level victory probabilities we consider in Sections \ref{sec:PDtwotypecooperativefraction} and \ref{sec:grouplocalupdate} coincide for the average payoff $G(x)$ on the defector-cooperator edge of the simplex. Furthermore, the effects of group-level competition are additively separable and can be described in terms of the net group-level reproduction rate $\mc{G}(x)$, resulting in multilevel competition following the PDE
\begin{equation}
\dsdel{f(t,x)}{t} =  c \dsdel{}{x} \left[ x(1-x) f(t,x)\right] + \lambda f(t,x) \left[ \mc{G}(x) - \int_0^1 \mc{G}(y) f(t,y) dy \right],
\end{equation}

We can then apply the results described in Section \ref{sec:existingresults} to see that the threshold selection strength required to achieve steady-state cooperation is given by
\begin{equation}
\lambda^* = c \theta,
\end{equation}
while the average level of cooperation and average fraction of cooperation achieved in the long-time limit are respectively given by
\begin{equation} \label{eq:DCedgecooplongtime}
\lim_{t \to \infty} \int_0^1 \mc{G}(x) f(t,x) dx = \lim_{t \to \infty} \int_0^1 x f(t,x) dx =  \left\{
    \begin{array}{lr}
      0 & : \lambda \leq \lambda^*\\
      1 - \ds\frac{c \theta}{\lambda} & : \lambda > \lambda^*
    \end{array}
  \right. .
\end{equation}
and 
\begin{equation} \label{eq:DCedgepayofflongtime}
\lim_{t \to \infty} \int_0^1 \mc{G}(x) f(t,x) dx = \lim_{t \to \infty} \int_0^1 x f(t,x) dx =  \left\{
    \begin{array}{lr}
      0 & : \lambda \leq \lambda^*\\
      b-c \left( 1 - \ds\frac{c \theta}{\lambda} \right) & : \lambda > \lambda^*
    \end{array}
  \right. .
\end{equation}
}

In Figure \ref{fig:DPvsDCcomparison}, we compare the average payoff achieved at steady state for multilevel selection on the defector-cooperator edge of the simplex with the average payoff achieved on the defector-punisher edge of the simplex. We explore cases of both per-interaction punishment costs and fixed punishment costs, and explore how the difference in average payoff depends on the relative strength of between-group selection $\lambda$. We see that, for the case of per-interaction punishment costs, the {multilevel dynamics feature a lower threshold selection strength $\lambda^*$ and greater average payoff achieved on the defector-punisher edge of the simplex than on the defector-cooperator edge for the case of low per-interaction punishment cost $k = 0.1$ (Figure \ref{fig:DPvsDCcomparison}, bottom-left). However, we see the opposite behavior for the case of stronger punishment cost $k = 50.25$ ((Figure \ref{fig:DPvsDCcomparison}, bottom-right), with the defector-cooperator edge featuring a lower threshold selection strength and greater steady-state average payoff than the defector-punisher. This difference in outcomes in the two cases arises from the fact that the minimal possible group payoff $G_*$ decreases with the punishment cost $k$, and therefore the collective advantage of an all-punisher group will decrease with $k$ and result in an increase in threshold selection strength and a decrease in the average steady-state payoff.} %

The situation is more nuanced when altruistic punishment features fixed punishment costs $q > 0$, {with the relative outcomes featuring a stronger possible dependence on the strength of group-level competition}. In this case, we see in Figure \ref{fig:DPvsDCcomparison}(top-right) that is possible for the multilevel dynamics to achieve greater average payoffs on the defector-cooperator edge of the simplex than on the defector-punisher edge of the simplex for all $\lambda$ when the cost $q$ of punishment is sufficiently large. When the cost $q$ of punishment is lower, we see in Figure \ref{fig:DPvsDCcomparison}(top-left) that it is possible for multilevel dynamics on the defector-punisher edge of the simplex to achieve a greater average payoff than achieved on the defector-punisher edge of the simplex for lower values of between-group selection strength $\lambda$, while multilevel dynamics on defector-cooperator achieve a greater average payoff when between-group selection is sufficiently strong. This suggests that costly punishment can help to sustain cooperative behavior in concert with multilevel competition when competition between groups is relatively weak, but that punishment with fixed costs can be unnecessary -- and actually detrimental -- when between-group selection is strong enough for group-level competition to successfully promote cooperation on its own. 

\begin{figure}[!ht]
    \centering
    \includegraphics[width = 0.48\textwidth]{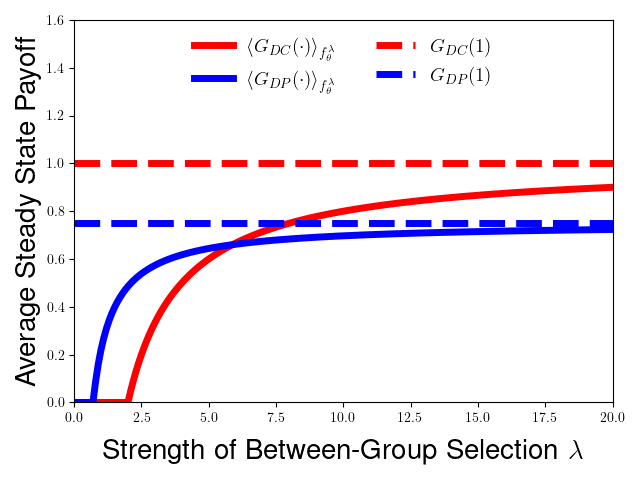}
    \includegraphics[width = 0.48\textwidth]{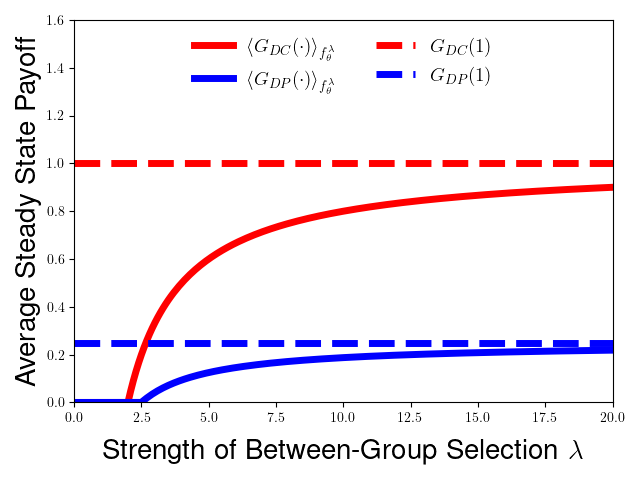}
    \includegraphics[width = 0.48\textwidth]{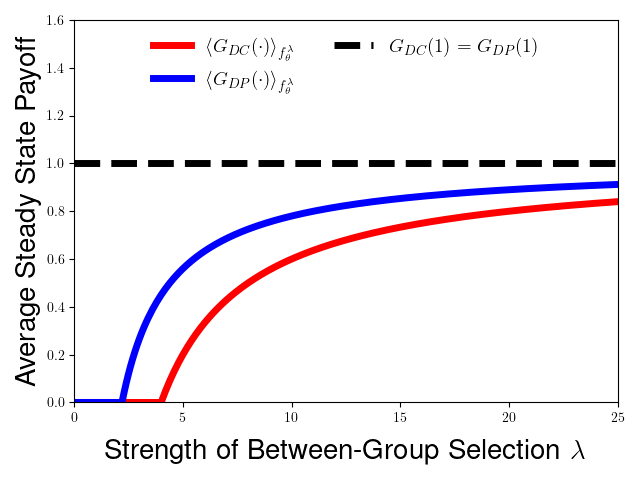}
    \includegraphics[width = 0.48\textwidth]{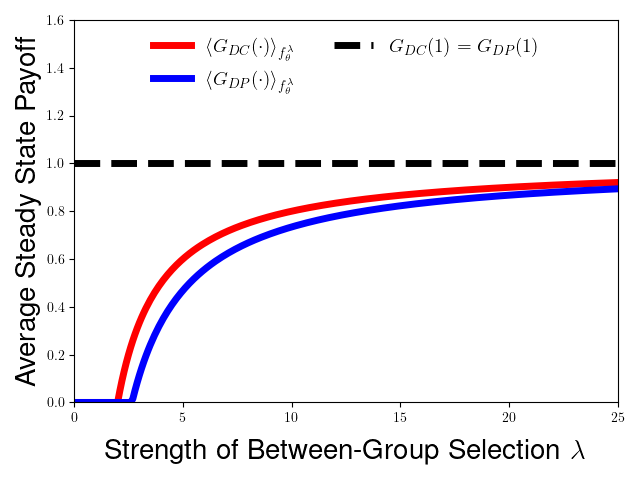}
    \caption{Comparison of the average payoff at steady state achieved by multilevel dynamics on the defector-punisher edge of the simplex ($\langle G_{DP}(\cdot) \rangle_{f^{\lambda}_{\theta}})$) and the defector-cooperator edge of the simplex $\langle G_{DP}(\cdot) \rangle_{f^{\lambda}_{\theta}})$), plotted as a function of the strength of between-group competition $\lambda$. We consider the case of fixed punishment costs with $k = 0$ for both $q = 0.25$ (top-left) and $q = 0.75$ (top-right), as well as the case of per-interaction punishment costs with $q = 0$ for both $k = 0.15$ (bottom-left) and $k = 50.25$ (bottom-right). The average payoffs on the defector-punisher edge and the defector-cooperator edge are plotted with solid blue and red lines, respectively. For the case of fixed punishment costs (top-left and top-right panels), we plot the average payoff achieved in the limit as $\lambda \to \infty$ on the defector-punisher edge (dashed red line) and on the defector-cooperator edge (dashed blue line). For the case of per-interaction punishment costs (bottom-left and bottom-right panels), we plot the average payoff achieved on each edge in the limit as $\lambda \to \infty$ with a dashed black line (as this average payoff agrees for this form of punishment cost). The other payoff parameters are fixed at $b = 2$, $c = 1$, and $p = 0.9$, and we calculate the average payoffs at steady-state with H{\"o}lder exponent $\theta = 2$ near the all-cooperator equilibrium (on the defector-cooperator edge) or near the all-punisher equilibrium (on the defector-punisher edge).}  
    \label{fig:DPvsDCcomparison} 
\end{figure}

\section{Multilevel Competition with Nonlinear Between-Group Replication Rate}
\label{sec:nonlineargroup}

{In this section, we extend our analysis of the two-strategy multilevel dynamics to take into account group-level victory probabilities that are not additively separable, studying the behavior of} for three group-level victory probabilities  $\rho(z,u)$ that are not additively separable functions of the fraction $z$ of altruistic punishers, exploring group-level competition corresponding to the group-level Fermi update rule (Equation \eqref{eq:groupFermi}),  to a group-level pairwise-comparison rule based on locally normalized differences in average payoffs (Equation \eqref{eq:grouppairwiselocal}),  and to a group-level Tullock contest function (Equation \eqref{eq:groupvictoryTullock}). {We start our analysis by applying the preliminary analytical results discussed in Section \ref{sec:pairwiseresults} to understand each of the group-level victory probabilities under consideration (Section \ref{sec:pairwiseanalyticalapplied}), and then we use numerical simulations to validate these analytical predictions and pursue further exploration on the average payoffs achieved under the multilevel dynamics with non-additive probabilities for victory in group-level conflict (Section \ref{sec:pairwisenumerical}.}

\subsection{Analytical Exploration of Non-Additive Victory Probabilities}
\label{sec:pairwiseanalyticalapplied}

{
We can start our application of the analytical predictions from Section \ref{sec:pairwiseresults} to study our model altruistic punishment by discussing how the parameters for punishment impact the long-time average collective success $\int_0^1 \rho(z,1) f(t,z) dz$ of the population under pairwise competition with the all-punisher group. Using the conjectured expression from Equation \eqref{eq:pairwiserholongtime}, we can apply our expressions for the individual-level incentive to defect $\Pi(z)$ and the conditions for stability of the all-punisher equilibrium to obtain our anticipated analytical expression for this measurement of long-time collective success, which is given by the following piecewise characterization}
{
\renewcommand{\arraystretch}{2}
\begin{equation} \label{eq:rhoz1longtime}
\begin{aligned}
\lim_{t \to \infty} \int_0^1 \rho(z,1) f(t,z) dz 
&=  \left\{ \begin{array}{cl}
    \rho(0,1) &: p < c + q,  \lambda \leq \lambda^*_{PW}  \\
     \ds\frac{1}{2} - \frac{\theta \left( c + q - p \right)}{2 \lambda}  &: p < c + q,  \lambda > \lambda^*_{PW}  \\
 \ds\frac{1}{2} &: p > c + q
\end{array}
\right. .
\end{aligned}
\end{equation}
\renewcommand{\arraystretch}{1}
}
{Therefore we expect the collective success to increase linearly with the strength of punishment $p$, decrease linearly with the fixed cost of punishment $q$, and have no dependence on the per-interaction cost of punishment $k$. In addition, we see that the collective success of the steady-state population will take the same form for each group-level victory probability $\rho(z,u)$ provided that $\lambda > \lambda^*_{PW}$. Therefore the key differences in the conjectured behavior will depend primarily on the values $\rho(0,1)$ and $\lambda^*$, so we we will use these quantities as a way to compare these different models for pairwise competition between groups. }

{In Table \ref{tab:rhotable}, we provide a comparison of the qualitative behaviors in our conjectured formulas for the threshold selection strength $\lambda^*_{PW}$ and the collective success for different group-level victory probabilities, using the value of $\rho(0,1)$ as a measurement distinguishing the long-time collective collective success for different models of group-level competition. Although we have already obtained more complete results for the long-time behavior for the cases of additively separable group-level victory probabilities in Section \ref{sec:analytical}, we include the calculations for $\lambda^*_{PW}$ and $\rho(0,1)$ in Table \ref{tab:rhotable} for the purpose of comparison with our non-additive victory probabilities. }

\renewcommand{\arraystretch}{2} 
\begin{table}[!ht]
\caption{ {Summary of analytical characterization of properties for multilevel selection for a variety of group-level victory probabilities. For each victory probability considered, we present the predicted threshold selection strength for the survival of altruistic punishment from Proposition \ref{prop:pairwisesteady} and the probability of group-level victory for an all-defector group when paired with an all-punisher group in pairwise conflict. } }
\begin{center}
\footnotesize
\begin{tabular}{|c|c|c|c|}
\hline
\makecell{Group \\ Competition \\ Scenario} & \makecell{Victory Probability \\ $\rho(z,u)$}  &   \makecell{Threshold Selection \\
Strength $\lambda^*_{PW}$}  & \makecell{All-Defector Collective \\ Success $\rho(0,1)$} \\
    \hline
\makecell{Fraction of \\ Altruistic \\ Punishers} 
& $\ds\frac{1}{2} + \ds\frac{z-u}{2}$ & $\theta(c+q-p)$ & 0 \\ 
\hline
\makecell{Local Update \\ (Globally \\ Normalized)} & $\ds\frac{1}{2} \left[ 1 + \ds\frac{G(z) - G(u)}{G^* - G_*} \right]$ & $\ds\frac{\theta (c+q-p) (G^* - G_*)}{b - c - q}$  & $\ds\frac{1}{2} \left[ 1 - \frac{b-c-q}{G^* - G_*} \right]$ \\
\hline
\makecell{Local Update \\ (Pairwise \\ Normalized)} & $\ds\frac{G(z) - G(u)}{|G(z)| + |G(0)|}$ & $\theta (c+q -p)$ & 0 
\\
\hline 
Fermi Rule & $\ds\frac{1}{2} + \ds\frac{1}{2} \tanh\left( s \left[ G(z) - G(u) \right] \right)$ & $\ds\frac{\theta (c+q-p)}{\tanh\left( s \left[ b-c-q\right]\right)}$ & $\ds\frac{1}{2} - \frac{1}{2} \tanh\left( s \left[ b - c - q\right] \right)$\\
\hline
\makecell{Tullock  \\ Contest \\ Function} & $\frac{\left(G(x) - G_{*}\right)^{1/a}}{\left(G(x) - G_{*}\right)^{1/a} + \left(G(x) - G_{*}\right)^{1/a}}$ & $\frac{\theta (c+q-p) \left[ \left(b-c-q - G_{*}\right)^{1/a} + \left(- G_{*}\right)^{1/a} \right] }{\left( b - c - q - G_{*} \right)^{1/a} - \left(- G_{*}\right)^{1/a}}$ & $\frac{\left(- G_{*}\right)^{1/a}}{\left(b-c-q - G_{*}\right)^{1/a} + \left(- G_{*}\right)^{1/a}}$
\\
   \hline 
\end{tabular}
\end{center}
\label{tab:rhotable}
\end{table}
\renewcommand{\arraystretch}{1}

{In particular, we see that the threshold selection strength $\lambda^*$ always decreases with the strength of punishment $p$ and increases with the fixed punishment cost $q$ for the group-level Fermi rule and the group-level local update rule with pairwise normalization, while $\lambda^*$ is independent of the per-interaction punishment cost $k$ for these two victory probabilities. However, we ee that that the threshold selection strength $\lambda^*$ can have a more subtle dependence of the punishment parameters for the Tullock contest function due to the presence of the minimal possible average payoff $G_*$ in the group-level victory probability. This motivates comparing the average collective payoff achieved. We also see  that the per-interaction punishment cost $k$ does not appear in the threshold selection strength

While the threshold selection strength $\lambda^*$ and the collective success $\int_0^1 \rho(z,1) f(z) dz$ against the all-punisher group are only derived as necessary conditions for a possible steady state solution to our PDE model for pairwise competition from Equation \eqref{eq:multilevelPDEtworho}, we can try to compare these conjectured analytical formulas with the long-time behavior for numerical simulations of our model. In addition, our characterization of quantities associated with the average group-level victory probability $\rho(z,u)$ do not necessarily correspond to conclusions we can draw about the average long-time payoff $\lim_{t \to \infty} \int_0^1 G(z) f(t,z) dz$, so we will use numerical simulations in Section \ref{sec:pairwisenumerical} to explore the long-time behavior of the average payoff of the population under multilevel selection with a range of functions determining victory in group-level conflict.
}

\subsection{Behavior of Numerical Solutions for Pairwise Group-Level Competition with Non-Additive Group-Level Victory Probabilitie}
\label{sec:pairwisenumerical}

{We will now begin our numerical exploration of solutions for the multilevel dynamics of Equation \eqref{eq:multilevelPDEtworho} for various group-level victory probabilities that are not additively separable. We perform numerical simulations using an upwind finite-volume scheme that we describe in Section \ref{sec:twostrategyfv}, studying numerical solutions for our PDE model of multilevel selection starting from a uniform initial distribution of strategy compositions within groups.} In Figure \ref{fig:tanhtrajectory}, we present snapshots of the time-dependent solutions to the multilevel dynamics of Equation \eqref{eq:multilevelPDEtworho} with the group-level victory probability given by {the Fermi group-level update rule of Equation \eqref{eq:groupFermi}}. For a small value of the relative between-group selection strength $\lambda = 0.1$, we see that the population competition quickly concentrates towards the all-defector composition {$z = 0$} and the altruistic cooperators are eliminated from the population (Figure \ref{fig:tanhtrajectory}, left). For a large value of the between-group selection strength $\lambda = 2$, we see that the numerical solutions starting from an initial uniform distribution appear to converge to a steady-state density supporting many groups with high proportions of altruistic punishers.

 \begin{figure}[!htb]
 	\centering
 	\includegraphics[width = 0.48\textwidth]{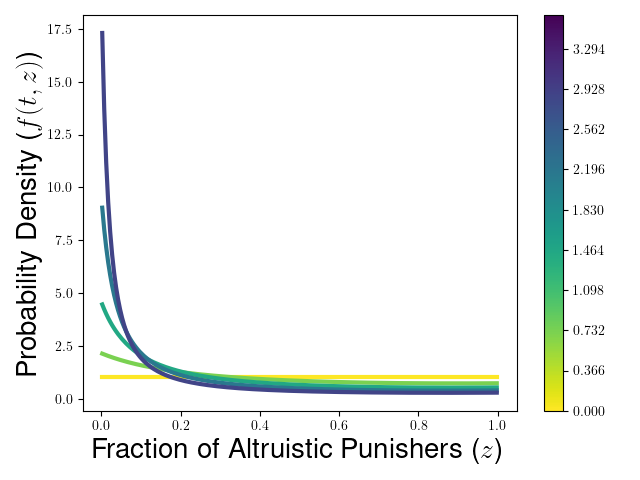}
  \includegraphics[width = 0.48\textwidth]{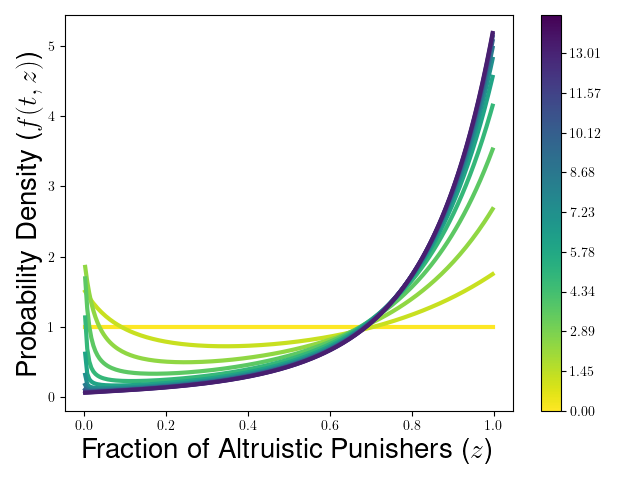}
	\caption{Snapshots of densities obtained at various times $t$ using upwind finite volume simulation of multilevel PDE model starting from uniform initial distribution for the case per-interaction punishment costs. We consider values of $\lambda = 0.1$ (left) and $\lambda = 2$ (right) of the relative strength of between-group competition. The color of the densities plotted corresponds to the time at which the solution was evaluated, with the values of time given in the colorbar next to each panel. The payoff parameters for both panels were fixed at $b = 2$, $c = 1$, $p = 0.5$, $k = 0.1$, and $q = 0$, {the group-level Fermi rule used the payoff sensitivity parameter $s = 2$,} and the numerical simulations were run for $600$ time-steps (left) and $2400$ time-steps (right) with step-size $\Delta t = 0.006$.  
 }
 \label{fig:tanhtrajectory}
 \end{figure}

In Figure \ref{fig:tanhnumericalsteady}, we look to compare the long-time behavior of our model of multilevel selection with the Fermi group-level victory probability for different strengths of punishment $p$. We present the steady state densities achieved under the multilevel dynamics after 9,600 timesteps for the case of per-interaction punishment costs $k > 0$, assuming the uniform strategic distribution of groups for each strength of punishment. We see that increasing the strength of punishment results in steady-state densities featuring increasing fractions of altruistic punishers. We see that the proportion of groups with high fractions of altruistic punishers increases with the strength of punishment $p$, with many group concentrated near the all-punisher equilibrium for the largest value of $p$ considered in Figure \ref{fig:tanhnumericalsteady}.  

 \begin{figure}[!htb]
 	\centering
 	\includegraphics[width = 0.6\textwidth]{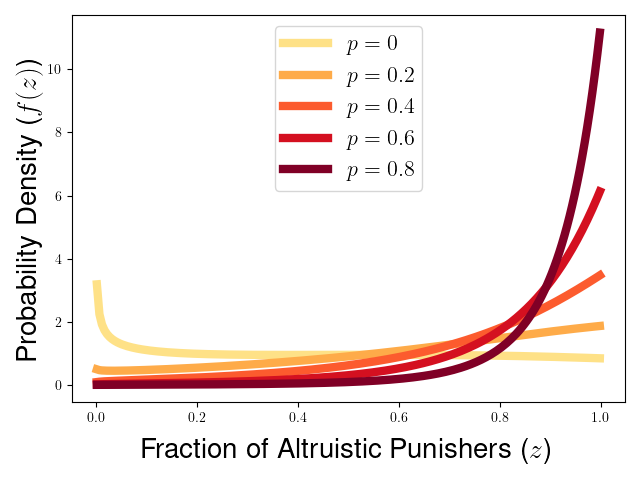}
	\caption{Comparison of numerically computed solutions after 9,600 time-steps of step size $\Delta t = 0.006$ under Fermi group-level victory probability for various values of the strength of punishment $p$. The other payoff parameters were fixed at $b = 2$, $c = 1$, {$k =0.1$}, and $q = 0$, the relative strength of between-group selection is fixed at $\lambda = 1.55$, the group-level Fermi rule used the payoff sensitivity parameter $s = 2$, and the population of groups was initialized with a uniform strategic distribution.}
 \label{fig:tanhnumericalsteady}
 \end{figure}

{We can also use our numerical simulations of the multilevel dynamics starting from a uniform initial distribution to explore the conjectured form of the long-time collective success $\lim_{t \to \infty} \int_0^1 \rho(z,1) f(t,z) dz$ against the all-punisher group for pairwise competition against the all-cooperator group. In Figure \ref{fig:steadyrho1xcompare}, we provide comparisons of numerical calculations for the collective outcome $\int_0^1 \rho(z,1) f(t,z) dz$ after many time-steps and the analytical prediction from Equation \eqref{eq:rhoz1longtime} for the long-time behavior of the population for the case of H{\"o}lder exponent $\theta = 1$ near $z=1$, exploring how these quantities change as we increase the strength of punishment $p$. We provide these comparisons for the group-level Fermi rule (Figure \ref{fig:steadyrho1xcompare}, top-left) and the local update rule with pairwise normalization (\ref{fig:steadyrho1xcompare}, top-right) for the case of fixed punishment cost, as well as the Tullock constant function for two values of the sensitivity parameter $a$ (\ref{fig:steadyrho1xcompare}, bottom-left and bottom-right) in the case of per-interaction punishment cost. We find good agreement between the numerical solutions and the predicted analytical formulas in all cases, and we see that all four cases describe a phase in which the long-time collective success is decribed by $\rho(0,1)$ for low strength of punishment, before increasing linearly with $p$ at intermediate punishment strength (for $p$ allowing $\frac{1}{2} - \frac{(c+q-p)}{2 \lambda} > \rho(0,1)$), before leveling off at the collective success of an all-punisher group when $p > c + q$ (resulting in $\int_0^1 \rho(z,1) f(z) dz = \frac{1}{2}$). Notably, the example of the Tullock constant function with $a > 0$ features two phases of growth of collective success, as $\rho(0,1)$ increases with $p$ in this case before a steady-state density is achieved when $p$ is large enough to allow $\lambda > \lambda^*$ (Figure \ref{fig:steadyrho1xcompare}, bottom-right).

\begin{figure}[!htp]
    \centering
    \includegraphics[width=0.48\linewidth]{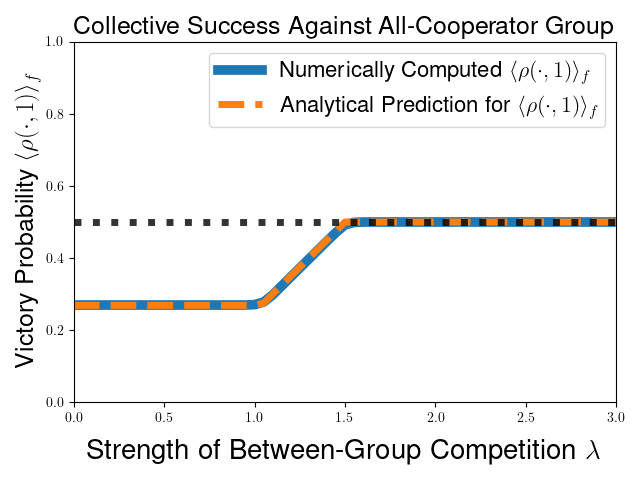}
    \includegraphics[width=0.48\linewidth]{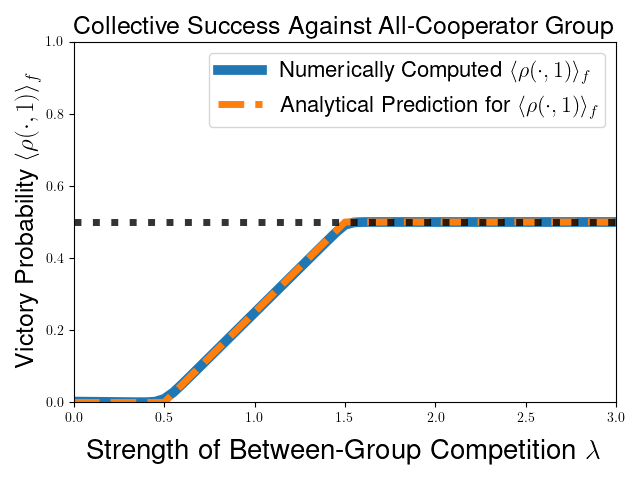}
        \includegraphics[width=0.48\linewidth]{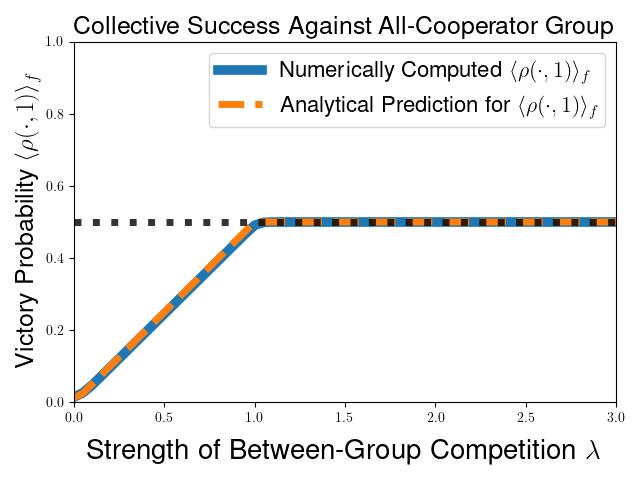}
    \includegraphics[width=0.48\linewidth]{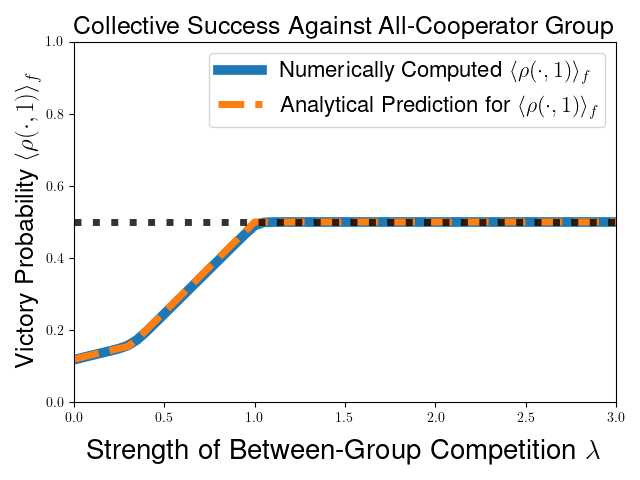}
    \caption{{Comparison of analytical prediction of the long-time average group-level victory probability $\lim_{t \to \infty} \int_0^1 \rho(z,1) f(t,z) dz$ for H{\"o}lder exponent $\theta = 1$ near $z=1$ (dashed orange line) and the numerical value of this collective success function calculated after 9,600 time-steps of step size $\Delta t = 0.01$ for the pairwise multilevel dynamics of Equation \eqref{eq:multilevelPDEtworho} (solid blue line), plotted as a function of the strength of punishment $p$. Simulations run for the group-level Fermi update rule with sensitivity parameter $s=1$ (top-left), the local victory probability with pairwise normalization (top-right), and the Tullock contest function for the cases of sensitivity parameter $a = 0.5$ (bottom-left) and $a = 1.1$ (bottom-right). The game-theoretic parameters were fixed at $b = 2$ and $c = 1$ in all panels, while the cases of the Fermi and local update rules considered punishment costs $q = 0.5$ and $k =0$ (top-left and top-right) and the cases of the Tullock contest function considered punsihment costs $q = 0$ and $k = 2$ (bottom-left and bottom-right).} {The black horizontal dotted lines indicate a probability of group-level victory of $\int_0^1 \rho(z,1) f(z) dz = \frac{1}{2}$, corresponding to a population that has a fifty-fifty chance of winning a pairwise conflict with an all-punishe.  group.} }
    \label{fig:steadyrho1xcompare}
\end{figure}
}

 We can also plot the average level of cooperation and the average payoff of group members after a large number of time-steps for various strengths of between-group selection $\lambda$ and strengths of punishment $q$ for defection {for the group-level Fermi update rule}. Using the plot of average long-time payoff (Figure \ref{fig:nonlinearDPavgGandCoop}, left), we see that there are strengths of fixed punishment costs $q$ and fixed values of $\lambda$ for which varying the strength of punishment {$p$} can allow for the possibility of convergence to a delta-function at the all-defector equilibrium {(for $p < c + q$ and $\lambda \leq \lambda^*_{PW})$}, convergence to a steady state density featuring coexistence of many different compositions of defectors and altruistic punishers within-groups {(for $p < c + q$ and $\lambda \geq \lambda^*_{PW}$)}, and convergence upon a delta-function at the all-punisher equilibrium {(for $p \geq c + q$)}. 

 \begin{figure}[!htb]
    \centering
    \includegraphics[width = 0.48\textwidth]{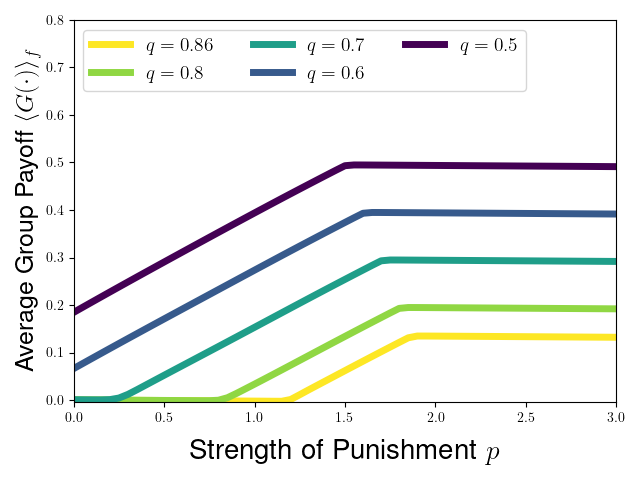}
    \includegraphics[width = 0.48\textwidth]{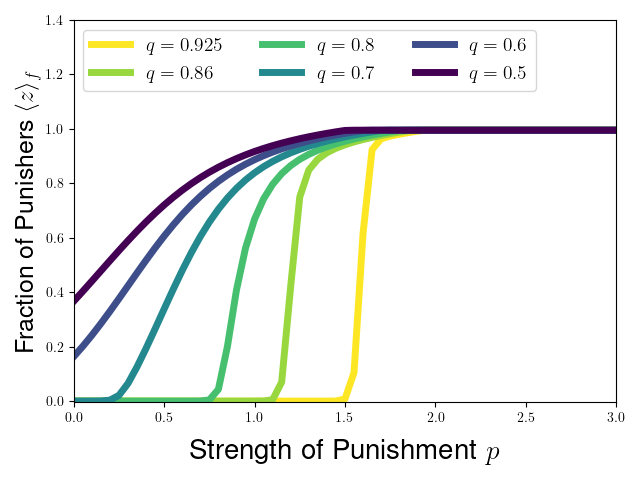}
    \caption{Average payoff (left) and fraction of altruistic punishers (right) achieved by numerical solutions of the model after 9600 time-steps with step-size $\Delta t = 0.006$ for the model with fixed cost of punishment $q > 0$ (and $k = 0$), plotted as a function of the strength of punishment $p$. Each line corresponds to a different fixed cost $q$ for punishment. The other payoff parameters were fixed at $b = 2$, $c = 1$,, the strength of between-group selection was $\lambda = 5$,  the sensitivity parameter for group payoff in the Fermi update rule was fixed at $s = 1$, and the initial strategic distribution of groups was a uniform density (corresponding to a strategic distribution with H{\"o}lder exponent $\theta = 1$ near $z = 1$). }
    \label{fig:nonlinearDPavgGandCoop}
\end{figure}
 
 For fixed punishment cost $q$, we see that the average payoff and fraction of punishers at steady state increases with the strength of punishment $p$ until reaching the values achieved for the steady state at which the population consists entirely of altruistic punishers. For each strength of punishment $p$, we see that average payoff and fraction of punishers at steady state decreases as we increase the fixed cost $q$ of punishing defectors. Unlike the case of the globally normalized local group-level update rule studied in Section \ref{sec:grouplocalupdate}, we see {that the multilevel dynamics with the Fermi group-level update rule can feature} monotonic dependence of the average payoff and average levels of cooperation on the punishment strength $p$ and punishment costs $q$, which aligns with an intuitive expectation that increasing the strength of punishment or decreasing the cost of punishment should be expected to ease the ability to achieve cooperation via multilevel selection. 

In Figure \ref{fig:nonlinearkmodel}, we also consider the long-time average payoff and average level of altruistic punishment achieved {under the group-level Fermi rule} in the case in which there are only per-interaction punishment costs (with $k > 0$ and $q = 0$). In this case, the average payoff at steady state appears to be a decreasing function of the punishment cost $k$ (Figure \ref{fig:nonlinearkmodel}, right). Despite the fact that average payoff appears to decrease with increasing per-interaction punishment cost $k$, we also see in Figure \ref{fig:nonlinearkmodel}(right) that the fraction of altruistic punishment achieved after 9600 time-steps increases with increasing punishment cost under multilevel selection with the Fermi group-level victory probability. This agrees with the behavior seen in the case of the density steady states plotted in Figure \ref{fig:group_local_density_k_model} for the globally normalized local group update rule, in which a greater level of altruistic punishment was seen at steady state in the panel featuring a higher cost of per-interaction punishment cost $k$. 

The observation that average payoff achieved under multilevel selection can decrease with increasing per-interaction punishment cost in the case of the group-level Fermi victory probability is consistent with the behavior seen in the case of the globally normalized group-level local update rule studied in Section \ref{sec:grouplocalupdate}. In the latter case, we saw above in Figure \ref{fig:group_local_density_k_model}(right) that the analytically computed average payoff decreased with punishment strength $k$. However, in Section \ref{sec:LMaltruisticpunishment} of the appendix, we explore the case in which multilevel selection follows a two-level replicator equation, and see that the average payoff at steady state is independent of the per-interaction punishment cost. This suggests that this dependence of long-time payoff on the per-interaction punishment cost may be a feature specific to our models of pairwise between-group competition, and that incorporating group-level frequency dependence in competition between groups can produce different qualitative behaviors from the long-time collective outcome seen in two-level replicator equation featuring frequency-independent group-level competition. 

\begin{figure}[!htb]
    \centering
    \includegraphics[width = 0.48\textwidth]{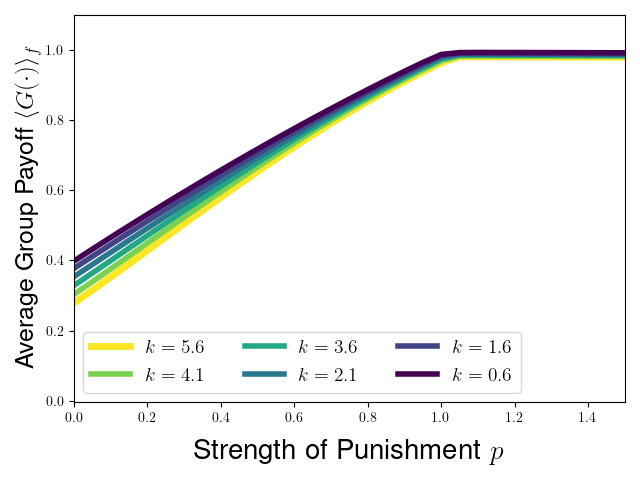}
     \includegraphics[width = 0.48\textwidth]{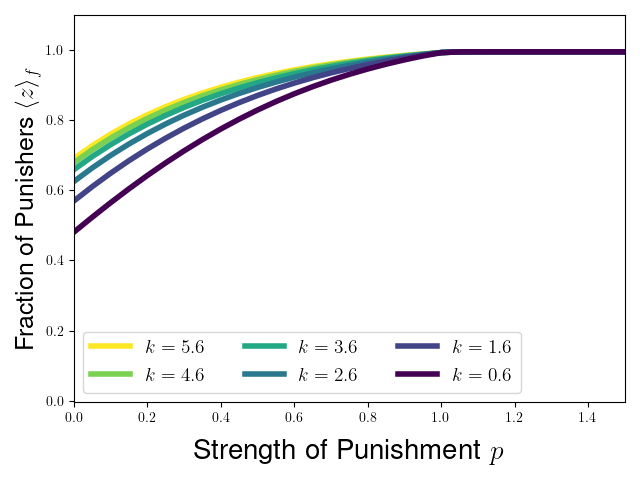}
    \caption{Average payoff (left) and average fraction of altruistic punishers (right) for numerical solutions after 9,600 time-steps with step-size $\Delta t = 0.006$ of multilevel model with Fermi group-level victory probability in the case of per-interaction costs of punishment (with $k > 0$ and $q = 0$). Each line corresponds to a different per-interaction cost of punishment $k$. The other parameters were fixed at the values $b = 2$, $c = 1$,  $s = 1$, and $\lambda = 2$,  and the simulations were run from a uniform initial strategic distribution of groups. } 
    \label{fig:nonlinearkmodel}
\end{figure}

We can also explore the behavior of numerical solutions for a variety of other pairwise group-level victory probabilities. In Figure \ref{fig:qmodellocalTullock}, {we plot the average payoff achieved by the multilevel dynamics of Equation \eqref{eq:PDEtrimorphicrho} after 9,600 time-steps in the case of fixed punishment costs ($q > 0$). for both the local group-level victory probability with pairwise normalization and for the Tullock contest function}.  Unlike the case of globally normalized group-level competition studied in Section \ref{sec:grouplocalupdate},  we see from Figure \ref{fig:qmodellocalTullock}(left)  that the average payoff achieved at steady state appears to be an increasing function of the punishment strength $p$ for all fixed punishment costs $q$ under consideration.  %
 {For the Tullock contest function, we see from  Figure \ref{fig:qmodellocalTullock}(right), that,  for sufficiently strong fixed costs of punishment $q$, the average payoff achieved after a large number of time-steps can feature a non-monotonic dependence of the strength of punishment $p$. In particular, as in the analytical results obtained the case of the local pairwise-comparison group update rule with global normalization studied in Section \ref{sec:grouplocalupdate}, we see that it is possible that strengthening punishment can cause a marginal decrease in the collective outcome for the population. Interesting, this non-monotonic behavior is not seen our analytical expression in the average group-level victory probability $\int_0^1 \rho(z,1) f(z) dz$ against the all-punisher group (which predicts a non-decreasing victory probability as $p$ increases), so we see that the numerical calculations of the long-time average payoff $\langle G(\cdot)\rangle_f$ provides additional insight into the behavior under the Tullock group-level contest function beyond the anayltical characterization from Section \ref{sec:pairwiseresults}. }

 \begin{figure}[!htpb]
     \centering
     \includegraphics[width = 0.48\textwidth]{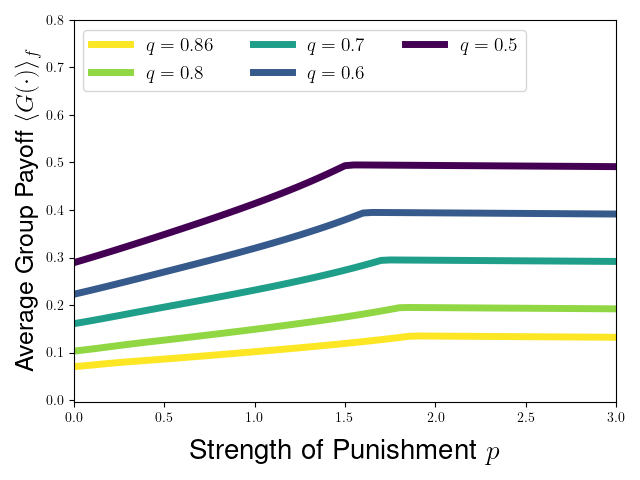}
      \includegraphics[width = 0.48\textwidth]{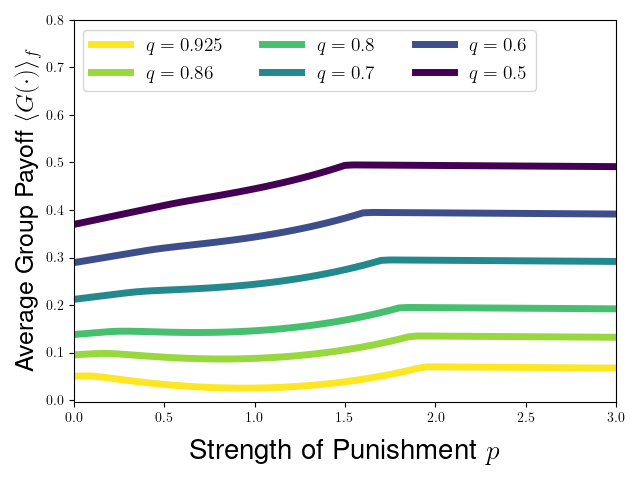}
     \caption{{Average payoff achieved by numerical solutions of the model after 9600 time-steps for model with fixed cost of punishment $q > 0$ (and $k = 0$), plotted as a function of the strength of punishment $p$. The simulations used the local group update rule with pairwise normalization (left) and the group-level Tullock contest function with sensitivity parameter $a = 0.5$ (right).} Each line corresponds to a different fixed cost $q$ for punishment.  The other parameters were fixed at the same values as in {Figure \ref{fig:nonlinearDPavgGandCoop}}, with $b = 2$, $c = 1$,, $\lambda = 5$, and the simulations were run from a uniform initial strategic distribution of groups.}
     \label{fig:qmodellocalTullock}
 \end{figure}

From studying multilevel dynamics with fixed costs of punishment with both the globally-normalized and pairwise-normalized local group update rules, we see that different qualitative behaviors appear to arise depending on the form we assume for group-level victory probabilities and the manner in which me normalize differences in collective average payoffs to produce a probability of winning a pairwise group conflict.  The discrepancy between the monotonic and non-monotonic dependence of average payoff on the strength of punishment between the globally and pairwise normalized local update rules raises questions as to whether one of these update rules is a better choice for understanding how differences in the average payoffs of groups may impact a group's ability in pairwise group-level competition. This suggests that further numerical and analytical study of such models of multilevel selection with pairwise group-level competition may be helpful for understanding the role of group-level frequency dependence on evolutionary dynamics operating across multiple levels of selection.

\begin{remark}
From our exploration of various group-level victory probabilities,  we saw a non-monotonic dependence of average long-time payoff on the strength of punishment $p$ only for the globally normalized local group update rule and the group-level Tullock contest function.  These were the two group-level victory probabilities we considered that depended on either the maximum or minimum possible average group payoffs $G^*$ and $G_{*}$,  which potentially suggests that the non-monotonic behavior is a consequence of the decrease in the minimum possible payoff $G_*$ that occurs when the strength of punishment increases.  This distinction between these two victory probabilities and the group-level victory probabilities  $\rho(z,u)$ depending only on the average payoffs $G(z)$ and $G(u)$ of the competing groups (such as in the group-level Fermi rule and the locally normalized local update rule) suggests that the modeling choice of a victory probability will depend on whether we would like to describe competition between groups based on absolute differences in payoffs or based on relative differences in collective payoffs relative set context by baseline reference payoffs. 
\end{remark}

\section{Multilevel Dynamics with Defectors, Cooperators, and Altruistic Punishers}
\label{sec:trimorphicnumerics}

So far, we have only considered the dynamics of multilevel selection when groups can be composed of two types of individuals, either considering groups featuring only defectors and cooperators or groups featuring only defectors and altruistic punishers. It is also natural to ask how our model of multilevel selection with within-group altruistic punishment behaves when groups can take any composition on the three-strategy simplex. In this section, we consider multilevel dynamics in the case of evolutionary competition in the presence of all three strategies. We present results of numerical simulations of an upwind finite-volume scheme for our trimorphic PDE model for multilevel selection, exploring how competition between-groups can combine with individual-level punishment to promote cooperative behavior in the long-run population. 

We will focus on the case of additively separable group-level victory functions of the form
\begin{equation}
\rho(x,y ; u,v) = \frac{1}{2} + \frac{1}{2} \left[ \mc{G}(x,y) - \mc{G}(u,v) \right]
\end{equation}
for a function $\mc{G}(x,y)$ satisfying $|\mc{G}(x,y)| \leq 1$ for all possible strategic compositions $(x,y), (u,v) \in \Delta^2$. For group-level replication rates of this form, we see that 
\begin{equation}
\rho(x,y;u,v) - \rho(u,v; x,y) = \mc{G}(x,y) - \mc{G}(u,v),
\end{equation}
and therefore the multilevel dynamics of Equation \eqref{eq:PDEtrimorphicrho} take the following form
\begin{equation} \label{eq:PDEtrimorphicGdiff}
\begin{aligned}
\dsdel{f(t,x,y)}{t} &= - \dsdel{}{x} \left[ x \left\{ (1-x)  \left(\pi_C(x,y) - \pi_P(x,y) \right) - y \left( \pi_D(x,y) - \pi_P(x,y) \right) \right\} f(t,x,y) \right] \\
&-   \dsdel{}{y} \left[ y \left\{ (1-y)  \left(\pi_D(x,y) - \pi_P(x,y) \right) - x \left( \pi_C(x,y) - \pi_P(x,y) \right) \right\} f(t,x,y) \right]  \\
&+ \lambda f(t,x,y) \left[ \mc{G}(x,y) - \int_0^1 \int_0^{1-u} G(u,v) f(t,u,v) dv du \right].
\end{aligned}
\end{equation}
This PDE is in the form of a two-level replicator equation for multilevel selection with three types of individuals, which has previously been studied in a model for multilevel dynamics of protocell evolution featuring three types of genetic replicators \citep{cooney2022pde}. To compute numerical solutions for this PDE, we adapt the finite volume approach used to study the existing model for protocell evolution. We present the details of this numerical scheme in Section \ref{sec:FVtrimorphic}.

In the remainder of this section, we focus on two additively separable group-level victory probabilities, highlighting group-level victory probabilities that we previously studied when restricting to the case of two strategies. In Section \ref{sec:trimorphicdefectdiff}, we study numerical solutions for the pairwise group-level victory probability introduced by Boyd and coauthors that describes the chances of group-level success in terms of the difference in the fraction of defectors between two competing groups. In Section \ref{sec:trimorphiclocalgroup}, we consider trimorphic multilevel dynamics in the case of the group-level victory determined by the difference in average payoffs of competing groups normalized by the maximum possible difference in average group payoff $G^* - G_{*}$. 

{
\begin{remark}
Although we studied a more general family of group-level victory probabilities in the case of two-strategy multilevel competition, we focus on the case of additively separable group-level victory probabilities for the three-strategy case for simplicity and computational considerations. When discretizing our PDE model with pairwise group-level conflict for a group featuring $d$ strategies, we describe the composition of groups as points $\vec{x} = (x_1,x_2,\cdots,x_{d-1})$on a $(d-1)-$dimensional simplex $\Delta^{d-1}$ and the group-level victory probabilities for a conflict between groups with composition vectors $\vec{x}$ and $\vec{u}$ of groups as a function $\rho(\vec{x},\vec{y}) : \Delta^{d-1} \times \Delta^{d-1} \to [0,1]$. This means that we need a $2(d-1)$-dimensional description of the group-level victory probability, and discretizing our corresponding PDE model to use a finite volume simulation will require representing this discretized victory probability as a $2(d-1)$th-order tensor. This approach was applied in the case of $d=2$ strategies for the numerical simulations in Section \ref{sec:pairwisenumerical} (see Section \ref{sec:twostrategyfv} for details on implementation), but applying this approach for three or more strategies is more difficulty due to the increased dimensionality the group-level victory probability. For the case of additively separable group-level victory probabilities, we are able to represent our $2(d-1)$-dimensional group-level victory function $\rho(\vec{x},\vec{u})$ into a $(d-1)$-dimensional net group-level reproduction rate $\mc{G}(\vec{x})$, which allows a much more efficient numerical implementation of our PDE models in this case. 
\end{remark}
}

\subsection{Trimorphic Multilevel Competition Depending on Differences in Fraction of Defectors}
\label{sec:trimorphicdefectdiff}

We first look to understand the dynamics of multilevel selection in the version of the PDE model with group-level victory probability given by
\begin{equation}
\rho\left(x,y ; u,v \right) = \frac{1}{2} + \frac{1}{2} \left[ \left( 1 - y\right) - \left( 1 - v \right) \right] = \frac{1}{2} + \frac{1}{2} \left[  v - y\right]
\end{equation}
in which the group-level victory probability depends on the difference in fraction of individuals displaying cooperative behavior within the two groups (namely we are considering the fraction of individuals who are either pure cooperators ($C$) or altruistic punishers ($P$) in each group). This is the version of pairwise between-group competition considered in the stochastic model employed by Boyd and coauthors \cite{boyd2003evolution}. This choice of group-level victory probability corresponds to a net group-level reproduction rate 
\begin{equation} \label{eq:nettrimorphiccoop}
\mc{G}(x,y) = 1 - y,
\end{equation}
and we can study multilevel competition with this group-level victory probability by studying numerical solutions of Equation \eqref{eq:PDEtrimorphicGdiff} with this net group-level reproduction rate. 

In Figure \ref{fig:trimorphicnumericscoop}, we present snapshots of numerical solution to the multilevel dynamics of Equation \eqref{eq:PDEtrimorphicGdiff} in the case of per-interaction punishment costs ($p > 0$, $q=0$) and group-level competition featuring the net reproduction rate $\mc{G}(x,y) = 1 -y$. Starting from a uniform initial density of group compositions, we plot how the density of strategic compositions changes over time. After only 10 time-steps, we see that the population of groups has moved towards compositions featuring many cooperators and altruistic punishers. As time progresses the fraction of defectors further decreases, but the distribution of strategic compositions also places more weight on groups close to the all-punisher equilibrium. By 20,000 time-steps, we see that the vast majority of groups have achieved a composition mostly composed of altruistic punishers, indicating that multilevel selection can support the persistence of a high level of altruistic punishment in the case of per-interaction punishment costs and pairwise group-level competition depending only differences in the fraction of defectors in each group.

\begin{figure}[htbp]
    \centering
    \includegraphics[width = 0.48 \textwidth,height = 0.42\textwidth]{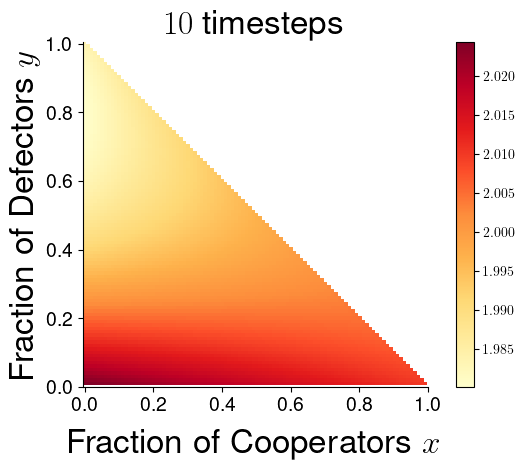}
\includegraphics[width = 0.48 \textwidth,height = 0.42\textwidth]{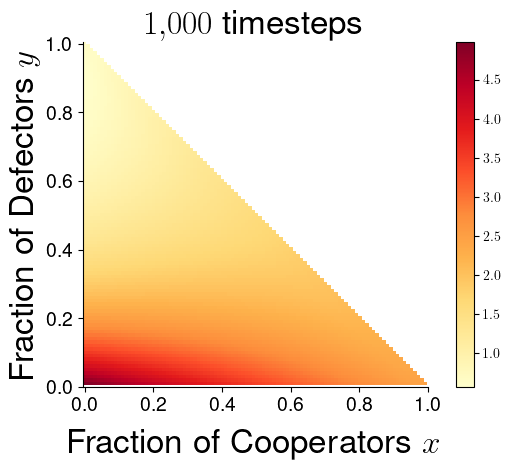}
 \includegraphics[width = 0.48 \textwidth,height = 0.42\textwidth]{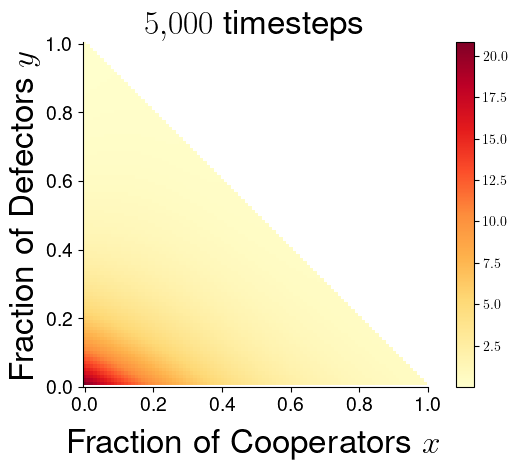}
 \includegraphics[width = 0.48 \textwidth,height = 0.42\textwidth]{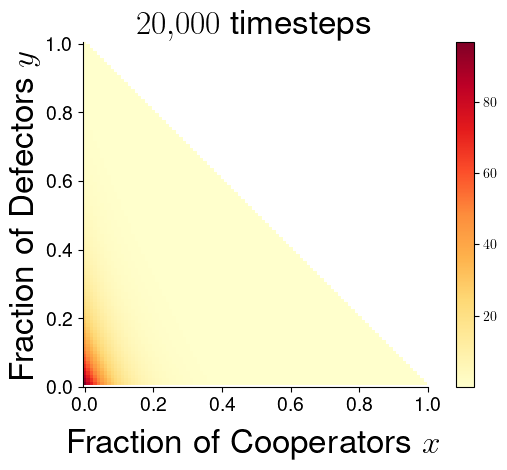}
    \caption{Numerical solutions for the finite volume approximation of the multilevel dynamics of cooperator-defector-punisher competition when between-group competition is determined by difference in fractions of cooperative individuals. Solutions displayed for the case of per-interaction punishment ($k = 0.1$, $q = 0$), punishment strength of $p = 0.5$, relative selection strength $\lambda = 4$, and payoff parameters $b = 3$ and $c = 1$ for the donation game. Each panel corresponds the the approximate density of $\rho(t,x,y)$ after 10 (top-left),  1,000 (top-right), 5,000 (bottom-left), and 20,000 (bottom-right) time-steps with a time increment $\Delta t = 0.015$. 
    }
    \label{fig:trimorphicnumericscoop}
\end{figure}

We can also consider how the fraction of individuals displaying cooperative behavior changes under the trimorphic dynamics as we change the strength of between-group selection $\lambda$ and the strength of punishment $p$. For the case of net group-level reproduction rate $G(x,y) = 1 -y$, we plot in Figure \ref{fig:trimorphiccooperationcompare} the average sum of the fractions of cooperators and altruistic punishers (adding up to the fraction of non-defectors $1-y$) achieved after 20,000 time-steps as a function of group-level selection strength. In Figure \ref{fig:trimorphiccooperationcompare}(left), we compare the average fraction of non-defectors to analytical results from Section \ref{sec:PDtwotypecooperativefraction} for the levels of altruistic punishment and cooperation achieved at steady state starting from a uniform initial strategy distribution for multilevel dynamics restricted to the defector-punisher and defector-cooperator edges of the simplex, respectively. We see that the average level of non-defection strategies for the trimorphic dynamics matches well with the average fraction of altruistic punishers at steady state on the defector-dimer edge of the simplex, achieving  a greater degree of cooperative behavior in the underlying donation game than is seen in multilevel competition featuring groups with only defectors and non-punishing cooperators. This agrees qualitatively with the prediction from the stochastic model by Boyd and coauthors, who showed that the presence of altruistic punishers allowed the achievement of greater levels of cooperative behavior via multilevel selection than would be achieved by defectors and pure cooperators alone. 

In Figure \ref{fig:trimorphiccooperationcompare}(right), we plot the average fraction of non-defectors achieved after 40,000 time-steps of the trimorphic multilevel dynamics for various strengths of punishment $p$. We see that increasing the strength of punishment increases the level of cooperative behavior achieved, which is also consistent with the behavior of the stochastic model of Boyd and coauthors \cite{boyd2003evolution}.  Consistent with the result from Figure \ref{fig:trimorphiccooperationcompare}, we see that the level of non-defection achieved for each punishment strength consider exceeds the analytically calculated level of cooperation achieved for multilevel dynamics on the cooperator-defector edge of the simplex.
\begin{figure}[!ht]
    \centering
    \includegraphics[width = 0.48\textwidth]{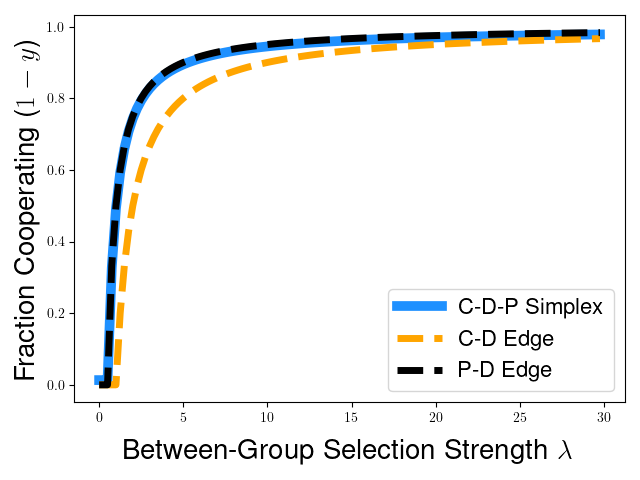}
  \includegraphics[width = 0.48\textwidth]{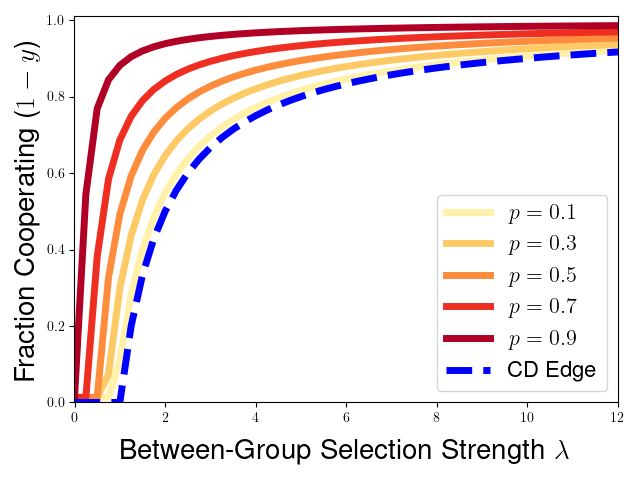}
    \caption{Comparison of total fraction of individuals cooperating in the donation {game} under numerical simulations of multilevel dynamics after 40,000 time-steps of step size $\Delta t = 0.015$ when pairwise group-level competition depends on the difference in the number of cooperative individuals between the two groups. Average fraction of non-defectors $1-y$ achieved under numerical simulations are plotted as functions of the strength $\lambda$ of between-group competition. Left: Comparison of average fraction of non-defectors $1-y$ for numerical simulation on the full cooperator-defector-punisher simplex (solid blue line) with analytical results for the average fraction of altruistic punishers at steady the defector-punisher edge of {the} simplex from Equation \eqref{eq:avgcoopDPedge} (dashed black line) and the average fraction of cooperators achieved at steady state on the cooperator-defector edge of the simplex (dashed orange line) from Equation \eqref{eq:DCedgecooplongtime}. Right: Comparison of fraction of non-defectors $1-y$ achieved under numerical simulations of trimorphic dynamics for different values of the strength of punishment $p$ (solid lines), as well as the average fraction of cooperators achieved under multilevel competition on cooperator-defector edge of the simplex (dashed blue line). Other payoff parameters are given by $b = 3$, $c = 1$, $k = 0.1$, and $q = 0$, and the trimorphic numerical simulations are started from an initial uniform distribution of strategies. he analytical results for multilevel dynamics on the defector-cooperator and defector-punisher edges of the simplex are calculated under the assumption of an initial H{\"o}lder exponent of $\theta = 1$ near the all-cooperator or all-punisher equilibrium, respectively.}
    \label{fig:trimorphiccooperationcompare}
\end{figure}

\subsection{Trimorphic Multilevel Competition Depending on Normalized Differences in Average Payoffs}
\label{sec:trimorphiclocalgroup}

Now we will look to consider numerical simulations for the three-strategy generalization of the pairwise group-level victory function $\mc{G}(z) = \frac{G(z)}{G^*-G_{*}}$ that we studied in Section \ref{sec:grouplocalupdate}. To do this, we consider pairwise between-group competition depending on the local update rule normalized by global payoff differences, which is characterized by the group-level victory probability given by
\begin{equation} \label{eq:trimorphicrholocal}
\rho(x,y; u,v) = \frac{1}{2} \left[1 + \frac{G(x,y) - G(u,v)}{G^*(\Delta^2) - G_*(\Delta^2)} \right],
\end{equation}
where the quantities $G^*(\Delta^2) := \max_{(x,y) \in \Delta^2} G(x,y)$ and $G_{*}(\Delta^2) = \min_{(x,y) \in \Delta^2} G(x,y)$ allow us to study the maximum possible difference $G^*(\Delta^2) - G_*(\Delta^2)$ in average group payoffs. This choice of group-level victory probability allows us to incorporate the differences between average payoff of group members in the two competing groups, and will highlight differences in effects of per-interaction and fixed costs for punishing defectors.

In Figure \ref{fig:trimorphiclongtimeavgG}, we plot numerical solutions achieved by the trimorphic multilevel dynamics when the group-level victory probability follows the local group-level update rule from Equation \eqref{eq:trimorphicrholocal} with global normalization of average payoff differences. We compare numerical solutions for this model for examples of both fixed and per-interaction punishment costs and for two different values of between-group selection strength $\lambda$. For the case of per-interaction punishment costs $k > 0$, we see that the long-time solutions of the trimorphic dynamics features densities primarily supporting strategic compositions close to the all-punisher equilibrium for both group-level selection strengths considered. However, for the case of fixed interaction costs $q > 0$, we see that the trimorphic dynamics feature a large proportion of all-punisher groups for the lower value of $\lambda = 3$, but the trimorphic dynamics primarily support compositions close to the all-cooperator state when $\lambda = 10$. This behavior is consistent with the observations seen in Figure \ref{fig:DPvsDCcomparison} when comparing the collective payoffs achieved at steady state for multilevel competition on the defector-cooperator and defector-punisher edges of the simplex in the case of fixed punishment costs. In this case, the long-time outcome seen in Figure \ref{fig:trimorphiclongtimeavgG}(bottom-left) and \ref{fig:trimorphiclongtimeavgG}(bottom-right) is consistent with the observation that a group-structured population on the defector-cooperator edge of the simplex can achieve higher average payoffs at steady state than achieved on the defector-punisher edge for lower strengths $\lambda$ of between-group selection, while groups on the defector-cooperator edge achieve better collective outcomes in the presence of stronger between-group selection. 

\begin{figure}[!htbp]
    \centering
    \includegraphics[width = 0.48 \textwidth,height = 0.42\textwidth]{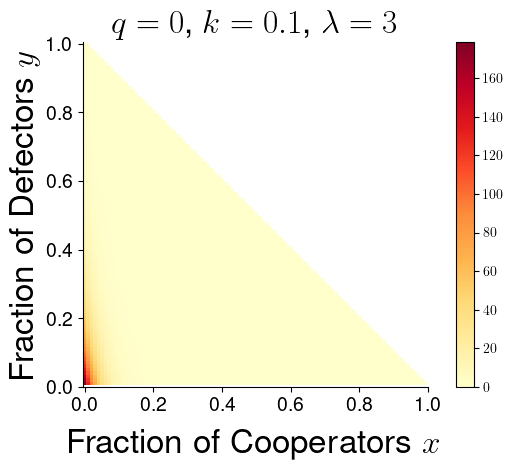}
\includegraphics[width = 0.48 \textwidth,height = 0.42\textwidth]{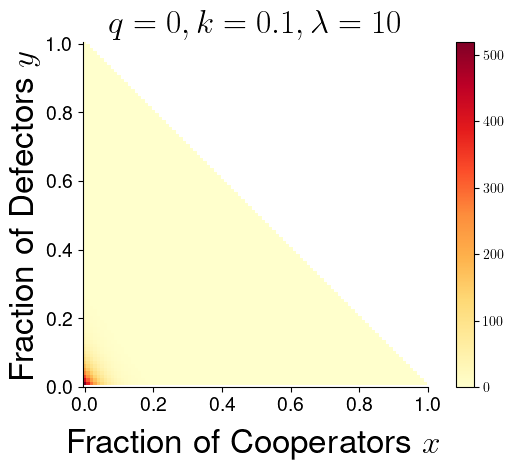}
 \includegraphics[width = 0.48 \textwidth,height = 0.42\textwidth]{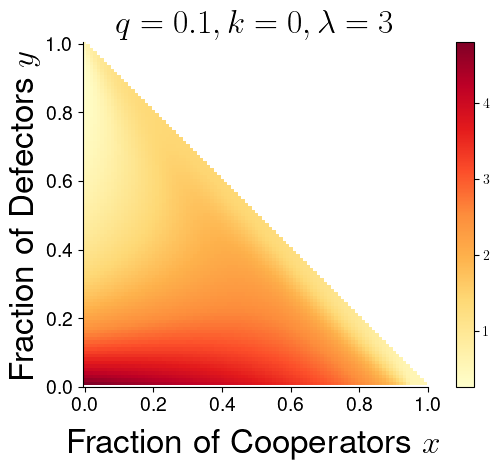}
 \includegraphics[width = 0.48 \textwidth,height = 0.42\textwidth]{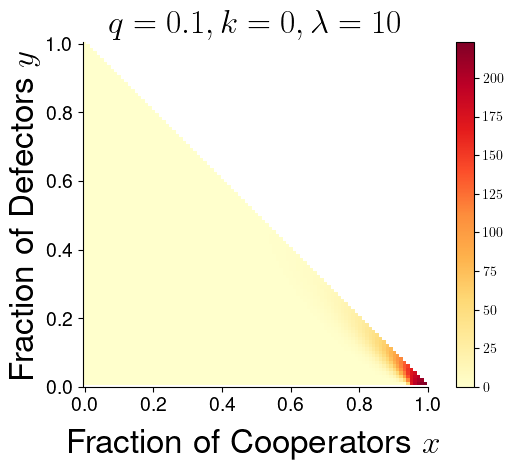}
    \caption{Numerical solutions achieved after 40,000 time-steps of the finite volume approximation of the multilevel dynamics of cooperator-defector-punisher competition based on normalized group-level update rule. We consider cases of punishment with per-interaction costs alone (with $q = 0$, $k = 0.1$) for between-group selection strength $\lambda = 3$ (top-left) and $\lambda = 10$ (top-right), as well as a case of fixed cost of punishment alone (with $q = 0.1$, $k = 0$) for $\lambda = 3$ (bottom-left) and $\lambda = 10$ (bottom-right). Other payoff parameters were fixed at $b = 3$, $c = 1$, and $p = 0.5$, and the initial strategy composition was the uniform density $u(0,x,y) = 1$.}
    \label{fig:trimorphiclongtimeavgG}
\end{figure}

In Figure \ref{fig:trimorphicpayoffcompare}, we explore the long-time average payoff achieved under the trimorphic multilevel dynamics in the case group-level victory probability described by Equation \eqref{eq:trimorphicrholocal}. We plot the average payoff achieved for various relative strengths of between-group competition $\lambda$ for both a case of per-interaction punishment costs alone (Figure \ref{fig:trimorphicpayoffcompare}, left) and for a case of fixed punishment costs alone (Figure \ref{fig:trimorphicpayoffcompare}, right). For comparison, we also plot the analytical expressions for the average payoff at steady state for multilevel selection with pairwise between-group selection on both the D-C and D-P edges of the simplex, exploring the extent to which including all three strategies can alter the long-time behavior of multilevel selection relative to competition occurring only between defectors and either cooperator or altruistic punishers. For the case of per-interaction punishment costs, we find that the average payoff for numerical solutions to the trimorphic dynamics after 20,000 time-steps have good agreement with the analytical expression for the average payoff received at steady state for multilevel dynamics in groups featuring only defectors and altruistic punishers for all values of $\lambda$ considered.

For the case of fixed cost of punishment, we see from Figure \ref{fig:trimorphicpayoffcompare} (right) that the average payoff achieved under the trimorphic multilevel dynamics has a more complicated comparison with the analytical results for average payoff achieved on the defector-cooperator and defector-punisher edges of the simplex. The long-time average payoff has good agreement with the formula for average payoff achieved on the defector-punisher edge of the simplex for lower values of $\lambda$, and agrees with the average payoff achieved on the defector-cooperator edge of the simplex for larger values of $\lambda$. In particular, the average payoff achieved by the numerical solutions of the trimorphic dynamics seems to have relatively good agreement with the analytical expression for average payoff on the edge of the simplex that features a higher payoff at steady state for the given strength of between-group selection. Because we have seen in Section \ref{sec:existingresults} that the multilevel competition for groups composed of two strategies seems to be determined as a tug-of-war between the individual and collective incentives determined by the payoffs achieved at the monomorphic states of the population, the agreement in collective payoff achieved in the trimorphic model with the behavior on the edges of the simplex perhaps suggests that multilevel competition featuring all three strategies may depend on a tug-of-war between the collective success of all-cooperator, all-defector, and all-punisher groups. 

\begin{figure}[htbp]
    \centering
    \includegraphics[width = 0.48\textwidth]{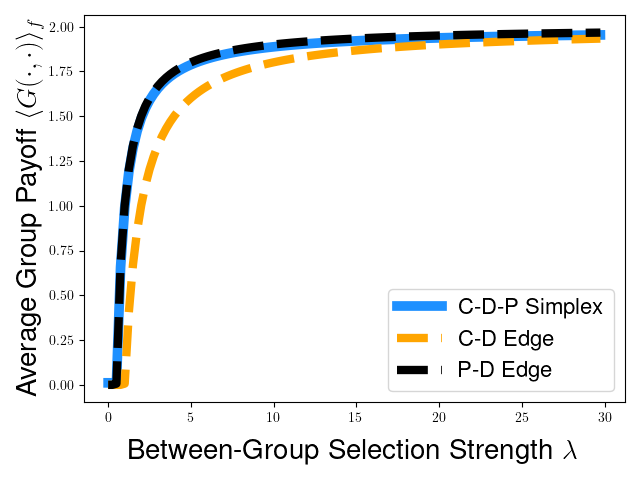}
  \includegraphics[width = 0.48\textwidth]{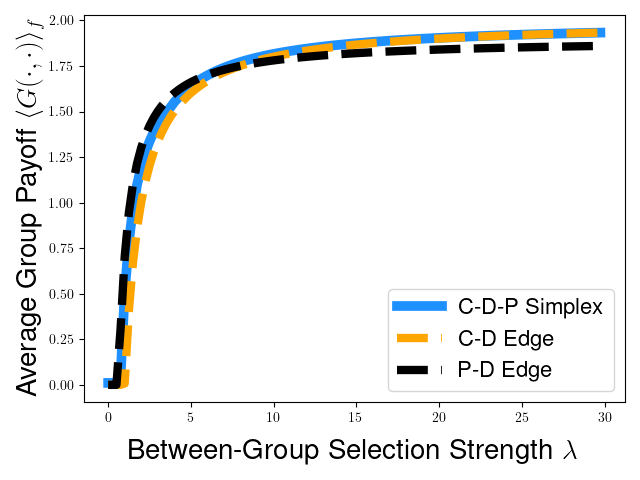}
    \caption{Comparison of long-time behavior for numerical simulations of multilevel dynamics when pairwise between-group competition depends on differences in average payoff in groups. Comparison of average payoffs for case of per-interaction costs for punishing defectors (left) and case of fixed costs for punishing defectors (right). Solid blue lines describe the average payoff computed for numerical solutions of trimorphic multilevel dynamics on defector-cooperator-punisher simplex after 24,000 timesteps with time increment of $\Delta t = 0.015$. The orange dashed lines describes analytical expression average payoff on the defector-punisher edge of the simplex ({from Equation \eqref{eq:averageGgrouplocal} evaluated with $q = 0$ in the left panel and with $k = 0$ in the right panel}), while the black dashed lines corresponds to analytically computed average payoffs for multilevel competition on the defector-punisher edge of the simplex (computed from Equation \eqref{eq:DCedgepayofflongtime}). Payoff parameters were $b =3$ and $c = 1$, strength of punishment was set to $p = 0.5$, and the costs of punishment were given by $k = 0.1$ and $q = 0$ (left) or $q = 0.1$ and $k = 0$ (right). }
    \label{fig:trimorphicpayoffcompare}
\end{figure}

\begin{remark}
The relatively good correspondence seen between the numerically calculated average payoff for the trimorphic multilevel dynamics and analytical results for dynamics on the edges of the simplex appears to hold for large strengths of fixed punishment $k$, but appears to break down in the case of stronger fixed punishmetn strengths like $q = 0.4$. We explore this behavior in Section \ref{sec:LMtrimorphic}, in which we also compare the behavior of trimorphic and dimorphic PDE models for multilevel selection based on two-level replicator equations describe frequency-independent group-level competition. 

Given the similarities and differences seen between the numerical simulations for the trimorphic multilevel dynamics and existing analytical results for two-strategy composition on the edges of the three-strategy simplex, there appears to be substantial room for further analytical exploration for three-strategy multilevel dynamics as well as future work on more comprehensive numerical approaches for studying solutions of PDE models for multilevel selection. One potential direction for expanding numerical exploration is to incorporate more explicitly the possibility that the strategic composition of a population can accumulate upon delta-measures at equilibria of the within-group dynamics, which has recently been explored in related models of Wright-Fisher-Kimura diffusion equations describing the dynamics individual-level selection in the presence of genetic drift \citep{carrillo2022optimal,zhao2013complete,waxman2011comparison,chalub2014frequency,epstein2010wright}. 
\end{remark}

\section{Discussion}
\label{sec:discussion}

In this paper, we have formulated PDE models for multilevel selection incorporating pairwise between-group competition and a mechanism of altruistic punishment within groups. Building off of a stochastic approach introduced by \citep{boyd2003evolution}, we see that altruistic punishment and group-level competition can work in concert to help promote the evolution of cooperative behavior via multilevel selection. Studying a deterministic PDE model for cultural multilevel selection allows us to obtain an analytical characterization of many of the qualitative predictions made by Boyd and coauthors using stochastic individual-based models, showing how the collective success of groups under multilevel selection can depend on the strength of altruistic punishment and the modes by which cooperative individuals incur costs to punish defectors in their groups. By introducing more general formulations of group-level victory probabilities that depend on the difference in average payoff between groups, we were further able to gain further insight into the role of the group-level frequency dependence on the evolution of cooperative behaviors under cultural multilevel selection. Notably, we found a surprising long-time behavior in which the average steady-state payoff achieved in our model could have a non-monotonic dependence on the strength of altruistic punishment, a behavior that does not appear to hold for the case of group-level victory probabilities based only on the difference in the fraction of cooperative individuals \citep{boyd2003evolution} or when considering PDE models of multilevel selection featuring frequency-independent group-level competition \citep{cooney2022long,cooney2022assortment}.

The formulation of PDE models for multilevel selection with pairwise between-group competition and the within-group mechanism of altruistic punishment serves as one step in trying to obtain an analytically tractable understanding of the synergistic effects of within-group and group-level mechanisms for the promotion of cooperation. While the particular form of group-level competition used in the model by Boyd and coauthors \cite{boyd2003evolution} can be described in terms of PDE models with known long-time behavior, the more general class of potential models for pairwise between-group competition motivates new directions for understanding the role of group-level frequency-dependence on evolutionary dynamics with group-level competition. By considering existing models of  pairwise-comparison dynamics typically used to study individual-level social learning, we can explore families of functions describing group-level frequency-dependence motivated by the Fermi update rule \citep{traulsen2007pairwise}, the local update rule \citep{traulsen2005coevolutionary,morgan2003pairwise}, and individual-level logit dynamics \citep{hofbauer2009stable,hommes2012multiple}. The pairwise group-level victory probabilities also resemble the analogues of logit or pairwise-comparison dynamics for continuous-strategy games \citep{lahkar2015logit}, and considering group-level competition that are not additively separable motivates connections to stochastic models for multilevel selection in which group-level competition is governed by nonlinear replication or death rates \citep{traulsen2008analytical,bottcher2016promotion}. By considering this broader class of functional forms describing between-group competition, we get to explore how robust our predictions are for different assumptions for the dynamics of competition between groups.
{This may be particularly relevant as a way to understand how the size of groups would impact the feasibility of achieving cooperation through a combination of altruistic punishment, as mechanisms like direct reciprocity \citep{boyd1989evolution,schmid2021unified} or centralized institutional incentives \citep{duong2021cost,gavrilets2021evolving} may be more effective means of achieving cooperative behavior respectively in scenarios featuring sufficiently small or sufficiently large group size. }

While we have focused on modeling multilevel selection with pairwise between-group competition in this paper, it is also natural to ask how altruistic punishment and group-level competition can work in concert to promote cooperation in the case of frequency-independent group-level competition. In Section \ref{sec:LMaltruisticpunishment} of the appendix, we study the long-time behavior of a two-level replicator equation for multilevel selection, exploring how altruistic punishment interacts with existing PDE models with group-level replication rates depending only on the composition of the replicating group. Considering both analytical results for a two-level replicator equation for dimorphic defector-punisher competition and numerical simulations for trimorphic multilevel dynamics featuring all three strategies, we find that altruistic punishment and frequency-independent group-level competition work synergistically to promote greater collective payoff and facilitate the evolution of cooperative behavior. However, unlike the results in Sections \ref{sec:twostrategypayoffs} and \ref{sec:nonlineargroup}, we see that collective payoff at steady state is always independent of per-interaction punishment costs $k$ and never features a non-monotonic dependence on the strength of punishment $p$. This discrepancy between two-level replicator models and models with pairwise group conflicts potentially suggests that these interesting behaviors seen in Sections \ref{sec:twostrategypayoffs} and \ref{sec:nonlineargroup} could be attributed to the effects of group-level frequency dependence and the forms of group-level victory probabilities considered in this paper. 

{We further summarize some of the main results of our paper in Table \ref{tab:summarytable}, explaining how key quantities for multilevel election like the threshold selection strength $\lambda^*$ or measures of the can depend on the different group-level victory probability probabilities considered in the paper. These results highlight the key sensitivity of the evolution of cooperative behavior and the emergence of altruistic punishment depend on how we model the impact of strategic composition of competing groups on the chance of victory under group-level conflict. In the remainder of the discussion, we further present an outlook on how our framework for multilevel selection with pairwise group-level competition can be used to further explore the evolution of cooperative social norms via cultural evolution (Section \ref{sec:culturalevolution}) and the extent to which our results suggest directions for future work on PDE models of multilevel selection with frequency-dependent group-level competition (Section \ref{sec:PDEoutlook}). }

\renewcommand{\arraystretch}{2}
\begin{table}[!htbp]
\caption{ {Summary of main results in the paper, comparing behavior of multilevel selection for different group-level victory probabilities. We summarize the status of analytical and numerical results for the two-strategy multilevel competition in groups featuring defectors and punishers, explore how the threshold selection strength $\lambda^*$ to sustain altruistic punishment via multilevel selection depends on the parameters for the costs and strength of punishment, describe how the collective outcome depends on these parameters, and how the numerical results for the three-strategy dynamics compare to results for two-strategy multilevel dynamics for the same group-level victory probability. We measure the collective outcome of the population as the average fraction of punishers or cooperative individuals (cooperators plus punishers) for the victory probability based on the difference in the fraction of cooperative individuals, and we measure the collective outcome as the average payoff at steady-state for the four other victory probabilities. We note that the numerical simulations for multilevel selection with all three strategies were only performed for the two additively separable group-level victory probabilities. We use the arrows $\uparrow$ or $\downarrow$ to indicate that a quantity has been shown to always be increasing or decreasing with an increase of a given parameter. }}
\begin{center}
\footnotesize
\begin{tabular}{|c|c|c|c|c|}
\hline
\makecell{Group-Level \\ Victory \\ Probability \\ $\rho(u,z)$ } & \makecell{Status of \\ Two-Strategy \\ Dynamics }  &   \makecell{Threshold \\ Selection \\
Strength $\lambda^*$}  & \makecell{Behavior of \\ Collective \\ Outcome for \\Two-Strategy Case} & \makecell{Status of \\ Three-Strategy \\ Dynamics} \\
   \hline 
\makecell{$\ds\frac{1}{2} + \ds\frac{z-u}{2}$ \\ \\ Difference in \\ Punisher Fractions} & \makecell{Characterized \\
Analytically \\ Section \ref{sec:longtimeadditive}} & \makecell{$\downarrow p$, $\uparrow q$ \\
$k$-independent} & \makecell{Average \\ Punishers $\langle z \rangle_{f^{\lambda}_{\theta}}$: \\ \\ $\uparrow p, \downarrow q$ \\ $k$-independent} & \makecell{Simulated \\ Numerically: \\ Long-Time Average \\ Cooperating \\ $\langle x+ z \rangle_f$ \\ Corresponds to \\ $\langle z \rangle_{f^{\lambda}_{\theta}}$ on DP Edge}
\\
\hline
\makecell{$\ds\frac{1}{2} + \ds\frac{1}{2}\left[\ds\frac{G(z) - G(u)}{G^* - G_*}\right]$  \\ \\ Difference in \\ Average Payoff \\ (Globally Normalized)} & \makecell{Characterized \\
Analytically \\ Section \ref{sec:longtimeadditive}} & \makecell{Complicated  \\ Dependence \\ on $p$, $q$, and $k$} & \makecell{Average Payoff $\langle G(\cdot) \rangle_{f^{\lambda}_{\theta}}$: \\ \\ Can decrease with $k$ \\ Figure \ref{fig:lambda_thresh_avg_G_k_model}(right) \\ \\ Non-monotonic \\ effect of $p$ \\ Figure \ref{fig:lambda_thresh_avg_G_q_model}(right) } & \makecell{Simulated \\ Numerically: \\  Payoff $\langle G(x,y) \rangle_f$ \\ Agrees With \\ Larger of  \\
$\langle G(\cdot) \rangle_{f^{\lambda}_{\theta}}$ Between \\  DP and DC Edges}
\\
\hline
\makecell{$\ds\frac{1}{2} + \ds\frac{1}{2} \tanh\left( G(z) - G(u) \right)$ \\ \\ Group-Level Fermi \\ Rule} & \makecell{Analytical  \\
Conjectures: \\ Section \ref{sec:pairwiseresults}; \\ \\ Numerical \\ Support: \\ Section \ref{sec:pairwisenumerical}} & \makecell{$\downarrow p$, $\uparrow q$, \\
$k$-independent} & \makecell{Average Payoff $\langle G(\cdot) \rangle_{f^{\lambda}_{\theta}}$: \\ \\ Can decrease with $k$ \\ Figure \ref{fig:nonlinearkmodel}(left) \\ \\ Can increase with $p$ \\ Figure \ref{fig:nonlinearDPavgGandCoop}(left)} & N/A \\
\hline
\makecell{$\ds\frac{1}{2} + \ds\frac{1}{2}\left[\ds\frac{G(z) - G(u)}{|G(z)| + |G(u)|}\right]$ \\ Difference in \\ Average Payoff \\ (Pairwise Normalized)} & \makecell{Analytical  \\
Conjectures: \\ Section \ref{sec:pairwiseresults}; \\  \\ Numerical \\ Support: \\ Section \ref{sec:pairwisenumerical}} & \makecell{$\downarrow p$, $\uparrow q$ \\
$k$-independent}
& \makecell{Average Payoff $\langle G(\cdot) \rangle_{f^{\lambda}_{\theta}}$:  \\ \\ Can increase with $p$ \\ Figure \ref{fig:qmodellocalTullock}(left) }   & N/A
\\
\hline
\makecell{${\scriptstyle{\frac{1}{2} + \frac{1}{2}\left[\frac{\left(G(z) - G_{*}\right)^{1/a}}{\left(G(z) - G_{*}\right)^{1/a} + \left( G(u) - G_{*} \right)^{1/a}}\right]}}$ \\ \\ Group-Level \\ Tullock Contest \\ Function} & \makecell{Analytical  \\
Conjectures: \\ Section \ref{sec:pairwiseresults}; \\ \\ Numerical \\ Support: \\ Section \ref{sec:pairwisenumerical}}  & \makecell{Complicated  \\ Dependence \\ on $p$, $q$, and $k$} & \makecell{Average Payoff $\langle G(\cdot) \rangle_{f^{\lambda}_{\theta}}$:  \\ \\ Non-monotonic \\ effect of $p$ \\ Figure \ref{fig:qmodellocalTullock}(right) } & N/A \\
\hline
\end{tabular}
\end{center}
\label{tab:summarytable}
\end{table}
\renewcommand{\arraystretch}{1}

\subsection{Connections to the Literature on Cultural Evolution and Cultural Multilevel Selection}
\label{sec:culturalevolution}

{
The PDE framework used in this paper allowed for the exploration of a wide range of assumptions for how the strategic composition of groups and the average payoff of group members can impact the probability of victory in group-level conflict. While prior work on stochastic models of cultural group selection has often focused on group-level competition based on difference in the fractions individuals following strategies that are beneficial to the group \citep{boyd2003evolution,choi2007coevolution}, it may be interesting to explore how different possible group-level victory probabilities would impact the behavior of traditional agent-based models of cultural group selection. In particular, one question of interest is whether the non-monotonic dependence of average payoff on the strength of punishment $p$ seen in our PDE model could also occur in a modified version of the model by Boyd and coauthors \cite{boyd2003evolution} in the case of the group-level local update rule or Tullock contest function. As recent work by Han and coauthors has shown a misalignment between the maximization of collective payoff and maximizing the fraction of cooperative individuals in models of costly peer punishment and peer rewards in finite populations under individual-level selection \cite{han2024evolutionary},  it appears that exploring such finite-population modes of multilevel selection with costly punishment and rewards would provide a nice setting for exploring cultural multilevel selection based on a variety of rules for victory in group-level conflict.} 

{
Although the analysis in our paper has focused on PDE models that have been derived in the limit of infinitely many groups and infinite group size, it is likely more realistic to study the evolution of cooperative social norms in human populations featuring relatively small group size. In particular, empirical studies of human hunter-gatherer populations have observed strong within-group cooperation in bands consisting of tens of individuals and consisting of both related and unrelated individuals \citep{hill2011co,hamilton2018scaling}, suggesting that finite group size may be a relevant setting for exploring the emergence of human cooperation via cultural multilevel selection. To explore these finite size effects, it could be helpful to perform stochastic simulations of the individual-based model discussed in Section \ref{sec:PDEderivation} that each have finitely many members. The derivation of our PDE model from the individual-based stochastic model provided an intermediate scaling limit featuring a system of ODEs that describes multilevel selection in an infinite population of groups that each have finite size. This system of ODEs could serve as an interesting model for future study both because it may more accurately reflect real-world cultural evolution due to competition between many small groups of individuals than the PDE model and because the deterministic nature of this ODE system may allow for a more tractable or extensive exploration of parameter space than the stochastic model for multilevel selection in the case of finitely many groups.
}

Many existing PDE models of multilevel selection have made a variety of simplifying assumptions about the dynamics of selection at multiple levels of organization, from the assumption that groups produce exact copies of themselves to the assumption that group-level success does not depend on the composition of the population of groups other than the focal group. This paper focuses on relaxing the assumption of group-level success depending only on the composition of the given group, allowing group-level competition to depend on the relative strategic compositions of groups that engage in a pairwise competition for group-level reproduction. However, it is also possible to relax other simplifying assumptions in this model to see how pairwise between-group competition can be used to understand evolutionary competition incorporating effects of group-level fission or extinction events \citep{simon2010dynamical,simon2013towards,simon2016group,puhalskii2017large,bottcher2016promotion}, density-dependent population dynamics within groups with nonconstant size \citep{lerchflexible,simon2016group}, {acculturation behavior after groups have won pairwise conflicts \citep{henriques2019acculturation}}, or including additional evolutionary forces like migration, mutation, strategic exploration, or group-level randomness \citep{boyd2003evolution,velleret2020individual,velleret2023two,dawson2018multilevel}. Such extensions can be useful for understanding the robustness of results obtained from our simplified PDE model, and to understand how the dynamics of multilevel selection may depend on different factors imposed by incorporating increasing realism into models of biological and cultural evolution. {Incorporating these more realistic assumptions of individual migration or group-level fission and fusion events may also be particularly relevant for connecting these PDE models of multilevel selection to the literature on cultural evolution and cultural evolution, as these behaviors have been discussed in the context of the emergence of cooperative norms in early human societies.}

{\subsection{Future Directions for PDE Models of Multilevel Selection}
\label{sec:PDEoutlook}
}

In addition to our approach for modeling pairwise between-group competition in PDE models of multilevel selection, the analysis in our paper differs from prior work on PDE models of multilevel selection due to differences between the behavior of altruistic punishment and other within-group mechanisms like assortment, other-regarding preference, within-group network structure, and forms of direct and indirect reciprocity based on conditional cooperation \citep{cooney2022assortment,cooney2023evolutionary}. While we were able to see scenarios in which average payoff of group members was minimized for an intermediate fraction of altruistic punishers, each of the previously studied mechanisms produces average payoff functions that are strictly increasing in the fraction of cooperative individuals when game-theoretic interactions consist of a donation game. To explore this difference, we consider in Section \ref{sec:indirectreciprocity} of the appendix an analysis of multilevel selection with pairwise between-group competition when within-group interactions follow a model for the image-scoring rule of indirect reciprocity \citep{nowak1998evolution,nowak2006five} and the group-level victory probability depends on a globally normalized difference in average payoff of group members.  Compared to the results of Section \ref{sec:twostrategypayoffs}, we see threshold selection strength and average payoff at steady state always have a monotonic dependence on the parameter $Q$ corresponding to the probability that a cooperative individual successfully identifies a defector and punishes them with defection. This suggests that perhaps some of the interesting behaviors seen in our model for multilevel selection with pairwise group conflict are most salient when individual-level interactions feature altruistic punishment, motivating future consideration of studying how multilevel selection and altruistic punishment can work in concert to promote the evolution of cooperative behaviors.

Our focus in this paper was on game-theoretic models of cooperative behavior based on the donation game, in which the average payoff of group members is always maximized by having a group composed entirely of cooperators. However, previous work on multilevel selection in evolutionary games has also explored cases of games in which the average payoff of group members is maximized by intermediate fractions of cooperation. In existing PDE models featuring frequency-independent group-level competition, games with an intermediate group-level optima have displayed an interesting behavior called the ``shadow of lower-level selection''. For such games, the long-time average payoff achieved by the population under a two-level replicator equation cannot exceed the average payoff of the all-cooperator group, even in the limit of infinitely strong between-group competition, preventing the achievement of the socially optimal level of cooperation that maximizes the average payoff of group members. {A preliminary analysis of this behavior has been explored by Alexiou and Cooney \cite{alexiou2024steady}, who used the necessary conditions discussed in Section \ref{sec:pairwiseresults} to show that the collective success $\int_0^1 \rho(z,1) f(z) dz$ is limited to an equal chance to defeating an all-punisher group in pairwise conflict. An interesting direction for future research is to further characterize the exten to which this behavior holds for the case of pairwise-group level competition and whether this shadow of lower-level seelction} seen in PDE models with replicator-like group-level competition holds as well for a wider variety of models featuring group-level frequency dependence.

So far, most analytical results for PDE models of multilevel selection have focused on the case of group-structured populations featuring two types of individuals. By considering multilevel competition for three-strategy games, we are able to extend the scope of strategic interactions that can be described by such models, and we can explore how the group-level tug-of-war based on average payoff of all-defector, all-cooperator, and all-punisher groups can impact the evolution of cooperative and punishing behaviors via multilevel selection. So far, we have only considered numerical simulations of our PDE model for multilevel selection with three strategies, but the good agreement seen in comparison of our numerical simulations with analytical results obtained for two-strategy multilevel dynamics on the edges of the three-strategy simplex suggest that the long-time behavior of the trimorphic PDE models may depend on the tug-of-war between the collective advantage of all-cooperator or all-punisher groups over the all-defector group and the individual advantage of defectors within any group. This correspondence between numerical solutions for three-strategy multilevel dynamics and analytical results for two-strategy dynamics was also seen in PDE models for multilevel selection for protocell evolution \citep{cooney2022pde}, providing further motivation to pursue future analytical work on PDE models for multilevel selection in groups featuring three or more strategies. 

In this paper, we primarily incorporate group-level frequency dependence in our PDE model for multilevel selection by assuming that pairwise group-level competition depends on differences in the average payoffs obtained by the members of two competing groups. However, it is also possible to consider group-level interactions that allow for cooperative interactions between groups \citep{tverskoi2021dynamics}, in which collective interactions between groups help to determine a group-level replication rate based on some combination of payoffs obtained by individual interactions within groups and collective interactions between groups. Such social dilemma arising from incentives to compete and cooperate both within and between groups have been studied in the context of hierarchical public goods games, and interesting directions for future work are to explore how nested models of within-group and between-group competition can be used to explore how effort taken in within-group interactions can impact cooperation and conflict between groups \citep{rusch2020logic,baik2008contests,crowley2010variable,reeve2007emergence} and understanding how group-level conflict and competition can impact within-group behaviors and interactions \citep{preston2021network,preston2021dynamics,thompson2017causes,philson2023evolution}. Such hierarchical social dilemmas provide scenarios in which a cooperative behavior within a group may correspond to a selfish behavior within a population of groups \citep{fujimoto2017hierarchical}, further producing tensions between levels of selection and motivating further work on multilevel selection with group-level game-theoretic interactions. 

Overall, both considering the within-group mechanism of altruistic punishment and the formulation of a PDE model for multilevel selection with pairwise between-group selection have allowed us to see new behaviors relative to prior work on two-level replicator equation models for multilevel selection. Analytical and numerical exploration of these PDE models for cultural multilevel selection also provide a complementary role to existing simulation studies describing cross-scale features in the evolution of social norms or altruistic punishment. However, there is substantial room for further study on the dynamics of the PDE models considered in this paper and efforts to connect these PDE models with stochastic, individual-based approaches, allowing us to achieve a deeper understanding of the role of group-level frequency dependence can play in establishing the evolution of cooperative traits and behaviors in both biological and cultural evolution. We hope that this approach for understanding multilevel selection with pairwise group-level competition can be applied to further understand evolutionary tensions between levels of selection across a range of systems featuring frequency-dependent competition occurring at multiple levels of organization. 

\vspace{10mm}
\renewcommand{\abstractname}{Acknowledgments}
\begin{abstract} 
A portion of this work supported by the Math+X grant from the Simons Foundation. I would like to thank Erol Akçay, Joshua Plotkin, and Yoichiro Mori for helpful discussions.  I am also grateful to two anonymous reviewers for helpful comments on the manuscript.
\end{abstract}

\renewcommand{\abstractname}{Statement on Code Availability}
\begin{abstract} 
All code used to generate figures is archived on Github (\href{https://github.com/dbcooney/Multilevel-Altruistic-Punishment-Paper-Code}{https://github.com/dbcooney/Multilevel-Altruistic-Punishment-Paper-Code}) and licensed for reuse, with appropriate
attribution/citation, under a BSD 3-Clause Revised License.
\end{abstract}

\bibliographystyle{unsrt}
\bibliography{references}

\appendix
\changelocaltocdepth{1}

\section{Derivation of PDE Model for Multilevel Selection with Pairwise Between-Group Competition}
\label{sec:PDEderivation}

In this section, we present a heuristic derivation of the PDE model from Equation \eqref{eq:multilevelPDErhodiff} for multilevel selection with pairwise between-group selection for the case of two-strategy games. We first consider the derivation model for the case of groups composed of two strategies in Section \ref{sec:dimorphicderivation}. The derivation of the terms corresponding to within-group competition will be analogous to the derivation in previous models for PDE models for multilevel selection in evolutionary games \citep{cooney2019replicator}, but we modify the derivation of the group-level terms to account for pairwise between-group competition events. 

Next, we sketch a derivation of the PDE model from Equation \eqref{eq:PDEtrimorphicrho} for three-strategy multilevel dynamics in groups featuring defectors, cooperators, and altruistic punishers in Section \ref{sec:trimorphicderivation}. We will follow the approach used to derive a three-strategy PDE model for multilevel selection in the context of protocell evolution with frequency-independent individual-level dynamics \citep{cooney2022pde}, and our derivation in this section will show how to modify this approach to incorporate frequency-dependent competition at the individual and group levels. 

\subsection{Derivation of PDE Model for Two-Strategy Case}
\label{sec:dimorphicderivation}

We consider a population with $m$ groups, each composed on $n$ members. We denote the number of altruistic punishers in the group by $i$, and we describe the probability of having a group with $i$ cooperators at time $t$ by $f_i^{m,n}(t)$. Within each group, we assume that the dynamics of individual-level selection follow the rules of a continuous-time Moran process, with defectors and altruistic punishers in an $i$-punisher group respectively producing copies of themselves at rates $1 + w_I \pi_D(\frac{i}{n})$ and $1 + w_I \pi_P\left(\frac{i}{n} \right)$, where the parameter $w_I$ characterizes the intensity of individual-level selection. To keep the size of groups constant over time, we assume that the offspring in individual-level reproduction replace a randomly chosen member of their parent's group. 

We consider group-level competition in which each group engages in pairwise competitions with other groups at rate $\Lambda$, and we assume that the group-level opponent is randomly sampled from the empirical distribution of groups in the population. This means that each of the $m f_i^{m,n}(t)$ $i$-punisher groups will have pairwise interactions with a $j$-punisher group with rate $\Lambda f_j(t)$. We then assume that the $i$-punisher group wins the pairwise competition with probability $\rho\left(\frac{i}{n},\frac{j}{n} \right)$, while the $j$-cooperator group wins the pairwise conflict with probability $\rho\left(\frac{j}{n},\frac{i}{n} \right) = 1 - \rho\left(\frac{i}{n},\frac{j}{n} \right)$. We then assume that that the winner of the pairwise competition produces an exact copy of itself, with the offspring group replacing the group that lost the pairwise conflict. 

Following the approach introduced by Luo and coauthors \citep{luo2014unifying,van2014simple}, we note that the fraction of $i$-punisher groups increases by $\frac{1}{m}$ due to individual-level selection when one of two events occurs:
\begin{itemize}
    \item A punisher reproduces in an $(i-1)$-punisher group and a defector is chosen to die, which occurs with rate
    \[ m f^{m,n}_{i-1}(t) \left( i - 1 \right) \left( 1 + w \pi_P\left(\frac{i}{n} \right) \right) \left( 1 - \frac{i-1}{n} \right) \]
    \item A defector is born in an $(i+1)$-punisher group and a punisher is chosen to die, which occurs with rate
      \[ m f^{m,n}_{i+1}(t) \left( n - i - 1 \right) \left( 1 + w \pi_D\left(\frac{i}{n} \right) \right) \left( \frac{i+1}{n} \right).\]
\end{itemize}
Similarly, we can see that the fraction of $i$-punisher groups can decrease by $\frac{1}{m}$ due to individual-level selection when one of the following two events occurs:
\begin{itemize}
    \item A punisher is born in an $i$-punisher group and a defector is chosen to die, which occurs with rate
    \[ m f_{i}^{m,n}(t) i \left( 1 + w \pi_P\left(\frac{i}{n} \right) \right) \left( 1 - \frac{i}{n} \right)\]
    \item A defector is born in an $i$-punisher group and a punisher is chosen to die, which occurs with rate
    \[ m f_{i}^{m,n}(t) \left( n - i\right) \left( 1 + w \pi_P\left(\frac{i}{n} \right) \right) \left(\frac{i}{n} \right)\]
\end{itemize}

Next, we can describe the impact of pairwise between-group competition on the changing fractions of $i$-punisher groups in our group-structured population. The fraction of groups with $i$ altruistic punishers increases by $\frac{1}{m}$ when an $i$-punisher group defeats a group with another number $j \ne i$ of altruistic punishers in a pairwise competition. Such an event occurs with rate
\begin{equation}
\Lambda m f_i^{m.n}(t) \sum_{\substack{j =0 \\ j \ne i}}^n f_j^{m,n}(t) \rho\left( \frac{i}{n} , \frac{j}{n} \right).
\end{equation}
Similarly, the fraction of groups with $i$ altruistic punishers decreases by $\frac{1}{m}$ when a $j$-punisher group defeats an $i$-cooperator group in pairwise competition, which occurs with rate
\begin{equation}
\Lambda m f_i^{m,n}(t) \sum_{\substack{j =0 \\ j \ne i}}^n f_j^{m,n}(t) \rho\left( \frac{j}{n} , \frac{i}{n} \right).
\end{equation}

For groups featuring $i$ altruistic punishers with $i \in \{1,\cdots,n-1\}$, we can then combine the rates at which the fraction of $i$-punisher groups change due to individual-level and group-level replication events to write that the conditional mean of the change in the fraction $f^{m,n}_i(t)$ of $i$-punisher groups is given by
\begin{equation}
\begin{aligned}
& E \left[f^{m,n}_i(t+\Delta t) - f^{m,n}_i(t) \bigg| f^{m,n}_i(t)  \right]  \\
&= \frac{1}{m} P\left(f^{m,n}_i(t+\Delta t) - f^{m,n}_i(t)  = \frac{1}{m} \right) - \frac{1}{m} P\left(f^{m,n}_i(t+\Delta t) - f^{m,n}_i(t)  = -\frac{1}{m} \right) + o(\Delta t)\\
&= \frac{1}{m} \left[ m f^{m,n}_{i-1}(t) \left( i - 1 \right) \left( 1 + w \pi_P\left(\frac{i}{n} \right) \right) \left( 1 - \frac{i-1}{n} \right) \right] \Delta t \\
&+ \frac{1}{m} \left[m f^{m,n}_{i+1}(t) \left( n - i - 1 \right) \left( 1 + w \pi_D\left(\frac{i}{n} \right) \right) \left( \frac{i+1}{n} \right) \right] \Delta t \\
&- \frac{1}{m} \left[m f_{i}^{m,n}(t) i \left( 1 + w \pi_P\left(\frac{i}{n} \right) \right) \left( 1 - \frac{i}{n} \right) \right] \Delta t \\
&- \frac{1}{m} \left[ m f_{i}^{m,n}(t) \left( n - i\right) \left( 1 + w \pi_P\left(\frac{i}{n} \right) \right) \left(\frac{i}{n} \right) \right]  \Delta t\\
&+ \frac{1}{m} \left\{ \Lambda m f_i^{m.n}(t) \sum_{\substack{j =0 \\ j \ne i}}^n f_j^{m,n}(t) \rho\left( \frac{i}{n} , \frac{j}{n} \right)\right\}  \Delta t
- \frac{1}{m} \left\{ \Lambda m f_i^{m,n}(t) \sum_{\substack{j =0 \\ j \ne i}}^n f_j^{m,n}(t) \rho\left( \frac{j}{n} , \frac{i}{n} \right) \right\} \Delta t \\ &+ o(\Delta t)
\end{aligned}
\end{equation}

By introducing the first-order forward and backward difference quotients
\begin{equation}
\begin{aligned}
D_1^{+} \left( u\left(\frac{i}{n} \right)\right) := \frac{u\left( \ds\frac{i + 1}{n} \right) - u \left( \ds\frac{i}{n} \right)}{\ds\frac{1}{n}} \\
D_1^{-} \left( u\left(\frac{i}{n} \right)\right) := \frac{u\left( \ds\frac{i}{n} \right) - u \left( \ds\frac{i-1}{n} \right)}{\ds\frac{1}{n}} 
\end{aligned}
\end{equation}
and the second-order difference quotient
\begin{equation}
D_2\left( u\left(\frac{i}{n} \right)\right) := \frac{u\left( \ds\frac{i + 1}{n} \right)  - 2 u\left( \ds\frac{i}{n} \right) + u\left( \ds\frac{i - 1}{n} \right)  }{\ds\frac{1}{n^2}},
\end{equation}
we may rewrite our expression for the infinitesimal mean of the two-level birth-death process as
\begin{equation}
\begin{aligned}
\frac{E \left[f^{m,n}_i(t+\Delta t) - f^{m,n}_i(t) \bigg| f^{m,n}_i(t) \right]}{\Delta t}
&= \frac{1}{n} D_2\left(\frac{i}{n} \left( 1- \frac{i}{n} \right) f^{m,n}_{i}(t) \right) \\
&+ w_I D_1^+\left( \frac{i}{n} \left( 1- \frac{i}{n} \right) \pi_{D}\left( \frac{i}{n} \right) f^{m,n}_{i}(t) \right) \\
&- w_I D_1^{-} \left( \frac{i}{n} \left( 1- \frac{i}{n} \right) \pi_P\left( \frac{i}{n} \right) f^{m,n}_{i}(t) \right) \\
&+ \Lambda f^{m,n}_{i,j}(t)  \sum_{j=0}^{n} \left\{ \left[ \rho\left( \frac{i}{n}, \frac{j}{n} \right) - \rho\left( \frac{j}{n},\frac{i}{n} \right) \right] f^{m,n}_{j}(t) \right\} + \frac{o(\Delta t)}{\Delta t},
\end{aligned}  
\end{equation}
where we used the fact that $\rho\left( \frac{i}{n} , \frac{i}{n} \right) = \frac{1}{2}$ to consolidate the two sums describing the impacts of between-group competition events. This allows us to see that the infinitesimal mean of our stochastic process takes the form
\begin{equation} \label{eq:infmeantwotype}
\begin{aligned}
 & \lim_{\Delta t \to 0} \left( \frac{1}{\Delta t} E \left[f^{m,n}_i(t+\Delta t) - f^{m,n}_i(t) \bigg| f^{m,n}_i(t) \right] \right) \\ 
&= w_I D_1^+\left( \frac{i}{n} \left( 1- \frac{i}{n} \right) \pi_{D}\left( \frac{i}{n} \right) f^{m,n}_{i}(t) \right) - w_I D_1^{-} \left( \frac{i}{n} \left( 1- \frac{i}{n} \right) \pi_P\left( \frac{i}{n} \right) f^{m,n}_{i}(t) \right) \\
&+ \Lambda f^{m,n}_{i,j}(t)  \sum_{j=0}^{n} \left\{ \left[ \rho\left( \frac{i}{n}, \frac{j}{n} \right) - \rho\left( \frac{j}{n},\frac{i}{n} \right) \right] f^{m,n}_{j}(t) \right\} + \frac{1}{n} D_2\left(\frac{i}{n} \left( 1- \frac{i}{n} \right) f^{m,n}_{i}(t) \right).
\end{aligned}
\end{equation}

Next, we can follow the approach used by Luo \citep{luo2014unifying} and by Czuppon and Traulsen 
\citep[Appendix 1]{czuppon2021understanding}, using the the form of the infinitesimal mean computed above to study the infinitesimal variance our process. We find that the infinitesmimal variance of the two-level birth-death process is given by
\begin{equation}
\begin{aligned}
&\lim_{\Delta t \to 0} \frac{1}{\Delta t} \var  \left[ f^{m,n}_i(t+\Delta t) - f^{m,n}_i(t)  \bigg| f^{m,n}_{i}(t) \right] \\ &= \lim_{\Delta t \to 0} \left[ \frac{1}{\Delta t} E\left[ \left(f^{m,n}_i(t+\Delta t) - f^{m,n}_i(t) \right)^2 \bigg| f^{m,n}_i(t) \right] - \frac{1}{\Delta t} \underbrace{\left( E \left[ f^{m,n}_i(t+\Delta t) - f^{m,n}_i(t) \bigg| f^{m,n}_i(t)\right] \right)^2}_{= O\left((\Delta t)^2\right)} \right] \\
&= \lim_{\Delta t \to 0} \frac{1}{\Delta t}  E\left[ \left(f^{m,n}_i(t+\Delta t) - f^{m,n}_i(t) \right)^2 \bigg| f^{m,n}_i(t) \right] \\
&= \lim_{\Delta t \to 0} \left\{ \frac{1}{\Delta t} \left(\frac{1}{m^2} \right) \left[ P\left( f^{m,n}_i(t+\Delta t) - f^{m,n}_i(t)  = \frac{1}{m} \right) +  P\left( f^{m,n}_i(t+\Delta t) - f^{m,n}_i(t)  = - \frac{1}{m} \right)  \right] \right\}
\end{aligned}
\end{equation}
Noting that the rates of replication events at the individual and group level are linear in $m$, we see that the infinitesimal variance may be written in the form
\begin{equation}
\begin{aligned}
\lim_{\Delta t \to 0} \frac{1}{\Delta t} \var \left[ f^{m,n}_i(t+\Delta t) - f^{m,n}_i(t)  \bigg| f^{m,n}_{i}(t) \right] = \lim_{\Delta t \to 0} \left[ \frac{1}{\Delta t} \frac{O(m) \Delta t}{m^2} \right] &= \frac{O(m)}{m^2} \\
& \to 0 \: \: \mathrm{as} \: \: m \to \infty.
\end{aligned}
\end{equation}
As the infinitesimal variance of the two-level birth-death process goes to $0$ as the number of groups $m$ tends to infinity, we expect the distribution of the composition $f^{m,n}_i(t)$ to behave deterministically in the limit as $m \to \infty$ and for $f^{m,n}_i(t)$ to agree with its mean $E[f^{m,n}_i(t)]$. In this limit, we then define $f^{n}_i(t) = \lim_{m \to \infty} f^{m,n}(t) = \lim_{m \to \infty} E[f^{m,n}_i(t)]$ and use linearity of expectation to heuristically write that
\begin{equation}
\lim_{ \Delta t \to 0} \frac{E\left[f^{n}_i(t+\Delta t) - f^{n}_t(t) \bigg| f^{n}_i(t) \right]}{\Delta t} = \lim_{\Delta t \to 0} \frac{f^{n}_i(t+\Delta t) - f^{n}_i(t)}{\Delta t} = \dsddt{f^{n}_i(t)}.
\end{equation}
Applying this to our expression for infinitesimal mean and taking the limit as $m \to \infty$ on both sides  Equation \eqref{eq:infmeantwotype} then allows us to expect that the fraction of $i$-punisher groups $f^{n}_i(t)$ satisfies the following system of ODEs in the case of infinitely many groups ($m \to \infty$) with finite group size $n$: 

\begin{equation} \label{eq:multilevelODEstwotype}
\begin{aligned}
\dsddt{f_i^n} &= \frac{1}{n} D_2\left( \frac{i}{n} \left( 1 - \frac{i}{n} \right) f_i^n \right) + \Lambda f_{i}^{n}(t) \left( \sum_{j=0}^{n} f_j^{n}(t) \left[ \rho\left( \frac{i}{n} , \frac{j}{n} \right) -  \rho\left( \frac{j}{n} , \frac{i}{n} \right) \right]\right) \\
\\ &+ w_I D_1^+ \left( \frac{i}{n} \left( 1 - \frac{i}{n} \right) \pi_D\left(\frac{i}{n} \right) f_i(t) \right) - w_I D_1^- \left( \frac{i}{n} \left( 1 - \frac{i}{n} \right) \pi_P\left(\frac{i}{n} \right) f_i(t) \right)
\end{aligned}
\end{equation}
for each $i \in \{1,\cdots,n-1\}$.
Following a similar approach, it is also possible to derive ODEs for the fraction groups with all-defector $(i=0$) and all-punisher ($i=n$) compositions, which are given by 
\begin{subequations} \label{eq:multilevelODEedge}
\begin{align}
\dsddt{f_0^{n}(t)} &= \left( \frac{n-1}{n} \right) \left( 1 + w_I \pi_D\left(\frac{1}{n} \right) \right) f_1^n(t) + \Lambda f_0(t)  \left( \sum_{j=0}^{n} f_j^{n}(t) \left[ \rho\left( \frac{i}{n} , \frac{j}{n} \right) -  \rho\left( \frac{j}{n} , \frac{i}{n} \right) \right]\right) \\
\dsddt{f_n^n(t)} &= \left( \frac{n-1}{n} \right) \left( 1 + w_I \pi_P\left(\frac{n-1}{n} \right) \right) f_1^n(t) + \Lambda f_0(t)  \left( \sum_{j=0}^{n} f_j^{n}(t) \left[ \rho\left( \frac{i}{n} , \frac{j}{n} \right) -  \rho\left( \frac{j}{n} , \frac{i}{n} \right) \right]\right).
\end{align}
\end{subequations}
We can then study the system of ODEs described by Equations \eqref{eq:multilevelODEstwotype} and \eqref{eq:multilevelODEedge} to explore the dynamics of multilevel selection with pairwise between-group selection for the case of infinitely many groups that each contain $n$ members. 

Next, we will continue to derive our PDE model for multilevel selection with pairwise between-group competition by considering the limit as the size of the groups also tends to infinity ($n \to \infty$). In the limit of infinite group size, we will look to describe the strategic composition of our group-structured population in terms of the probability density $f(t,z)$ of groups featuring a fraction $z$ of altruistic punishers (and a corresponding fraction $1-z$ of defectors). 

Multiplying both sides of Equation \eqref{eq:multilevelODEstwotype} by $n$ further allows us to rewrite our ODEs in terms of the quantity $n f_i^n(t)$, yielding
\begin{equation} \label{eq:ODEprelimit}
\begin{aligned}
\dsddt{}\left(n f_i^n \right) &= \frac{1}{n} D_2\left( \frac{i}{n} \left( 1 - \frac{i}{n} \right) n f_i^n \right) + \Lambda n f_{i}^{n}(t) \left( \sum_{j=0}^{n} \frac{1}{n} \left( n f_j^{n}(t) \right) \left[ \rho\left( \frac{i}{n} , \frac{j}{n} \right) -  \rho\left( \frac{j}{n} , \frac{i}{n} \right) \right]\right) \\
\\ &+ w_I D_1^+ \left( \frac{i}{n} \left( 1 - \frac{i}{n} \right) \pi_D\left(\frac{i}{n} \right) n f_i(t) \right) - D_1^- w_I \left( \frac{i}{n} \left( 1 - \frac{i}{n} \right) \pi_P\left(\frac{i}{n} \right) n f_i(t) \right)
\end{aligned}
\end{equation}
\begin{remark}
The goal of writing these ODEs in terms of the quantity $n f_i^n(t)$ is to convert a description of $f_i^n(t)$ as the empirical probability of groups with $i$ punishers into a description that will allow us to derive a probability density $f(t,z)$ over the fraction of altruistic punishers for any $z \in [0,1]$. While a uniform mix of group compositions is described by $f^n_i(t) = \frac{1}{n}$ for each $i \in \{1,\cdots,n\}$ for the case of groups of finite size $n$, we would want to analogously define a uniform distribution of groups of infinite size by a probability density satisfying $f(t,z) = 1$ for each $z \in [0,1]$.  
\end{remark}
We then look to take the limit as $n \to \infty$ to describe the dynamics of our group-structured population as group size becomes infinite. In this case, we see that the sum on the righthand side of Equation \eqref{eq:ODEprelimit} can be written as the following integral satisfying
\begin{equation}
\sum_{j=0}^{n} \frac{1}{n} \left( n f_j^{n}(t) \right) \left[ \rho\left( \frac{i}{n} , \frac{j}{n} \right) -  \rho\left( \frac{j}{n} , \frac{i}{n} \right) \right]\to \int_0^1 \left[ \rho(z,u) - \rho(u,z) \right] f(t,u) dy
\end{equation}
as $n \to \infty$, where we view the sum on the lefthand side as a Riemann approximation of the integral on the righthand side with the quantity $n f_i^n(t)$ approximating the values of $f(t,z)$ on the gridpoints $z \in \{0,\frac{1}{n},\cdots,\frac{n-1}{n},1\}$.
Further using the definition of our difference quotients, we see that, in the limit as $n \to \infty$, the probability density $f(t,z)$ will satisfy the following PDE
\begin{equation}
\dsdel{f(t,z)}{t} = -w_I \dsdel{}{z} \left( z (1-z) \left( \pi_P(z) - \pi_D(z) \right) f(t,z) \right) + \Lambda f(t,z) \left( \int_0^1 \left[ \rho(z,u) - \rho(u,z) \right] f(t,u) du \right).
\end{equation}
Finally, we can introduce the rescaling of time $\tau := \frac{t}{w_I}$ and introduce the new parameter $\lambda = \frac{\Lambda}{w_I}$ to write our PDE in the following form
\begin{equation}
\dsdel{f(\tau,z)}{\tau} = - \dsdel{}{z} \left( z (1-z) \left( \pi_P(z) - \pi_D(z) \right) f(\tau,z) \right) + \lambda f(\tau,z) \left( \int_0^1 \left[ \rho(z,u) - \rho(u,z) \right] f(\tau,u) du \right).
\end{equation}

\subsection{Sketch of Derivation for Trimorphic Multilevel Dynamics with Pairwise Between-Group Competition}
\label{sec:trimorphicderivation}

The derivation of our PDE model for the multilevel dynamics featuring three strategies is more involved due to the greater number of individual-level birth-death events that can occur. In this section, we will sketch the heuristic derivation of our PDE model from Equation \eqref{eq:PDEtrimorphicrho} for multilevel competition with pairwise between-group competition from our stochastic model of selection at two levels. We will follow the general approach used to derive a three-type PDE model for multilevel selection used to describe protocell evolution \citep{cooney2022pde}, extending this approach to incorporate within-group frequency-dependent interactions and pairwise between-group competition.

We start by considering a population featuring $m$ groups each composed of $n$ members, and we assume that each group member can be an cooperator, defector, or altruistic punisher. We denote the number of cooperators in a group by $i$, the number of defectors by $j$, and the the number of altruistic punishers by $n - i - j$. We describe the group composed of $i$ cooperators and $j$ defectors as an $(i,j)$-group, and we denote the fraction of groups with the composition $(i,j)$ at time $t$ by $f^{m,n}_{i,j}(t)$. 

We assume that individuals following strategy $X$ reproduce at rate $1 + w_I \pi_X\left(\frac{i}{n},\frac{j}{n}\right)$, producing an offspring with the same strategy, with the offspring replacing a randomly chosen member of the group. For our derivation of the PDE model for multilevel selection, we will focus our attention on group compositions $(i,j)$ for which all three strategies are present, satisfying the constraints $i \geq 1$, $j \geq 1$, and $n - i - j \geq 1$. For groups with such compositions, we see that the fraction of $(i,j)$-groups will increase by $\frac{1}{m}$ when one of the following six individual-level reproduction events occurs
\begin{itemize}
    \item A cooperator reproduces in an $(i-i,j+1)$ group and a defector is chosen to die, which occurs with rate
    \begin{equation}
 m f^{m,n}_{i-1,j+1}(t) \left( i - 1\right) \left(1 + w_I \pi_C\left( \frac{i-1}{n} , \frac{j+1}{n} \right)  \right) \left( \frac{j+1}{n} \right)
    \end{equation}
    \item  A cooperator reproduces in an $(i-i,j)$ group and an altruistic punisher is chosen to die, which occurs with rate
    \begin{equation}
 m f^{m,n}_{i-1,j}(t) \left( i-1 \right) \left(1 + w_I \pi_C\left( \frac{i-1}{n} , \frac{j}{n} \right)\right)  \left( 1 - \left[ \frac{i-1}{n} + \frac{j}{n} \right] \right)
    \end{equation}

    \item A defector reproduces in an $(i+1,j-1)$ group and a cooperator is chosen to die, which occurs with rate
    \begin{equation}
 m f^{m,n}_{i+1,j-1}(t) \left(j-1\right) \left(1 + w_I \pi_D\left( \frac{i+1}{n} , \frac{j-1}{n} \right)  \right) \left( \frac{i+1}{n} \right)
    \end{equation}

    \item   A defector reproduces in an $(i,j-1)$ group and an altruistic punisher is chosen to die, which occurs with rate
    \begin{equation}
 m f^{m,n}_{i,j-1}(t) \left( j - 1\right) \left(1 + w_I \pi_D\left( \frac{i}{n} , \frac{j-1}{n} \right)\right)  \left( 1 - \left[ \frac{i}{n} + \frac{j-1}{n} \right] \right)
    \end{equation}

    \item An altruistic punisher reproduces in an $(i+1,j)$ group and a cooperator is chosen to die, which occurs with rate
    \begin{equation}
 m f^{m,n}_{i+1,j}(t) \left( n - i - j - 1 \right) \left(1 + w_I \pi_P\left( \frac{i+1}{n} , \frac{j}{n} \right)  \right) \left( \frac{i+1}{n} \right)
    \end{equation}

    \item An altruistic punisher reproduces in an $(i,j+1)$ group and a defector is chosen to die, which occurs with rate
    \begin{equation}
 m f^{m,n}_{i,j+1}(t) \left( n - i - j - 1 \right) \left(1 + w_I \pi_P\left( \frac{i}{n} , \frac{j+1}{n} \right)\right)  \left( 1 - \left[ \frac{i}{n} + \frac{j+1}{n} \right] \right).
    \end{equation}
\end{itemize}

Similarly, the fraction of groups $f^{m,n}_{i,j}(t)$ with composition $(i,j)$ decreases by $\frac{1}{m}$ when one of the following six events occurs
\begin{itemize}
    \item A cooperator reproduces in an $(i,j)$ group and a defector is chosen to die, which occurs with rate
    \begin{equation}
 m f^{m,n}_{i,j}(t) i  \left(1 + w_I \pi_C\left( \frac{i}{n} , \frac{j}{n} \right)  \right) \left( \frac{j}{n} \right)
    \end{equation}
    \item  A cooperator reproduces in an $(i,j)$ group and an altruistic punisher is chosen to die, which occurs with rate
    \begin{equation}
 m f^{m,n}_{i,j}(t) i \left(1 + w_I \pi_C\left( \frac{i}{n} , \frac{j}{n} \right)\right)  \left( 1 - \left[ \frac{i}{n} + \frac{j}{n} \right] \right)
    \end{equation}

    \item A defector reproduces in an $(i,j)$ group and a cooperator is chosen to die, which occurs with rate
    \begin{equation}
 m f^{m,n}_{i,j}(t)  j \left(1 + w_I \pi_D\left( \frac{i}{n} , \frac{j}{n} \right)  \right) \left( \frac{i}{n} \right)
    \end{equation}

    \item   A defector reproduces in an $(i,j)$ group and an altruistic punisher is chosen to die, which occurs with rate
    \begin{equation}
 m f^{m,n}_{i,j}(t) j \left(1 + w_I \pi_D\left( \frac{i}{n} , \frac{j}{n} \right)\right)  \left( 1 - \left[ \frac{i}{n} + \frac{j}{n} \right] \right)
    \end{equation}

    \item An altruistic punisher reproduces in an $(i,j)$ group and a cooperator is chosen to die, which occurs with rate
    \begin{equation}
 m f^{m,n}_{i,j}(t) \left( n - i - j \right) \left(1 + w_I \pi_P\left( \frac{i}{n} , \frac{j}{n} \right)  \right) \left( \frac{i}{n} \right)
    \end{equation}

    \item An altruistic punisher reproduces in an $(i,j+1)$ group and a defector is chosen to die, which occurs with rate
    \begin{equation}
 m f^{m,n}_{i,j}(t) \left( n - i - j \right) \left(1 + w_I \pi_P\left( \frac{i}{n} , \frac{j}{n} \right)\right)  \left(  \frac{j}{n} \right).
    \end{equation}
\end{itemize}

Now we can examine the rates at which group compositions change due to pairwise between-group competition. The fraction of groups $f^{m,n}_{i,j}(t)$ increases by $\frac{1}{m}$ when an $(i,j)$-group wins a pairwise competition with a group with composition $(k,l)$, which occurs with rate
\begin{equation}
\Lambda m f^{m,n}_{i,j}(t) \sum_{\substack{k = 1 \\ (k,l) \ne (i,j)}}^{n} \sum_{l=1}^{n-i} \left[ \rho\left( \frac{i}{n} , \frac{j}{n} \right) f^{m,n}_{k,l}(t) \right] .
\end{equation}
The number of groups with composition $(i,j)$ decreases by $\frac{1}{m}$ when a group with composition 
\begin{equation}
\Lambda m f^{m,n}_{i,j}(t) \sum_{\substack{k = 1 \\ (k,l) \ne (i,j)}}^{n} \sum_{l=1}^{n-i} \left[ \rho\left( \frac{j}{n} , \frac{i}{n} \right) f^{m,n}_{k,l}(t) \right] .
\end{equation}

Combining the expressions for the rates at which the group compositions $f^{m,n}_{i,j}(t)$ of within-group and between-group selection events, we see that the conditional mean for the change in $f^{m,n}_{i,j}$ in the time-interval $[t,t+\Delta t]$ can be written as

\begin{equation} \label{eq:conditionalmeantrimorphic}
\begin{aligned}
& E \left[f^{m,n}_{i,j}(t+\Delta t) - f^{m,n}_{i,j}(t) \bigg| f^{m,n}_{i,j}(t)  \right]  \\
&= \frac{1}{m} P\left(f^{m,n}_{i,j}(t+\Delta t) - f^{m,n}_{i,j}(t)  = \frac{1}{m} \right) - \frac{1}{m} P\left(f^{m,n}_{i,j}(t+\Delta t) - f^{m,n}_{i,j}(t)  = -\frac{1}{m} \right) + o(\Delta t)\\
&= \frac{1}{m} \left[  m f^{m,n}_{i-1,j+1}(t) \left(i-1\right) \left(1 + w_I \pi_C\left( \frac{i-1}{n} , \frac{j+1}{n} \right)  \right) \left( \frac{j+1}{n} \right) \right] \Delta t \\
&+ \frac{1}{m} \left[m f^{m,n}_{i-1,j}(t) \left( i -1 \right) \left(1 + w_I \pi_C\left( \frac{i-1}{n} , \frac{j}{n} \right)\right)  \left( 1 - \left[ \frac{i-1}{n} + \frac{j}{n} \right] \right) \right] \Delta t \\
&+ \frac{1}{m} \left[  m f^{m,n}_{i+1,j-1}(t) \left( j - 1 \right) \left(1 + w_I \pi_D\left( \frac{i+1}{n} , \frac{j}{n} \right)  \right) \left( \frac{i+1}{n} \right) \right] \Delta t \\
&+ \frac{1}{m} \left[ m f^{m,n}_{i,j-1}(t) \left( j - 1\right) \left(1 + w_I \pi_D\left( \frac{i}{n} , \frac{j-1}{n} \right)\right)  \left( 1 - \left[ \frac{i}{n} + \frac{j-1}{n} \right] \right) \right] \Delta t \\
&+ \frac{1}{m} \left[  m f^{m,n}_{i+1,j}(t) \left( n - i - j - 1 \right) \left(1 + w_I \pi_P\left( \frac{i+1}{n} , \frac{j}{n} \right)  \right) \left( \frac{i+1}{n} \right) \right] \Delta t \\
&+ \frac{1}{m} \left[ m f^{m,n}_{i,j+1}(t) \left( n - i - j - 1 \right) \left(1 + w_I \pi_P\left( \frac{i}{n} , \frac{j+1}{n} \right)\right)  \left(  \frac{j+1}{n} \right) \right] \Delta t \\
&- \frac{1}{m} \left[ m f^{m,n}_{i,j}(t) i  \left(1 + w_I \pi_C\left( \frac{i}{n} , \frac{j}{n} \right)  \right) \left(1 -  \frac{i}{n} \right) \right] \Delta t \\
&-  \frac{1}{m} \left[ m f^{m,n}_{i,j}(t)  j \left(1 + w_I \pi_D\left( \frac{i}{n} , \frac{j}{n} \right)  \right) \left( 1 - \frac{j}{n} \right) \right] \Delta t \\
&-  \frac{1}{m} \left[  m f^{m,n}_{i,j}(t) \left( n - i - j \right) \left(1 + w_I \pi_P\left( \frac{i}{n} , \frac{j}{n} \right)  \right) \left( \frac{i}{n} + \frac{j}{n} \right) \right] \Delta t \\
&+ \frac{1}{m} \left\{ \Lambda m f^{m,n}_{i,j}(t) \sum_{\substack{k = 1 \\ (k,l) \ne (i,j)}}^{n} \sum_{l=1}^{n-i} \left[ \rho\left( \frac{i}{n} , \frac{j}{n} \right) f^{m,n}_{k,l}(t) \right] \right\} \Delta t \\
&- \frac{1}{m} \left[ \Lambda m f^{m,n}_{i,j}(t) \sum_{\substack{k = 1 \\ (k,l) \ne (i,j)}}^{n} \sum_{l=1}^{n-i} \left[ \rho\left( \frac{j}{n} , \frac{i}{n} \right) f^{m,n}_{k,l}(t) \right] \right] \Delta t + o\left( \Delta t\right) 
\end{aligned}
\end{equation}

We can then use an analogous approach to the case of two-strategy multilevel dynamics studied in Section \ref{sec:dimorphicderivation} to derive the infinitesimal mean and variance for the two-level birth-death process, and we can similarly deduce that the infinitesimal variance vanishes in the limit as $m \to \infty$. This will allow us to derive a deterministic system of ODEs for the fractions $f^{n}_{i,j}(t)$ of groups featuring a composition $(i,j)$ in the limit of infinitely many groups that are each composed of $n$ members. This derivation of a system of ODEs is also inspired by the approach taken to study trimorphic models of multilevel selection in the context of protocell evolution (see \citep[Section A.2.2]{cooney2022pde} for details on justifying the form taken by the system of ODEs for $f^{n}_{i,j}(t)$ in the limit as $m \to \infty$). 

In deriving this system of ODEs for $f^n_{i,j}(t)$, we look to simplify the terms in our expression for the conditional mean from Equation \eqref{eq:conditionalmeantrimorphic} by writing the righthand side in terms of the forward and backward first-order difference quotients in $x$ and $y$

\begin{subequations}
\begin{align}
D_{1,x}^{+} \left( u\left( \ds\frac{i}{n}, \ds\frac{j}{n} \right) \right) &= \frac{u \left( \ds\frac{i + 1}{n} , \ds\frac{j}{n}\right) - u \left( \ds\frac{i}{n} , \ds\frac{j}{n}\right)}{\frac{1}{n}} \: \: , \: \: D_{1,x}^{-} \left( u\left( \ds\frac{i}{n}, \ds\frac{j}{n} \right) \right) &= \frac{u \left( \ds\frac{i}{n} , \ds\frac{j}{n}\right) - u \left( \ds\frac{i-1}{n} , \ds\frac{j}{n}\right)}{\frac{1}{n}} \\
D_{1,y}^{+} \left( u\left( \ds\frac{i}{n}, \ds\frac{j}{n} \right) \right) &= \frac{u \left( \ds\frac{i}{n} , \ds\frac{j + 1}{n}\right) - u \left( \ds\frac{i}{n} , \ds\frac{j}{n}\right)}{\ds\frac{1}{n}} \: \: , \: \: D_{1,y}^{-} \left( u\left( \ds\frac{i}{n}, \ds\frac{j}{n} \right) \right) &= \frac{u \left( \ds\frac{i}{n} , \ds\frac{j}{n}\right) - u \left( \ds\frac{i}{n} , \ds\frac{j-1}{n}\right)}{\ds\frac{1}{n}},
\end{align}
\end{subequations}

the central second-order difference quotients in $x$ and $y$
\begin{subequations}
\begin{align}
D_{2,x}^{c} \left( u\left( \ds\frac{i}{n}, \ds\frac{j}{n} \right) \right) &= \frac{u\left( \ds\frac{i+1}{n}, \ds\frac{j}{n} \right) - 2 u\left( \ds\frac{i}{n}, \ds\frac{j}{n} \right) + u\left( \ds\frac{i-1}{n}, \ds\frac{j}{n} \right) }{\ds\frac{1}{n^2}} \\
D_{2,yy}^{c} \left( u\left( \ds\frac{i}{n}, \ds\frac{j}{n} \right) \right) &= \frac{u\left( \ds\frac{i}{n}, \ds\frac{j+1}{n} \right) - 2 u\left( \ds\frac{i}{n}, \ds\frac{j}{n} \right) + u\left( \ds\frac{i}{n}, \ds\frac{j-1}{n} \right) }{\ds\frac{1}{n^2}},
\end{align}
\end{subequations}
and a mixed second-order difference quotient 
\begin{equation}
D_{1,x}^c \left( D_{1,y}^{-} \left( u\left( \ds\frac{i}{n}, \ds\frac{j}{n} \right) \right)\right) = \frac{u\left( \ds\frac{i+1}{n}, \ds\frac{j}{n} \right) - u\left( \ds\frac{i-1}{n}, \ds\frac{j}{n} \right) - u\left( \ds\frac{i+1}{n}, \ds\frac{j-1}{n} \right) + u\left( \ds\frac{i-1}{n}, \ds\frac{j-1}{n} \right)}{\ds\frac{2}{n^2}}.
\end{equation}

We can then use these difference quotients to rewrite the righthand side of Equation \eqref{eq:conditionalmeantrimorphic}, divide both sides by $\Delta t$, and then take the limits as $\Delta t \to 0$ and $m \to \infty$, which allows us to obtain the following system of ODEs for the quantity $f^{n}_{i,j}(t) := \lim_{m \to \infty} f^{m,n}_{i,j}(t)$:
\begin{equation}
\begin{aligned} 
\dsddt{f^{n}_{i,j}(t)} &= \frac{1}{n} \left[ D_{2,xx}^c \left( \frac{i}{n} \left( 1 - \frac{i}{n} \right)  f^{n}_{i,j}(t) \right) - 2 D_{1,x}^c \left( D_{1,y}^{-} \left( \left[ \frac{i}{n} \right] \left[ \frac{j}{n} \right] f^{n}_{i,j}(t) \right) \right) \right] \\  &+ \frac{1}{n} \left[ D_{2,yy}^{c} \left( \frac{j}{n} \left( 1 - \frac{i+j}{n} \right)f^{n}_{i,j}(t) \right) + D_{2,yy}^c \left( \frac{j}{n} \left( \frac{i-1}{n} \right) \right) f^{n}_{i,j}(t) \right] \\
&+ w_I \left[ \left( \frac{i-1}{n} \right) D_{1,y}^+ \left( \frac{j}{n} \pi_C\left( \frac{i-1}{n}, \frac{j}{n} \right) f^{n}_{i-1,j}(t) \right) - D_{1,x}^+ \left( \frac{i}{n} \left( 1 - \frac{i}{n} \right) \pi_C\left( \frac{i}{n},\frac{j}{n} \right) f^n_{i,j}(t) \right) \right] \\
&+ w_I \left[ \left( \frac{j-1}{n} \right) D_{1,x}^+ \left( \frac{i}{n} \pi_D\left( \frac{i}{n}, \frac{j-1}{n} \right) f^{n}_{i,j-1}(t) \right) - D_{1,y}^+ \left( \frac{j}{n} \left( 1 - \frac{j}{n} \right) \pi_D\left( \frac{i}{n},\frac{j}{n} \right) f^n_{i,j}(t) \right) \right]  \\
&+ w_I \left[ D_{1,x}^+ \left( \left[ \frac{i}{n} \right] \left[ 1 - \frac{i+j}{n} \right] \pi_P\left( \frac{i}{n} ,\frac{j}{n} \right) f^n_{i,j}(t) \right) + D_{1,y}^+ \left( \left[ \frac{j}{n} \right] \left[ 1 - \frac{i+j}{n} \right] \pi_P\left( \frac{i}{n} ,\frac{j}{n} \right) f^n_{i,j}(t) \right)\right] \\ 
&+  \Lambda m f^{n}_{i,j}(t) \left\{ \left( \sum_{\substack{k = 1 \\ (k,l) \ne (i,j)}}^{n} \sum_{l=1}^{n-k} \left[ \left[ \rho\left( \frac{i}{n} , \frac{j}{n} ; \frac{k}{n}, \frac{l}{n} \right)  - \rho\left( \frac{k}{n} , \frac{l}{n} ; \frac{i}{n}, \frac{j}{n} \right)    \right] f^{n}_{k,l}(t) \right] \right) \right\}
\end{aligned}
\end{equation}

Next, we prepare to obtain our PDE model in the limit as $n \to \infty$ by multiplying both sides of our equation by $\left( \frac{(n+1)(n+2)}{2} \right)$, which allow us to write our ODE in terms of the quantity $\hat{f}^n_i(t) := \left( \frac{(n+1)(n+2)}{2} \right) f^n_{i,j}(t)$, yielding
\begin{equation} \label{eq:trimorphicODEprelimit}
\begin{aligned}
\dsddt{\hat{f}^n_{i,j}(t)} &= \frac{1}{n} \left[ D_{2,xx}^c \left( \frac{i}{n} \left( 1 - \frac{i}{n} \right)  \hat{f}^{n}_{i,j}(t) \right) - 2 D_{1,x}^c \left( D_{1,y}^{-} \left( \left[ \frac{i}{n} \right] \left[ \frac{j}{n} \right]  \hat{f}^{n}_{i,j}(t) \right) \right) \right] \\  &+ \frac{1}{n} \left[ D_{2,yy}^{c} \left( \frac{j}{n} \left( 1 - \frac{i+j}{n} \right) \hat{f}^{n}_{i,j}(t) \right) + D_{2,yy}^c \left( \frac{j}{n} \left( \frac{i-1}{n} \right) \right) \hat{f}^{n}_{i,j}(t) \right] \\
&+ w_I \left[ \left( \frac{i-1}{n} \right) D_{1,y}^+ \left( \frac{j}{n} \pi_C\left( \frac{i-1}{n}, \frac{j}{n} \right)\hat{f}^{n}_{i-1,j}(t) \right) - D_{1,x}^+ \left( \frac{i}{n} \left( 1 - \frac{i}{n} \right) \pi_C\left( \frac{i}{n},\frac{j}{n} \right) \hat{f}^n_{i,j}(t) \right) \right] \\
&+ w_I \left[ \left( \frac{j-1}{n} \right) D_{1,x}^+ \left( \frac{i}{n} \pi_D\left( \frac{i}{n}, \frac{j-1}{n} \right) \hat{f}^{n}_{i,j-1}(t) \right) - D_{1,y}^+ \left( \frac{j}{n} \left( 1 - \frac{j}{n} \right) \pi_D\left( \frac{i}{n},\frac{j}{n} \right) \hat{f}^n_{i,j}(t) \right) \right]  \\
&+ w_I \left[ D_{1,x}^+ \left( \left[ \frac{i}{n} \right] \left[ 1 - \frac{i+j}{n} \right] \pi_P\left( \frac{i}{n} ,\frac{j}{n} \right) \hat{f}^n_{i,j}(t) \right) + D_{1,y}^+ \left( \left[ \frac{j}{n} \right] \left[ 1 - \frac{i+j}{n} \right] \pi_P\left( \frac{i}{n} ,\frac{j}{n} \right) \hat{f}^n_{i,j}(t) \right)\right] \\ 
&+  \Lambda m \hat{f}^{n}_{i,j}(t) \left\{ \left( \sum_{k=1}^n \sum_{l=1}^{n-k} \left\{ \left(\frac{2}{(n+2)(n+1)}\right)\left[ \rho\left( \frac{i}{n} , \frac{j}{n} ; \frac{k}{n}, \frac{l}{n} \right)  - \rho\left( \frac{k}{n} , \frac{l}{n} ; \frac{i}{n}, \frac{j}{n} \right)  \right] \hat{f}^{n}_{k,l}(t) \right\} \right) \right\},
\end{aligned}
\end{equation}
where we used the fact that $\rho\left(\frac{i}{n},\frac{j}{n};\frac{i}{n},\frac{j}{n}\right) = \frac{1}{2}$ to simplify the sum in the term for between-group competition. As in the case of two-strategy multilevel dynamics, this approach allows us to compare the probabilities $f^{n}_{i,j}(t)$ of having $n$-member groups with compositions $(i,j)$with a corresponding pronability density $f(t,x,y)$ for groups having composition with fractions $(x,y)$ in the limit of infinite group size. 

We then look to take the limit of both sides as $n \to \infty$, considering the continuous variables $x = \frac{i}{n}$, $y = \frac{j}{n}$, $u = \frac{k}{n}$, and $v = \frac{l}{n}$ and the continuous density $f(t,x,y) = \lim_{n \to \infty} \hat{f}^{n}_{i,j}(t)$. For the terms with second-order difference quotients. We note that the diffusive effects are described by a discrete Kimura diffusion operator given by
\begin{equation}
\begin{aligned}
\mc{L}^{n}_{K} f^{n}_{i,j}(t) &=  D_{2,xx}^c \left( \frac{i}{n} \left( 1 - \frac{i}{n} \right)  \hat{f}^{n}_{i,j}(t) \right) - 2 D_{1,x}^c \left( D_{1,y}^{-} \left( \left[ \frac{i}{n} \right] \left[ \frac{j}{n} \right]  \hat{f}^{n}_{i,j}(t) \right) \right)    \\ &  \hspace{5mm}  + D_{2,yy}^{c} \left( \frac{j}{n} \left( 1 - \frac{i+j}{n} \right) \hat{f}^{n}_{i,j}(t) \right) + D_{2,yy}^c \left( \frac{j}{n} \left( \frac{i-1}{n} \right)\hat{f}^{n}_{i,j}(t)  \right)  ,
\end{aligned}
\end{equation}
which satisfies
\begin{equation}
\begin{aligned}
\lim_{n \to \infty} \mc{L}^N_K f^{n}_{i,j}(t) = 
&= \frac{\partial^2}{\partial x^2} \left[ x(1-x) f(t,x,y) \right] - 2 \frac{\partial^2}{\partial x \partial y} \left( x y f(t,x,y) \right) + \frac{\partial^2}{\partial y^2} \left( y(1-y) f(t,x,y) \right),
\end{aligned}
\end{equation}
and therefore we can deduce that 
\begin{equation}
\begin{aligned}
&\frac{1}{n} \left[ D_{2,xx}^c \left( \frac{i}{n} \left( 1 - \frac{i}{n} \right)  \hat{f}^{n}_{i,j}(t) \right) - 2 D_{1,x}^c \left( D_{1,y}^{-} \left( \left[ \frac{i}{n} \right] \left[ \frac{j}{n} \right]  \hat{f}^{n}_{i,j}(t) \right) \right) \right] \\  &+ \frac{1}{n} \left[ D_{2,yy}^{c} \left( \frac{j}{n} \left( 1 - \frac{i+j}{n} \right) \hat{f}^{n}_{i,j}(t) \right) + D_{2,yy}^c \left( \frac{j}{n} \left( \frac{i-1}{n} \right) \right) \hat{f}^{n}_{i,j}(t) \right]  \\
& \to 0 \: \: \mathrm{as} \: \: n \to \infty.
\end{aligned}
\end{equation}
This means that the diffusive effects vanish in the limit of infinite group size, and our PDE model will only depend on advection terms describing the effects of individual-level payoffs and a nonlocal term describing group-level competition. For the nonlocal term, we use a Riemann approximation to see that 
\begin{equation}
\begin{aligned}
 &\left\{ \left( \sum_{k=1}^n \sum_{l=1}^{n-k} \left\{ \left(\frac{2}{(n+2)(n+1)}\right)\left[ \rho\left( \frac{i}{n} , \frac{j}{n} ; \frac{k}{n}, \frac{l}{n} \right)  - \rho\left( \frac{k}{n} , \frac{l}{n} ; \frac{i}{n}, \frac{j}{n} \right)  \right] \hat{f}^{n}_{k,l}(t) \right\} \right) \right\} \\ &\to \int_0^1 \int_0^{1-u} \left[ \rho(x,y;u,v) - \rho(u,v;x,y) \right] f(t,u,v) dv du \: \: \mathrm{as} \: \: n \to \infty.
 \end{aligned}
\end{equation}
We can then see by taking the limit of both sides of Equation \eqref{eq:trimorphicODEprelimit} as $n 
\to \infty$ that our two-level birth-death process can be described through the following PDE for the probability density $f(t,x,y)$
\begin{equation}
\begin{aligned}
\dsdel{f(t,x,y)}{t} &= w_I \left[ x \dsdel{}{y} \left\{ y \pi_C(x,y) f(t,x,y) \right\} - \dsdel{}{x} \left\{ x(1-x) \pi_C(x,y) f(t,x,y) \right\} \right] \\
&+ w_I \left[y \dsdel{}{x} \left\{ x \pi_D(x,y) f(t,x,y) \right\} - \dsdel{}{y} \left\{ y(1-y) \pi_D(x,y) f(t,x,y)\right\} \right] \\
&+ w_I \left[  \dsdel{}{x} \left\{ x (1-x-y) \pi_P(x,y) f(t,x,y) \right\} + \dsdel{}{y} \left\{ y (1-x-y) \pi_P(x,y) f(t,x,y) \right\}\right] \\
&+ \Lambda f(t,x,y) \left[ \int_0^1 \int_0^{1-u} \left[\rho(x,y; u,v) - \rho(u,v;x,y) \right] f(t,x,y) dv du\right].
\end{aligned}
\end{equation}
After rearranging terms, dividing both sides by $w_I$, rescaling time as $\tau := \frac{t}{w_I}$, and introducing the new parameter $\lambda := \frac{\Lambda}{w_I}$, we can write our PDE model for multilevel selection with pairwise between-group selection as
\begin{equation}
\begin{aligned}
\dsdel{f(\tau,x,y)}{\tau} &= - \dsdel{}{x} \left[ x\left\{ (1-x) \left( \pi_C(x,y) - \pi_P(x,y) \right) - y \left( \pi_D(x,y) - \pi_P(x,y) \right) \right\} f(\tau,x,y) \right] \\ 
&- \dsdel{}{y} \left[ y\left\{ (1-y) \left( \pi_D(x,y) - \pi_P(x,y) \right) - x \left( \pi_C(x,y) - \pi_P(x,y)\right) \right\} f(\tau,x,y) \right] \\
&= \lambda f(\tau,x,y) \left[ \int_0^1 \int_0^{1-u} \left[\rho(x,y; u,v) - \rho(u,v;x,y) \right] f(\tau,x,y) dv du\right].
\end{aligned}
\end{equation}

\section{Deriving Steady-State Densities for Additively Separable Group-Level Victory Probabilities}
\label{sec:densityderivation}

In Section \ref{sec:analytical}, we characterized the long-time behavior of the dynamics of multilevel selection for the case of pairwise between-group competition featuring additively separable group-level victory probabilities. Proposition \ref{prop:steadycooperators} provided our characterization for the cases of a group-level victory probability depending on the difference in average fraction of altruistic punishers between groups (corresponding {to a group-level victory probability $\rho(z,u) = \frac{1}{2} + \frac{z-u}{2}$ and} a net group-level reproduction rate $\mc{G}(z) = z$), while Proposition \ref{prop:steadypayoff} described analogous results for the case in which group-level victory depended on the difference in average group payoffs normalized by the maximum possible difference in group payoffs $G^* - G_{*}$ (corresponding to {a group-level victory probability of $\rho(z,u) = \frac{1}{2} + \frac{1}{2} \left[\frac{G(z) - G(u)}{G^* - G_*} \right]$ and} a net group-level reproduction rate of $\mc{G}(z) = \frac{G(z)}{G*-G_{*}}$).

While the long-time behavior of both models is determined by direct application of the general results Theorems \ref{thm:longtimePD} and \ref{thm:longtimePDel}, calculating the explicit expression for the steady state densities can be useful for plotting and interpreting the form of steady state densities for each of these group-level victory scenarios. In this section, we present the details of the derivations of the steady state densities for these two group-level victory scenarios, applying the result of Theorem \ref{thm:longtimePD} to derive the form of our steady state densities for the net group-level reproduction rates $\mc{G}(z) = z$ (for Proposition \ref{prop:steadycooperators}) {in Section \ref{sec:altruisticdiffsteady}} and $\mc{G}(z) = \frac{G(z)}{G*-G_{*}}$ (for Proposition \ref{prop:steadypayoff}) 
{in Section \ref{sec:payoffdiffsteady}}.

\subsection{Steady-State Densities for Group-Level Victory Determined by Fraction of Altruistic Punishers}
\label{sec:altruisticdiffsteady}

We now derive the family of steady state densities $f^{\lambda}_{\theta}(z)$ from Proposition \ref{prop:steadycooperators} for pairwise group-level competition based on the difference of altruistic punishers between groups. 

\begin{proof}[Derivation of Steady-State Densities for Proposition \ref{prop:steadycooperators}]
For our choice of group-level victory probability, we have the net group-level reproduction rate $\mc{G}(z) = z$, and we therefore have that $\mc{G}(1) - \mc{G}(s) = 1 -s$ and $\mc{G}(s) - \mc{G}(0) = s$. We also consider $\Pi(s) = \pi_D(z) - \pi_P(z) = c + k + q - (p+k) s$, and find that
\begin{subequations}
\begin{align}
\Pi(s) - \Pi(0) &= - (p+k) s \\
\Pi(s) - \Pi(1) &=  (p+k) (1-s).
\end{align}
\end{subequations}
We can then use Equation \eqref{eq:flambdatheta} to see that steady states for this choice of $\mc{G}(z)$ and $\Pi(z)$ are given by 
\begin{equation} \label{eq:flambdathetaGcoop}
\begin{aligned}
f^{\lambda}_{\theta}(z) &= z^{\Pi(0)^{-1} \left( \lambda \left[ \mc{G}(1) - \mc{G}(0) \right] - \theta \Pi(1)\right) - 1} \left( 1 - z\right)^{\theta - 1} \frac{1}{\Pi(z)} \exp\left(-\lambda \int_z^1 \frac{C(s)}{\Pi(s)} ds \right) \\
&= z^{\frac{1}{c+k+q} \left[ \lambda - \theta \left(c+q -p\right) \right] - 1} \left( 1 - z \right)^{\theta - 1} \left(c + k + q - (p+k) z \right)^{-1} \exp\left( - \lambda \int_z^1 \frac{C(s)}{\Pi(s)} ds \right)
\end{aligned}
\end{equation}
where $-\lambda C(s)$ is given by
\begin{equation}
\begin{aligned}
- \lambda C(s) &= \lambda \left( \frac{\mc{G}(s) - \mc{G}(0)}{s} \right) + \left( \frac{\lambda \left[ \mc{G}(1) - \mc{G}(0) \right] - \theta \Pi(1)}{\Pi(0)} \right) \left( \frac{\Pi(s) - \Pi(0)}{s} \right) \\
&+ \lambda \left( \frac{\mc{G}(s) - \mc{G}(1)}{1-s} \right) - \theta \left( \frac{\Pi(s) - \Pi(1)}{1-s} \right) \\
&= \lambda - \left(  \frac{\lambda - \theta (c + q - p)}{c+q+k} \right) \left( p + k \right) - \lambda - \theta \left(p + k  \right) \\
&= -\left(p + k \right) \left( \frac{ \lambda + \theta (k+p)}{c+q+k} \right).
\end{aligned}
\end{equation}
We can further compute that
\begin{equation}
\begin{aligned}
- \lambda \int_z^1 \frac{C(s)}{\Pi(s)} ds &= - \int_z^1 \left[ \left(p + k \right) \left( \frac{ \lambda + \theta (k+p)}{c+q+k} \right) \left(\frac{1}{c+k + q - (p+k) s}  \right) \right] ds \\
&= \left( \frac{ \lambda + \theta (k+p)}{c+q+k} \right) \log\left( c+k + q - (p+k) s \right) \bigg|_z^1  \\
&= \left(  \frac{\lambda + \theta (k+p)}{c+q+k} \right) \left[ \log\left( c + q - p \right) - \log\left( c + k + q  - (p+k)z \right)  \right].
\end{aligned}
\end{equation}
We can combine this with Equation \eqref{eq:flambdathetaGcoop} to see that the steady state densities $f^{\lambda}_{\theta}(y)$ can be written in the form
\begin{equation}
\begin{aligned}
f^{\lambda}_{\theta}(z) &= \frac{1}{Z_f} z^{\frac{1}{c+k+q} \left[ \lambda - \theta \left(c+q -p\right) \right] - 1} \left( 1 - z \right)^{\theta - 1} \left(c + k + q - (p+k) z \right)^{-\left( \frac{ \lambda + \theta (k+p)}{c+q+k} \right) - 1} \\
Z_f &= \int_0^1 \left[  z^{\frac{1}{c+k+q} \left[ \lambda - \theta \left(c+q -p\right) \right] - 1} \left( 1 - z \right)^{\theta - 1} \left(c + k + q - (p+k) z \right)^{-\left( \frac{\lambda + \theta (k+p)}{c+q+k} \right) - 1} \right] dz.
\end{aligned}
\end{equation}

\end{proof}

\subsection{Steady State Density for Group-Level Victory with Globally Normalized Local Update Rule}
\label{sec:payoffdiffsteady}
{Before providing the derivation of steady-state densities for the case of globally normalized differences in average payoffs of competing groups, it is helpful for us to calculate the minimum and maximum possible average payoffs $G^*$ and $G_*$ of groups for our model of altruistic punishment. In Lemma \ref{lem:minmaxpayoff}, we show that the average payoff of groups is maximized by the all-punisher composition (provide that the benefit conferred cooperation exceeds the cost to cooperate and the fixed cost to punish defectors), and that the minimum possible average group payoff can either be achieved by the all-defector group or for an intermediate mix of defectors and punishers depending on the parameters describing the costs and strength of punishment.  }

{
\begin{lemma}
\label{lem:minmaxpayoff}
Recall that the average payoff of a group with a fraction $z$ altruistic punishers a fraction $1-z$ of defectors is given by
\begin{equation}
G(z) = z \left[ b - \left( c + p + q +k \right) + \left (p+k\right) z \right].
\end{equation}
If $b - c > q$, then the maximum possible average payoffs of group members $G^* = \max_{z \in [0,1]} G(z)$ is given by
\begin{equation}
G^* = b - c -q
\end{equation}
and the minimum possible average payoff of group members  $G_* = \min_{z \in [0,1]} G(z)$ is given by
\begin{equation}
G_* = \left\{
\begin{array}{cc}
   0   &: b \geq c + p + k + q  \\
   - \ds\frac{\left( b - \left( c + p + k + q \right)\right)^2}{4 (p+k)}  &  b < c + p + k + q 
\end{array}\right. \\
\end{equation}
\end{lemma}
}

\begin{proof}
{
We can differentiate the average payoff function $G(z)$ to find that
\begin{equation}
G'(z) = b - \left( c + p + q + k \right) + 2 \left( p + k \right) z.
\end{equation}
By noting that $G(1) = b - c - q$, we see that this derivative for the all-punisher composition $z = 1$ satisfies
\begin{equation}
G'(1) = b - c - q + p + k = G(1) + p + k > G(1),
\end{equation}
and therefore we see that the average payoff of group members will be an increasing function of $z$ for $z$ sufficiently close to $1$ whenever the payoff parameters satisfy $b > c + q$. Because this assumption is required to have the average payoff of an all-punisher group exceed that of the all-defector group ($G(1) > G(0)$), we can combine the facts that $G(y)$ is a quadratic function, $G'(1) > 0$, and $G(1) > G(0)$ to deduce that $G(y)$ is maximized by all-punisher groups. This allows us to deduce that the maximum possible payoff of a group is given by
\begin{equation}
G^* = \max_{z \in [0,1]} G(z) = G(1)  \Longrightarrow G^* = b - c - q. 
\end{equation}

To characterize the minimal achievable average payoff of group members, we study the derivative of $G(z)$ at the all-defector composition 
\begin{equation}
G'(0) = b - \left( c + p + q + k\right),
\end{equation}
so $G(z)$ will be an increasing function on $[0,1]$ if $b \geq c + p + q + k$, while $G(z)$ will achieve a minimum value $G_*$ at an intermediate value of $z \in (0,1)$ when if $b < c + p +q + k$. In the latter case, we see that this minimum value $G_*$ of $G(z)$ is achieved when the fraction of altruistic punishers is given by
\begin{equation}
z_{min} = \frac{\left( c + p + k + q\right) - b}{2 (p+k)}. 
\end{equation}
By evaluating the average payoff $G(0)$ and $G(z_{min})$ for the two cases, we see that the minimum achievable average payoff is the smaller of $G(0) = 0$ and 
}
{
\begin{equation}
G(z_{min}) = - \ds\frac{\left( b - \left( c + p + k + q \right)\right)^2}{4 (p+k)}.
\end{equation}
}
\end{proof}

\revision[Now that we have understood the normalization term $G^*-G_*$ in the group-level victory probability based on the normalized difference of average group payoffs, we can look to describe the steady-state densities achieved for this group-level victory probability. We now present the derivation of the family of steady state densities $f^{\lambda}_{\theta}(z)$ from Proposition \ref{prop:steadypayoff} for the case of pairwise-group level competition {given by $\rho(z,u) = \frac{1}{2} + \frac{1}{2} \left[ \frac{G(z) - G(u)}{G^* - G_*} \right]$}. 

\begin{proof}[Derivation of Steady-State Density for Proposition \ref{prop:steadypayoff}]

For our choice of pairwise group-level victory probability, our net-group reproduction rate is $\mc{G}(z) = \frac{G(z)}{G^* - G_{*})}$. To calculate the values of the net-reproduction rate, we recall that the average payoff of group members is given by
\begin{equation}
G(z) = \left[ b - \left( c + p + q + k \right) \right] z + \left(p+k\right) z^2
\end{equation}
and that the average payoffs at the all-punisher and all-defector states are given by $G(1) = b - c - q$ and $G(0) = 0.$ Using the net-reproduction rate $\mc{G}(z)$ and the average payoff expressions then allows us to see that $\mc{G}(1) = \frac{G(1)}{G^* - G_{*}}$ and $\mc{G}(0) = \frac{G(0)}{G^* - G_{*}}$, so we find that
\begin{subequations}
\begin{equation}
\begin{aligned}
\mc{G}(1) - \mc{G}(s) &= \left(\frac{1}{G^{*} - G_{*}} \right) \left( G(1) - G(s) \right) \\ &= \left(\frac{1}{G^{*} - G_{*}} \right) \left[ \left( b - c - q \right) \left( 1 -s \right) + \left(p+k\right) s \left( 1 - s\right) \right]
\end{aligned}
\end{equation}
and 
\begin{equation}
\begin{aligned}
\mc{G}(y) - \mc{G}(0)&= \left(\frac{1}{G^{*} - G_{*}} \right) \left( G(s) - G(0) \right) \\ &= \left(\frac{1}{G^{*} - G_{*}} \right) \left[ \left( b - (c+p+q+k) \right) s + \left( p + k \right) s^2 \right]
\end{aligned}
\end{equation}
\end{subequations}

We can then again apply Equation \eqref{eq:flambdatheta} to see that steady states for this choice of $\mc{G}(z)$ and $\Pi(z)$ are given by 
\begin{equation} \label{eq:flambdathetaGcooprecall}
\begin{aligned}
f^{\lambda}_{\theta}(z) &= z^{\Pi(0)^{-1} \left( \lambda \left[ \mc{G}(1) - \mc{G}(0) \right] - \theta \Pi(1)\right) - 1} \left( 1 - z\right)^{\theta - 1} \frac{1}{\Pi(x)} \exp\left(-\lambda \int_z^1 \frac{C(s)}{\Pi(s)} ds \right) \\
&= z^{\frac{1}{c+k+q} \left[\frac{\lambda (b-c-q)}{G^* - G_{*}} - \theta \left(c+q -p\right) \right] - 1} \left( 1 - z \right)^{\theta - 1} \left(c + k + q - (p+k) z \right)^{-1}  \exp\left( - \lambda \int_z^1 \frac{C(s)}{\Pi(s)} ds \right),
\end{aligned}
\end{equation}
where the quantity $-\lambda C(s)$ is given by
\begin{equation}
\begin{aligned}
- \lambda C(s) &= \lambda \left( \frac{\mc{G}(s) - \mc{G}(0)}{s} \right) + \left( \frac{\lambda \left[ \mc{G}(1) - \mc{G}(0) \right] - \theta \Pi(1)}{\Pi(0)} \right) \left( \frac{\Pi(s) - \Pi(0)}{s} \right) \\
&+ \lambda \left( \frac{\mc{G}(s) - \mc{G}(1)}{1-s} \right) - \theta \left( \frac{\Pi(s) - \Pi(1)}{1-s} \right) \\
&= \lambda  \left( \frac{1}{G^* - G_{*}} \right) \left[ b - c - q + (p+k) s \right] \\
&- \left(p+k\right) \left(  \frac{\left(\frac{\lambda}{G^* - G_*}\right)(b-c-q) - \theta (c+q-p)}{c+k+q}\right) \\
&- \lambda \left( \frac{1}{G^* - G_{*}} \right) \left( b - (c+p+q +k) + (p+k) s \right) -\theta (p+k) \\
&= - \left(\frac{p+k}{c+k+q}\right) \left[ \frac{\lambda (b+k)}{G^*-G_{*}} - \theta (k+p) \right].
\end{aligned}
\end{equation}
We may then further compute that
\begin{equation}
\begin{aligned}
- \lambda \int_z^1 \frac{C(s)}{\Pi(s)} ds &= - \left(\frac{p+k}{c+k+q} \right)\int_z^1 \left( \left[  \frac{\lambda (b+k)}{G^*-G_{*}} + \theta (k+p) \right] \left( \frac{1}{c+k+q - (p+k) s} \right) \right) ds \\
&= %
\left( \frac{\lambda (b+k) + \theta (k + p) (G^* - G_*)}{(G^* - G_*)(c+k+q)} \right)\log\left( c + k + q - (p+k) s \right) \bigg|_{s=z}^{s=1} \\ %
&= \left( \frac{\lambda (b+k) +  \theta (k + p) (G^* - G_*)}{(G^* - G_*)(c+k+q)} \right) \left[ \log\left( c + q - p \right) -  \log\left( c + k + q - (p+k) z \right)  \right],
\end{aligned}
\end{equation}
and we may exponentiate both sides of this equation to see that
\begin{equation}
\exp\left( - \lambda \int_z^1 \frac{C(s)}{\Pi(s)} ds\right) = \left(c + q - p\right)^{\frac{\lambda (b+k) + \theta (k + p) (G^* - G_*)}{(G^* - G_*)(c+k+q)} } \left( c + k + q - (p+k) z \right)^{- \left( \frac{\lambda (b+k) + \theta (k + p) (G^* - G_*)}{(G^* - G_*)(c+k+q)} \right) }
\end{equation}
We can then plug this in to our expression for the steady state density from Equation \eqref{eq:flambdathetaGcooprecall} to see that
\begin{equation}
\begin{aligned}
f^{\lambda}_{\theta}(z) &= \frac{1}{Z_f} z^{\frac{1}{c+k+q} \left[\frac{\lambda (b-c-q)}{G^* - G_{*}} - \theta \left(c+q -p\right) \right] - 1} \left( 1 - z \right)^{\theta - 1} \left(c + k + q - (p+k) z \right)^{-\left( \frac{\lambda (b+k) + \theta (k + p) (G^* - G_*)}{(G^* - G_*)(c+k+q)} \right)-1} \\
Z_f &= \int_0^1 z^{\frac{1}{c+k+q} \left[\frac{\lambda (b-c-q)}{G^* - G_{*}} - \theta \left(c+q -p\right) \right] - 1} \left( 1 - z \right)^{\theta - 1} \left(c + k + q - (p+k) z \right)^{-\left( \frac{\lambda (b+k) + \theta (k + p) (G^* - G_*)}{(G^* - G_*)(c+k+q)} \right)-1} dz. 
\end{aligned}
\end{equation}
\end{proof}

\section{Effect of Altruistic Punishment in PDE Model of Multilevel Selection With Group-Level Replicator Competition}
\label{sec:LMaltruisticpunishment}
While the focus of this paper has been on studying the evolution of cooperation in the presence of altruistic punishment under multilevel selection with pairwise between-group competition, it is also natural to explore how altruistic punishment can impact the dynamics of multilevel selection using existing group-level replicator models in which groups reproduce at a rate depending on the state of their own group. In Section \ref{sec:LMdimorphic}, we consider the dynamics of a two-level replicator equation model for multilevel selection featuring altruistic punishment between groups. This allows us to illustrate similarities and differences between the dynamics of multilevel selection under the model from the main paper featuring frequency-dependent group-level conflicts and the two-level replicator equation with frequency-independent group-level competition in which a group replicates at a rate depending only on the strategic composition of the replicating group itself. In Section \ref{sec:LMtrimorphic}, we consider a two-level replicator equation featuring all three strategies (defectors, cooperators, and altruistic punishers), and we show how the trimorphic dynamics of multilevel selection compare with the tug-of-war between collective and individual incentives seen in two-strategy, two-level replicator dynamics.

\subsection{Studying Multilevel Replicator Dynamics Between Defectors and Altruistic Punishers}
\label{sec:LMdimorphic}

In this section, we explore a PDE model for multilevel selection for groups featuring defectors and altruistic punishers in the absence of pairwise between-group selection, using instead the model of group-level replication rates used previously in which group-level competition increases the frequency of group compositions with collective payoffs that are above the average payoff of the population. By performing this analysis, we are able to gain some intuition about which of the results from the main paper are primarily due to the effects of within-group altruistic punishment and which results may arise primarily from our choice to use pairwise group-level conflicts to describe group-level competition. This analysis of two-level replicator equations will apply existing results for long-time behavior of these PDE models that we have summarized in Theorems \ref{thm:longtimePD} and \ref{thm:longtimePDel}. 

If we consider the same within-group dynamics studied in the main paper and group-level reproduction rates that depend only on the strategic composition of the reproducing group, we can apply the framework introduced by Luo and coauthors \citep{luo2014unifying,van2014simple,luo2017scaling} to study a two-level replicator equation for multilevel selection in the presence of altruistic punishment. In particular, in a population featuring only defectors and altruistic punishers, we can describe a two-level replicator equation for the density of groups $f(t,z)$ featuring a fraction $z$ of altruistic punishers and a fraction $1-z$ of defectors by the PDE
\begin{equation}
\dsdel{f(t,z)}{t} = - \dsdel{}{z} \left( z (1-z) \left( \pi_P(z) - \pi_D(z) \right) f(t,z) \right) + \lambda f(t,z) \left[ G_{DP}(z) - \int_0^1 G_{DP}(u) f(t,u) du \right],
\end{equation}
where $\pi_D(z)$ and $\pi_P(z)$ are the payoffs received by defectors and altruistic punishers in a $z$-punisher group and $G_{DP}(z)$ is the average payoff received by the members of a $z$-punisher group. This PDE is of the form of Equation \eqref{eq:twotypeLM} with $\mc{G}(z) = G_{DP}(z)$ and $\Pi(z) := \pi_D(z) - \pi_P(z)$. 

Using the payoffs defined in Section \ref{sec:twostrategypayoffs} and the assumption that $p < c + q$, we can apply Theorem \ref{thm:longtimePD} to see that the threshold selection strength required to achieve a steady-state density featuring altruistic punishers is given by
\begin{equation} \label{eq:LMlambdastar}
\lambda^* = \frac{\theta \Pi(1)}{\mc{G}(1) - \mc{G}(0)} = \frac{\theta \left( c + q - p \right)}{b-c}.
\end{equation}
This is a decreasing function of $p$ and an increasing function of $q$, so it is easier to achieve cooperation through the combined effects of altruistic punishment and multilevel selection when the punishments imposed on defectors are stronger and the fixed costs to confer punishment are lower. This threshold $\lambda^*$ does not depend on the per-interaction cost of punishment $k$, and therefore it is easier to achieve cooperative behavior in our model when altruistic punishers only pay a cost to punish for each interaction they actually have with a defector. 

We may also calculate that, when $c + q > p$, the average payoff achieved at steady state starting from an initial population with H{\"o}lder exponent $\theta$ near $z=1$ is given by
\begin{equation} \label{eq:GDPLM}
\begin{aligned}
\lim_{t \to \infty} \langle G_{DP} \rangle_f &= \left\{
    \begin{array}{cr}
      \mc{G}(0) & : \lambda < \lambda^*\\
     \mc{G}(1) - \theta \Pi(1) & : \lambda > \lambda^*.
    \end{array}
  \right. \\
  &= \left\{
    \begin{array}{cr}
      0 & : \lambda < \lambda^*\\
      b - c - q - \theta ( c + q - p) & : \lambda > \lambda^*.
    \end{array}
  \right.
  \end{aligned}
\end{equation}
For $\lambda > \lambda^*$, this average payoff is increasing in the punishment $p$ imposed on defectors and decreasing in the fixed cost $q$ paid by altruistic punishers. 

\begin{remark}
Notably, we do not see any non-monotonic dependence of the threshold selection strength $\lambda^*$ or the average long-time payoff $\langle G_{DP} (\cdot) \rangle_f$ on any of the parameters related to the strength or cost of punishment. This differs from the behavior we saw in Section \ref{sec:grouplocalupdate} for the model with pairwise between-group competition following the local group update rule normalized based on the maximum possible group-level payoff difference. This serves to highlight that the non-monotonic behavior found in that case really appears to be a consequence of the assumption that group-level competition is normalized against the greatest possible payoff difference in that case, rather than a generic feature of group-level competition depending on average payoff gained in the presence of a within-group mechanism of altruistic punishment. 
\end{remark}

\begin{remark}
We see from Equations \eqref{eq:LMlambdastar} and \eqref{eq:GDPLM} that the threshold selection strength $\lambda^*$ and the average steady-state payoff $\langle G_{DP}(\cdot) \rangle_f$ do not depend on the per-interaction cost $k$ to punish defectors for our two-level replicator equation model for multilevel selection. This stands in contrast to the behavior seen in Figures \ref{fig:group_local_density_k_model} and \ref{fig:nonlinearkmodel} for our models with pairwise between-group competition and per-interaction punishment costs, in which we saw that average payoff appeared to decrease as per-interaction punishment cost $k$ increased. 

We can further consider the discrepancy between the two-level replicator model and our multilevel selection model with pairwise between-group competition by noting from Equation \eqref{eq:GDPLM} that the average payoff at steady state for the two-level replicator equation model depends only on the group-level payoffs $G_{DP}(0)$ and $G_{DP}(1)$ and difference in individual-level $\pi_D(1) - \pi_P(1)$ evaluated at the all-defector and all-punisher compositions. As the payoff differences $G_{DP}(1) - G_{DP}(0)$ and $\pi_D(1) - \pi_P(1)$ are independent of per-interaction punishment cost $k$, the fact that average payoff under payoff-group level competition changes in $k$ suggests the possibility that the collective outcome under pairwise group conflicts may not simply be a tug-of-war between the collective outcome of the all-punisher and all-defector groups. An interesting question for future work is to provide an analytical characterization for the long-time behavior for pairwise group-level competition, as well as to understand the extent to which group-level and individual-level payoffs for interior compositions of altruistic punishers $z \in (0,1)$ may impact long-time average payoff when between-group competition occurs through pairwise conflicts. 
\end{remark}

\subsubsection{Analytical Results for Two-Level Replicator Equation Featuring Only Cooperators and Defectors}

In order to provide a comparison with the long-time behavior of multilevel replicator dynamics in groups featuring cooperators, defectors, and altruistic punishers, we will also review existing analytical results for the two-level replicator equation in the case of a group-structured population containing only defectors and cooperators. Describing a population of groups featuring these two strategies by the density $g(t,x)$ of groups at time $t$ featuring a fraction $x$ of cooperators and a fraction $1-x$ of defectors, we consider the following two-level replicator equation
\begin{equation}
\begin{aligned}
\dsdel{g(t,x)}{t} &= - \dsdel{}{x} \left[x(1-x) \left( \pi_C(x) - \pi_D(x) \right) g(t,x) \right] \\
&+ \lambda g(t,x) \left[ G_{DC}(x) - \int_0^1 G_{DC}(y) g(t,y) dy \right],
\end{aligned}
\end{equation}
where $\pi_C(x)$, $\pi_D(x)$, and $G_{DC}(x)$ describe the individual-level and group-average payoffs in an $x$-cooperator group. 

Because the individual-level and average payoffs for the donation game in an $x$-cooperator group can be written as $\pi_C(x) = bx - c$, $\pi_D(x) = bx$, and $G_{DC}(x) = (b-c) x$, we can write our PDE for the dynamics on the defector-cooperator edge of the simplex as
\begin{equation}
\dsdel{g(t,x)}{t} = c \dsdel{}{x} \left[ x(1-x) g(t,x) \right] + \lambda \left( b - c\right) g(t,x) \left[ x - \int_0^1 y g(t,y) dy \right].
\end{equation}

This PDE is a rescaled version of the original model for multilevel selection with frequency-independent within-group competition introduced by Luo and Mattingly \citep{luo2014unifying,luo2017scaling}, who characterized the long-time behavior of measure-valued solutions to this equation \citep[Theorem 3]{luo2017scaling}. We can apply Theorem \ref{thm:longtimePD} and use the fact that $G_{DC}(1) = b-c$, $G_{DC}(0) = 0$, and $\pi_D(1) - \pi_C(1) = c$ to see that the threshold selection strength required to sustain long-time cooperation is given by
\begin{equation}
\lambda^*_{DC} = \frac{c \theta}{b-c},
\end{equation}
while the average payoff achieved in the long-time limit will be given by
\begin{equation} \label{eq:avgGLMCD}
 \langle G_{DC} \rangle_{g^{\lambda}_{\theta}} := \lim_{t \to \infty} \int_0^1 G_DC(y) g(t,y) dy = \left\{
    \begin{array}{cr}
      0 & : \lambda < \lambda^*\\
     b-c - \theta c & : \lambda >  \lambda^*.
    \end{array}
  \right. 
\end{equation}

\subsection{Numerical Exploration of Two-Level Replicator Dynamics on Simplex Featuring Defectors, Cooperators, and Altruistic Punishers}
\label{sec:LMtrimorphic}

We can also formulate a two-level replicator equation for multilevel selection in groups featuring all three strategies. To do this, we study the probability density $f(t,x,y)$ of having groups featuring a fraction $x$ of cooperators, a fraction $y$ of defectors, and a fraction $z = 1 - x -y$ of altruistic punishers at time $t$. We can then study the change in this probability density over time through the following two-level replicator equation
\begin{equation}
\begin{aligned}
\dsdel{f(t,x,y)}{t} &= -\dsdel{}{x} \left( x \left[ (1-x) \left( \pi_C(x,y) - \pi_P(x,y) \right) - y \left( \pi_D(x,y) - \pi_P(x,y) \right) \right] f(t,x,y) \right) \\
&- \dsdel{}{y} \left(y \left[ (1-y) \left( \pi_D(x,y) - \pi_P(x,y) \right) - x \left(\pi_C(x,y)  - \pi_P(x,y) \right) \right] f(t,x,y) \right) \\
&+ \lambda f(t,x,y) \left[G(x,y) - \int_0^1 \int_{0}^{1-u} G(u,v) f(t,u,v) dv du\right]
\end{aligned}
\end{equation}
This PDE describes the dynamics of multilevel selection in which individuals give birth at a rate depending on their payoff from playing the game within their group, while group-level replication events depend on the average payoff of group members and correspond to group-level replicator dynamics in the form first introduced by Luo and coauthors \citep{luo2014unifying,luo2017scaling}.

\begin{remark}
This PDE model can be derived as the large-population limit of an individual-based model describing a population consisting of a finite number of groups $m$, with each group composed of $n$ members. We present a derivation of a PDE model from a two-level stochastic model in Section \ref{sec:PDEderivation} for the case of pairwise between-group competition, and an analogous derivation can be performed for the case of a two-level replicator equation (see also the derivation of a two-level replicator equation in prior work on multilevel selection in protocells \citep{cooney2022pde}). 
\end{remark}

We study the behavior of this three-strategy replicator equation by performing numerical simulations using a finite-volume discretization of the three-strategy simplex. In Figure \ref{fig:trimorphicLM}, we plot the average payoff achieved after 40,000 time-steps for the numerical solution to the trimorphic dynamics for the case of per-interaction punishment costs alone with $q = 0$ and $k = 0.1$ (Figure \ref{fig:trimorphicLM}, left), as well as the case of fixed punishment costs alone with $k = 0$ and $q = 0.1$ (Figure \ref{fig:trimorphicLM}, right). We compare these numerical results with the analytically computed average payoffs at steady state achieved on the defector-cooperator and defector-punisher edges of the simplex. In both panels, we see that the average payoff achieved for the trimorphic dynamics has good agreement with analytically calculated values of average payoff achieved for two-strategy multilevel dynamics on the edges of the simplex. For the case of fixed punishment costs with $q = 0.1$, we see that, for a given strength of between-group selection $\lambda$, the numerically computed average payoff for trimorphic dynamics agrees with the average payoff achieved on the edge of the simplex featuring a higher average payoff at steady state. This potentially suggests that the long-time behavior for the three-strategy, two-level replicator equation may depend on a tug-of-war between the individual incentive to defect and the collective advantage achieved by the high average payoffs achieved an all-cooperator or all-punisher group.

\begin{figure}[!htb]
    \centering
    \includegraphics[width = .48\textwidth]{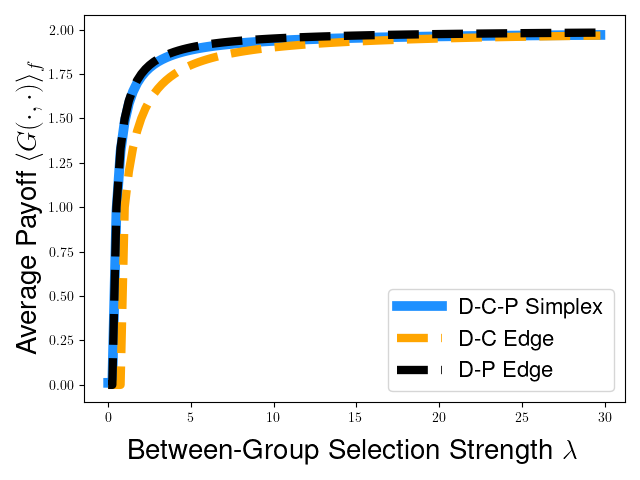}
        \includegraphics[width = .48\textwidth]{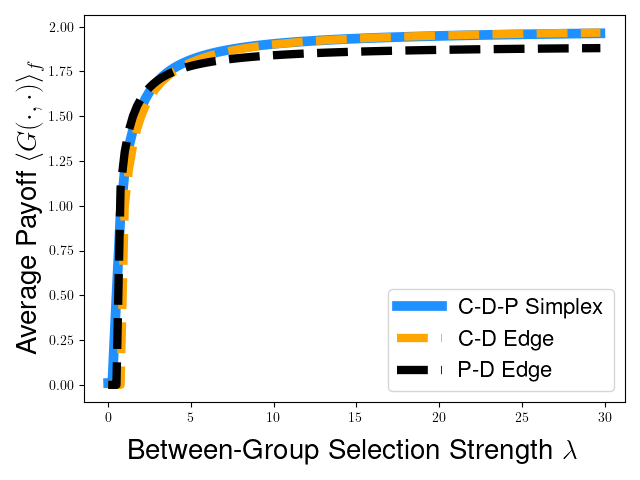}
    \caption{Plot of average payoff achieved under numerical simulation of trimorphic multilevel dynamics for two-level replicator equation, as well as analytically calculated average payoff at steady state on the defector-cooperator and defector punisher edges of the simplex for the two-level replicator equation, plotted as a function of the strength of between-group competition $\lambda$. We consider the case of per-interaction punishment cost $k = 0.1$ (left) as well as the case of fixed punishment cost $q = 0.1$ (right). Solid blue lines describe the average payoff computed for numerical solutions of trimorphic multilevel dynamics on defector-cooperator-punisher simplex after 40,000 timesteps with time increment of $\Delta t = 0.015$. Orange dashed line describes analytical expression average payoff on defector-punisher edge of the simplex (from Equation \eqref{eq:GDPLM}), while black dashed lines corresponds to analytically computed average payoffs for multilevel competition on the defector-punisher edge of the simplex (computed from Equation \eqref{eq:avgGLMCD}). Other payoff parameters were fixed at $b =3$ and $c = 1$ and the strength of punishment was set to $p = 0.5$ for each panel.}
    \label{fig:trimorphicLM}
\end{figure}

In Figure \ref{fig:trimorphiccooperationcompare}, we further explore the comparison between numerical solutions to the trimorphic dynamics and analytical results for multilevel dynamics on the edge of the simplex for cases with stronger per-interaction or fixed costs of punishment. We explore these effects both for the two-level replicator dynamics introduced in this section (left panels) and for our model of pairwise group-level competition based on globally normalized differences in the average payoffs of competing groups whose numerical solutions were first explored in Section \ref{sec:trimorphiclocalgroup} (right panels). We consider the case of per-interaction punishment costs with $k - 2$, we see in Figure \ref{fig:trimorphicq0p4compare}(top-left and top-right) that the numerical solutions for the trimorphic dynamics for replicator and pairwise group-level competition have good agreement with the average payoff achieved for the corresponding dynamics on the defector-punisher edge of simplex starting from a uniform initial strategy distribution. However, we see in Figure \ref{fig:trimorphicq0p4compare}(bottom-left and bottom-right) that the agreement between trimorphic multilevel dynamics and the corresponding two-strategy multilevel dynamics is not as good for the case of stronger fixed punishment costs with $q = 0.4$. 

\begin{figure}[!htb]
    \centering
      \includegraphics[width = .48\textwidth]{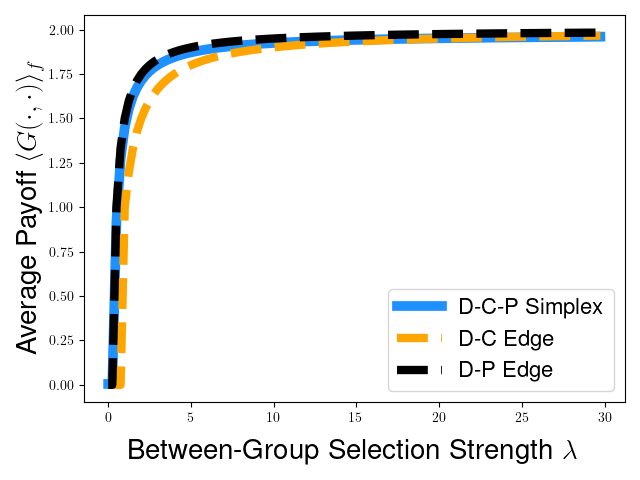}
        \includegraphics[width = .48\textwidth]{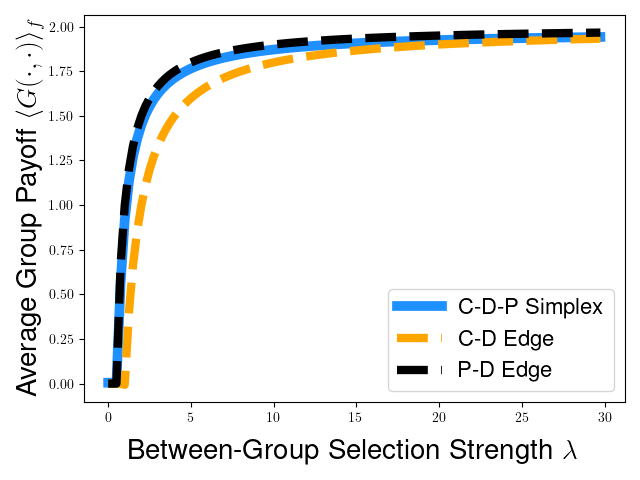}
    \includegraphics[width = .48\textwidth]{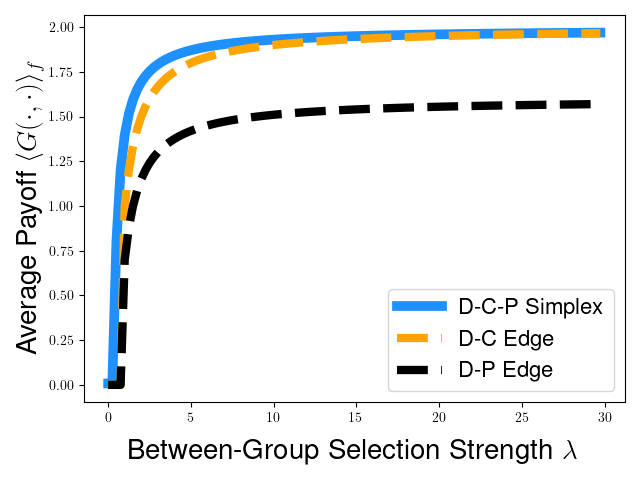}
        \includegraphics[width = .48\textwidth]{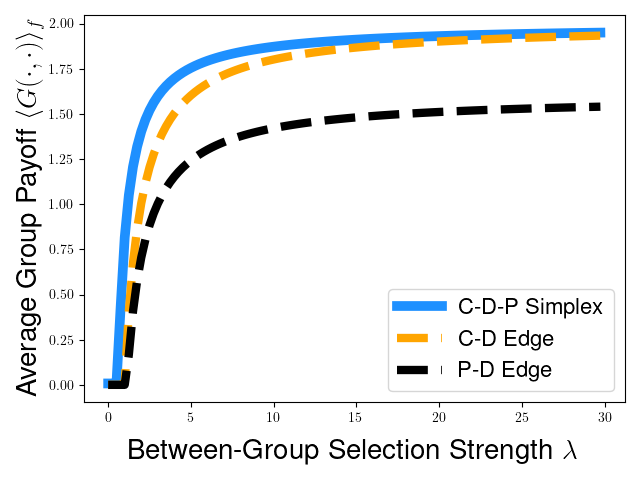}
    \caption{Comparison of average payoff achieved under numerical simulation of trimorphic multilevel dynamics and analytically calculated average payoff at steady state on the defector-cooperator and defector punisher edges of the simplex for higher costs of punishment. For the case of per-interaction punishment costs alone, we consider punishment costs of $q = 0$ and $k = 2$ for both a two-level replicator equation (top-left) and pairwise group-level competition based on globally normalized differences in average payoffs (top-right). We also consider the case of fixed punishment costs with $k = 0$ and $q = 0.4$ for both a two-level replicator equation (bottom-left) and pairwise group-level competition (bottom-right). All lines displayed follow the same conventions introduced in Figure \ref{fig:trimorphicLM} and all parameters other than the fixed cost of punishment $q$ are based on the values given in Figure \ref{fig:trimorphicLM}(right) as well (with $b = 3$, $c = 2$, and $p = 0.5$).}
    \label{fig:trimorphicq0p4compare}
\end{figure}

While the quantitative agreement between the average payoff achieved for the multilevel dynamics on the three-strategy simplex and analytical results for the edges of the simplex does not appear to be as good when we increase the fixed cost of punishment from $q = 0.1$ to $q = 0.4$, we still see that the numerically calculated average payoff for the trimorphic dynamics increases with the strength of between-group competition $\lambda$ and appears to interpolate between the average payoff of the all-defector group $G(0,1)$ to the average payoff of the all-punisher group $G(0,0)$ as $\lambda$ increases (Figure \ref{fig:trimorphicq0p4compare}, bottom-left and bottom-right). This discrepancy seen between numerical solutions of the trimorphic dynamics and the analytical results raises the questions as to whether this discrepancy should be expected for the true solutions to our PDE models of multilevel selection, or whether this discrepancy may be an artifact of the form of our finite-volume discretization of the strategic composition or our forward-Euler discretization in time. In any event, both the agreement and discrepancies seen between average payoff achieved in  the trimorphic and dimorphic models for multilevel selection motivate future work on numerical analysis of these PDE models, perhaps with more detailed numerical schemes that can distinguish between solutions to the PDEs that concentrate upon equilibrium points or edges of the simplex and those that are supported as densities on the simplex in the long-time limit. Such methods have recently been applied in the context of one-dimensional diffusion models with individual-level selection in population genetics and evolutionary games \citep{zhao2013complete,carrillo2022optimal}, and a natural direction is to ask whether similar approaches can be used to understand models of evolutionary dynamics featuring group-level competition or more than two strategies.

\section{Comparison of Multilevel Dynamics with Pairwise Between-Group Selection for Model of Indirect Reciprocity} \label{sec:indirectreciprocity}

In this section, we will consider the dynamics of multilevel selection pairwise between-group competition when within-group interactions follow a simplified model for indirect reciprocity. We use the model of image scoring \citep{nowak1998evolution,nowak2006five}, which was introduced as a mechanism for showing how punishment of individuals with a reputation for defecting can help to stabilize a cooperative population against invasion from defectors. The image scoring rule is one example of a social norm for maintaining cooperation via indirect reciprocity \citep{ohtsuki2006leading,santos2007multi,radzvilavicius2019evolution,santos2018social}, and it has been used in previous work to study synergistic effects of combining within-group indirect reciprocity with between-group competition to promote the evolution of cooperation via multilevel selection \citep{cooney2022assortment,cooney2023evolutionary}.

For this model of indirect reciprocity, we assume that game-theoretic interactions take place through a donation game. We assume that there are two types of individuals: defectors ($D$) and conditional cooperators ($C$). Defectors always defect in the donation game, paying no cost and conferring no benefit to their opponent. Conditional cooperators always cooperate with other conditional cooperators, but they may defect when interacting with defectors. Specifically, we assume that conditional cooperators defect when playing with defectors with probability $Q$ and cooperate in such interactions with probability $1-Q$, where the parameter $q$ represents the probability that a conditional cooperator recognizes a defector and punishes them with defection. Assuming that individuals play the donation game according to these rules, the payoffs $\pi_C^Q(x)$ and $\pi_D^Q(x)$ received from playing this game in a group featuring a fraction $x$ of conditional cooperators are given by
\begin{subequations}
\begin{alignat}{2}
\pi_C^Q(x) &= x (b-c) + (1-x) \left[(1-Q) (-c) + Q (0) \right] &&= (b - Qc) x - (1-Q) c\\
\pi_D^Q(x) &= x \left[(1-Q) b  + Q(0) \right] + (1-x) (0) &&= (1-Q)b x.
\end{alignat}
\end{subequations}
We can therefore see that the individual-level advantage payoff advantage defectors have over cooperators is given by
\begin{equation}
\Pi_Q(x) := \pi_D^Q(x) - \pi_C^Q(x) = (1-Q)c - Q(b-c) x.
\end{equation}
In particular, we note that $\Pi_q(x)$ is a decreasing function of $x$, and we have that
\begin{equation}
\Pi_Q(1) = c - b Q.
\end{equation}
Therefore we see that $\Pi_Q(x) > 0$ for all $x \in [0,1]$ when $Q < \frac{c}{b}$, while we have that there is an $x_{eq} < 1$ such that $\Pi_Q(x) < 0$ for $x \in (x_{eq},1)$ when $Q > \frac{c}{b}$. This means that our model of indirect reciprocity will take the form of a generalized Prisoners' Dilemma scenario when $Q < \frac{c}{b}$ and satisfies the assumptions of a generalized Stag-Hunt game when $Q > \frac{c}{b}$. 

Next, we look to describe the role of group-level competition in the presence this within-group assortment process. We can calculate that the average payoff of group members is given by
\begin{equation}
G_Q(x) := x \pi_C^{Q}(x) + (1-x) \pi_D^Q(x) = \left( b - c\right) \left[ (1-Q) x + Q x^2\right].
\end{equation}
Here we note that $G_Q'(x) = (b-c) \left[ 1 - Q + 2 Q x\right] > 0$ for all $x \in [0,1]$, and therefore we can deduce that $G^* := \max_{x \in [0,1]} = G_Q(1) = b - c$ and that $G_* := \min_{x \in [0,1]} = G_Q(0) = 0$. This means that we can write the normalized group-level reproduction rate $\mc{G}(x) = \frac{G_Q(x)}{G^* - G_*}$ as
\begin{equation}
\mc{G}(x) = \frac{G_Q(x)}{G^* - G_*} = \frac{(b-c) \left( (1-Q) x + Q x^2 \right)}{(b-c) - 0} = (1-Q) x + Q x^2. 
\end{equation}
We can then combine the effects of differences in individual payoff and average group payoff by considering the dynamics of multilevel selection with the group-level probability with the globally normalized pairwise group update rule given by
\begin{equation}
\rho(x,u) = \frac{1}{2} \left[ 1 + \frac{G_Q(x) - G_Q(u)}{G^* - G_{*}} \right]
\end{equation}
Under these assumptions, the density of groups $f(t,x)$ with a fraction $x$ of conditional cooperators at time $t$ evolves according to the PDE
\begin{equation} \label{eq:qmodelPDE}
\dsdel{f}{t} = \dsdel{}{x} \left[ x (1-x) \Pi_Q(x) f(t,x) \right] + \lambda \left[ \mc{G}(x) - \int_0^1 \mc{G}(y) f(t,y) dy \right].
\end{equation}
For the case in which $Q < \frac{c}{b}$, this PDE corresponds to the case of a generalized PD game, and we can apply Theorem \ref{thm:longtimePD} to identify the threshold selection strength $\lambda^*_q$ needed to sustain long-time cooperation and to characterize the long-time behavior for solutions to Equation \eqref{eq:qmodelPDE}. 

From the form of $\mc{G}(x)$ and $\Pi_Q(x)$ for our model of indirect reciprocity, we are able to see that $\mc{G}(1) = 1$, $\mc{G}(0) = 0$, and $\Pi_Q(1) = c - b q$. For $Q < \frac{c}{b}$, we can then apply Theorem \ref{thm:longtimePD} to see that conditional cooperation can be supported via multilevel selection when the strength of between-group competition $\lambda$ exceeds the threshold level
\begin{equation}
\lambda^*_Q := \frac{\theta \Pi_Q(z)}{\mc{G}(1) - \mc{G}(0)} = \theta \left( c - bQ \right).
\end{equation}
When $\lambda > \lambda^*$, the average payoff at steady state is given by
\begin{equation}
\langle G_Q(\cdot) \rangle_{f^{\lambda}_{\theta}} = \left( \frac{\lambda^*_Q}{\lambda} \right) G_Q(0) + \left( 1 - \frac{\lambda^*_Q}{\lambda}  \right) G_Q(1) =\left(b -c \right) \left( 1 - \left(c - b Q\right) \frac{\theta}{\lambda} \right).
\end{equation}
We therefore see that the threshold selection strength $\lambda^*_Q$ is always decreasing in the defector detection probability $Q$ and the average payoff at steady state increases with the defector detection probability $Q$. Unlike the form of punishment studied in the main paper, it is not possible to have a threshold selection strength or average long-time payoff with a non-monotonic dependence on the defector detection probability $Q$.

\section{Derivation of Numerical Schemes}
\label{sec:numericalderivation}

In this section, we provide additional background material on the finite volume simulations used in Section \ref{sec:nonlineargroup} and Section \ref{sec:trimorphicnumerics}. In Section \ref{sec:twostrategyfv}, we present a derivation of an upwind finite volume scheme used in Section \ref{sec:nonlineargroup} for our PDE model for multilevel selection with pairwise group-level competition in groups featuring only defectors and altruistic punishers. In Section \ref{sec:FVtrimorphic}, we summarize the finite volume method used to perform the numerical simulations of the trimorphic multilevel dynamics in Section \ref{sec:trimorphicnumerics} with groups featuring defectors, cooperators, and altruistic punishers. 

\subsection{Derivation of Finite Volume Scheme for Two-Strategy Dynamics}
\label{sec:twostrategyfv}

In this section, we present a derivation of the finite volume scheme used in Section \ref{sec:nonlineargroup} to study our PDE model for pairwise between-group competition in the case of group-level victory probabilities that are not additively separable. We start by considering the PDE model from Equation \eqref{eq:multilevelPDErhodiff} for pairwise group-level competition in groups featuring only defectors and altruistic punishers, which is given by
\begin{equation}
\begin{aligned}
\dsdel{f(t,z)}{t} &= - \dsdel{}{z} \left[ z (1-z) \left( \pi_P(z) - \pi_D(z) \right) f(t,z)  \right] + \lambda f(t,z) \left[ \int_0^1 \left\{ \rho(z,u) - \rho(u,z)  \right\} f(t,u) du \right].
\end{aligned}
\end{equation}

We can integrate our PDE to see that
\begin{equation} \label{eq:fvintegrated}
\begin{aligned}
\int_{z_j}^{z_{j+1}} \dsdel{f(t,z)}{t} dz &= - \int_{z_i}^{z_{i+1}} \left[ \dsdel{}{z} \left( z (1-z) \left(\pi_P(z) - \pi_D(z) \right) f(t,z) \right) \right] dz \\ &+ \lambda  \int_{z_i}^{z_{i+1}}  f(t,z) \left[ \int_0^1 \left[\rho(z,u) - \rho(u,z) \right] f(t,u) du \right] dz \\
&= z_{i+1} (1-z_{i+1}) \left[ \pi_D(z_{i+1}) - \pi_P(z_{i+1})\right] f(t,z_{i+1}) \\ &- z_{i} (1-z_{i}) \left[ \pi_D(z_{i}) - \pi_P(z_{i})\right] f(t,z_i) \\
&+ \lambda  \int_{z_j}^{z_{i+1}}  f(t,z) \left[ \int_0^1 \left[\rho(z,u) - \rho(u,z) \right] f(t,u) du \right] dz.
\end{aligned}
\end{equation}
We will now consider a formulation of our density $f(t,z)$ as a piecewise constant function by considering the average value that the densities takes on the $i$th cell
\begin{equation}
f_i(t) := \frac{1}{z_{i+1} - z_i} \int_{z_i}^{z_{i+1}} f(t,z) dz =  \frac{1}{\frac{i+1}{N} - \frac{i}{N}} \int_{z_i}^{z_{i+1}} f(t,z) dz = N \int_{z_i}^{z_{i+1}} f(t,z) dz,
\end{equation}
and approximate the solution $f(t,z)$ to our PDE with the piecewise constant description
\begin{equation}
f(t,z) = \sum_{i=0}^{N-1} f_i(t) 1_{z \in (z_i,z_{i+1}]}.
\end{equation}
We then note that, for a piecewise constant $f(t,z)$, the lefthand side of Equation \eqref{eq:fvintegrated} can be written as
\begin{equation}
\int_{z_i}^{z_{i+1}} \dsdel{f(t,z)}{t} dz = \dsddt{} \int_{z_i}^{z_{i+1}} f(t,z) dz = \dsddt{} \int_{\frac{i}{N}}^{\frac{i+1}{N}} f_i(t) dz = \frac{1}{N} \dsddt{f_i(t)}.
\end{equation}
Using the piecewise constant form of $f(t,z)$, we see that the integral term on the righthand side of Equation \eqref{eq:fvintegrated} can be written as
\begin{equation}
\begin{aligned}
 & \int_{z_j}^{z_{j+1}}  f(t,z) \left[ \int_0^1 \left[\rho(z,u) - \rho(u,z) \right] f(t,u) du \right] dz \\ &=  \int_{z_j}^{z_{j+1}} f_i(t) \left[ \int_0^1  \left\{ \left[ \rho(z,u) - \rho(u,z) \right] \left(\sum_{j=0}^{N-1} f_j(t) 1_{u \in (u_i,u_{i+1}]} \right) \right\} du \right] dz \\
 &= f_i(t) \int_{z_j}^{z_{j+1}} \left\{ \sum_{j=0}^{N-1} \int_{u_i}^{u_{i+1}} f_j(t) \left[\rho(z,u) - \rho(u,z) \right] du \right\}  dz \\
 &= \sum_{j=0}^{N-1} f_i(t) f_j(t) \int_{z_i}^{z_{i+1}} \int_{u_j}^{u_{j+1}} \left[\rho(z,u) - \rho(u,z) \right] du dz.
\end{aligned}
\end{equation}
By introducing $\rho_{ij}$ to denote the average group-level victory probability on the square $[z_i,z_j] \times [u_i,u_j]$, we have that
\begin{equation}
\rho_{ij} = \frac{\int_{z_i}^{z_{i+1}} \int_{u_j}^{u_{j+1}}\rho(z,u) du dz }{\int_{z_i}^{z_{i+1}} \int_{u_j}^{u_{j+1}} du dz} = N^2 \int_{u_j}^{u_{j+1}} \rho(z,u) du dz,
\end{equation}
and we can finally rewrite the integral term for the righthand side of Equation \eqref{eq:fvintegrated} as
\begin{equation}
\int_{z_j}^{z_{j+1}}  f(t,z) \left[ \int_0^1 \left[\rho(z,u) - \rho(u,z) \right] f(t,u) du \right] dz = \frac{1}{N^2}   \sum_{j=0}^{N-1} \left[ \rho_{ij} - \rho_{ji} \right] f_i(t) f_j(t) 
\end{equation}
For the advection terms, we will choose the values of the density $f(t,z_i)$ at each grid point using an upwind convention \citep{leveque2002finite}, using the following piecewise characterization for $f(t,z_i)$
\begin{equation}
 f(t,z_i) = f_i^{UW}(t) := \left\{
    \begin{array}{lr}
      f_{i}(t) & : z_i (1-z_i) \left[ \pi_D(z_i) - \pi_P(z_i) \right] > 0 \\
      f_{i-1}(t) & : z_i (1-z_i) \left[ \pi_D(z_i) - \pi_P(z_i) \right] \leq 0,
    \end{array}
  \right.
\end{equation}
and we similarly define the value $f(t,z_{i+1})$ with the upwind convention as
\begin{equation}
f(t,z_{i+1}) = f_{i+1}^{UW}(t) := \left\{
    \begin{array}{lr}
      f_{i+1}(t) & : z_{i+1} (1-z_{i+1}) \left[ \pi_D(z_{i+1}) - \pi_P(z_{i+1}) \right] > 0 \\
      f_{i}(t) & : z_{i+1} (1-z_{i+1}) \left[ \pi_D(z_{i+1}) - \pi_P(z_{i+1}) \right] \leq 0,
    \end{array}
  \right.
\end{equation}
Combining the different terms, we may rewrite Equation \eqref{eq:fvintegrated} in the form
\begin{equation}
\begin{aligned}
\frac{1}{N} \dsddt{f_i(t)}  &= z_{i+1} (1-z_{i+1}) \left[ \pi_D(z_{i+1}) - \pi_P(z_{i+1})\right] f_{i+1}^{UW}(t) \\ &- z_{i} (1-z_{i}) \left[ \pi_D(z_{i}) - \pi_P(z_{i})\right] f_i^{UW}(t) + \lambda   \frac{1}{N^2}   \sum_{j=0}^{N-1} \left[ \rho_{ij} - \rho_{ji} \right] f_i(t) f_j(t).
\end{aligned} 
\end{equation}
We can then multiply both sides by $N$ to see that the cell averages $f_i(t)$ satisfy the following system of ODEs
\begin{equation}
\begin{aligned}
\dsddt{f_i(t)}  &= -N z_{i+1} (1-z_{i+1}) \left[ \pi_D(z_{i+1}) - \pi_P(z_{i+1})\right] f_{i+1}^{UW}(t) \\ &+ N z_{i} (1-z_{i}) \left[ \pi_D(z_{i}) - \pi_P(z_{i})\right] f_i^{UW}(t) + \lambda f_i(t) \left( \frac{1}{N} \sum_{j=0}^{N-1} \left[ \rho_{ij} - \rho_{ji} \right] f_j(t) \right).
\end{aligned} 
\end{equation}
To perform numerical simulations, we then evaluate the integrals $\rho_{ij}$ for the group-level victory probability over each square $[z_i,z_{i+1}] \times [u_{j},u_{j+1}]$ using the trapezoidal rule. We can then solve the system of ODEs numerically by discretizing our ODE system in time as well. In particular, we use Euler's forward method to solve this ODE system to produce each of the figures in Section \ref{sec:nonlineargroup}. 

\subsection{Derivation of Finite Volume Scheme for Three-Strategy Dynamics}
\label{sec:FVtrimorphic}

In this section, we present details on the numerical scheme we use to analyze the dynamics of our PDE featuring groups with three strategies and group-level victory probabilities that are additively separable. To study multilevel dynamics for groups with compositions described by points three-strategy simplex, we adapt an approach that has been previously used to explore the dynamics of multilevel selection in protocells \citep{cooney2022pde}. 

We consider a PDE of the form
\begin{equation}
\begin{aligned}
\dsdel{f(t,x,y)}{t} &= -\dsdel{}{x} \left( x \left[ (1-x) \left( \pi_C(x,y) - \pi_P(x,y) \right) - y \left( \pi_D(x,y) - \pi_P(x,y) \right) \right] f(t,x,y) \right) \\
&- \dsdel{}{y} \left(y \left[ (1-y) \left( \pi_D(x,y) - \pi_P(x,y) \right) - x \left(\pi_C(x,y)  - \pi_P(x,y) \right) \right] f(t,x,y) \right) \\
&+ \lambda f(t,x,y) \left[\mc{G}(x,y) - \int_0^1 \int_{0}^{1-u} \mc{G}(u,v) f(t,u,v) dv du\right]
\end{aligned}
\end{equation}
and look to solve this PDE numerically using an upwind finite-volume discretization on a grid of cells $\mc{C}_{ij}$.

Following the approach used for the trimorphic protocell model \citep{cooney2022pde}, we discretize the three-strategy simplex into a grid composed of $\frac{N(N-1)}{2}$ square volumes with side length $\frac{1}{N}$ and $N$ isosceles triangle volumes with leg length $\frac{1}{N}$. We use the indices $(i,j)$ to describe a cell $\mc{C}_{i,j}$ whose bottom-left corner is the point $(x_i,y_j) = \left(\tfrac{i}{N},\tfrac{j}{N}\right)$ where $i \geq 0$, $j \geq 0$, and $i + j \leq N -1$. Square volumes $\mc{C}_{i,j}$ consist of the set of points $\mc{C}_{i,j} = [x_i,x_{i+1}] \times [y_j,j_{j+1}]$ for $i + j < N - 1$, and the triangular volumes are isosceles triangles with legs given by $[x_{i},x_{i+1}]$ and $[y_{N-i-1},y_{N-i}]$. We will then use these definition of the square and triangular grid volumes to characterize the flux across boundaries corresponding individual-level dynamics and the integrals over grid volumes that describe group-level dynamics. 

We can introduce the average value $f_{i,j}(t)$ of the density $f(t,x,y)$ on the grid cell $\mc{C}_{i,j}$ as
\begin{equation}
f_{i,j}(t) := \iint_{\mc{C}_{i,j}} f(t,x,y) dy dx = \left\{
    \begin{array}{lr}
      \ds\int_{x_i}^{x_{i+1}} \ds\int_{y_j}^{y_{j+1}} f(t,x,y) dy dx & : i + j < N - 1\\
       \ds\int_{x_i}^{x_{i+1}} \ds\int_{y_j}^{1-x} f(t,x,y) dy dx & : i + j = N-1,
    \end{array}
  \right.
\end{equation}
and we can write the average value $\mc{G}_{i,j}$ of the net group-level replication rate $\mc{G}(x,y)$ of groups in the grid cell $\mc{C}_{i,j}$ as 
\begin{equation}
\mc{G}_{i,j} := \iint_{\mc{C}_{i,j}} \mc{G}(x,y) dy dx = \left\{
    \begin{array}{lr}
      \ds\int_{x_i}^{x_{i+1}} \ds\int_{y_j}^{y_{j+1}} G(x,y) dy dx & : i + j < N - 1  \vspace{2mm}\\
       \ds\int_{x_i}^{x_{i+1}} \ds\int_{y_j}^{1-x} \mc{G}(x,y) dy dx & : i + j = N-1,
    \end{array}
  \right.
\end{equation}
For convenience, we write the characteristic curves for our PDE in the form
\begin{subequations}
\begin{align}
\dsddt{x} &= F_1(x,y) \\
\dsddt{y} &= F_2(x,y)
\end{align}
\end{subequations}
where the righthand-side functions satisfy
\begin{subequations} \label{eq:F1F2}
\begin{align}
F_1(x,y) &= x \left[ (1-x) \left( \pi_C(x,y) - \pi_P(x,y) \right) - y \left( \pi_D(x,y) - \pi_P(x,y) \right) \right] \\
F_2(x,y) &= y \left[ (1-y) \left( \pi_D(x,y) - \pi_P(x,y) \right) - x \left( \pi_C(x,y) - \pi_P(x,y) \right) \right].
\end{align}
\end{subequations}
Having defined the characteristic curves and average net group-level replication rate, we can follow the approach used to model multilevel dynamics with three types of individuals, we may write our upwind finite-volume scheme as the following system of ODEs of the form
\begin{equation}
\begin{aligned}
\frac{1}{\beta_{i,j}} \dsddt{\rho_{i,j}(t)} &=  \left(\int_{y_j}^{y_{j+1}} F_1(x_i,y) dy \right) U\left( x_i,y_j \right) - \alpha_{i,j} \left(\int_{y_j}^{y_{j+1}} F_1(x_{i+1},y) dy \right)  U\left( x_{i+1},y_j \right) \\ 
&+ \left(\int_{x_i}^{x_{i+1}} F_2(x,y_j) dx \right) V\left( x_i,y_j \right) - \alpha_{i,j} \left(\int_{x_i}^{x_{i+1}} F_2(x,y_{j+1}) dx \right)  V\left( x_i,y_{j+1} \right)\\
&+ \frac{\lambda}{\beta_{i,j}} \rho_{i,j}(t) \left[ \mc{G}_{i,j} - \sum_{l,m > 0}^{l + m \leq N - 1} \mc{G}_{l,m} \rho_{l,m}(t) \right],
\end{aligned}
\end{equation}
where the quantities $\alpha_{i,j}$, $\beta_{i,j}$, $U\left(x_i,x_j\right)$, and $V\left(x_i,y_j\right)$ are given by
\begin{subequations}
\begin{align}
\alpha_{i,j} &= \left\{
    \begin{array}{lr}
      1 & : i + j < N - 1\\
      0 & : i + j = N - 1
    \end{array}
  \right.  \\
\beta_{i,j} &= \left\{
    \begin{array}{lr}
      1 & : i + j < N - 1\\
      2 & : i + j = N - 1
    \end{array}
  \right. \\
U\left( x_i,y_j \right) &= \left\{
    \begin{array}{lr}
      \rho_{i-1,j}(t) & : \ds\int_{y_j}^{y_{j+1}} F_1(x_i,y) dy \geq 0 \\
       \rho_{i,j}(t) & : \ds\int_{y_j}^{y_{j+1}} F_1(x_i,y) dy < 0 
    \end{array}
  \right. \\
V\left( x_i,y_j \right) &= \left\{
    \begin{array}{lr}
       \rho_{i,j-1}(t) & : \ds\int_{x_i}^{x_{i+1}} F_2(x,y_j) dx \geq 0 \\
       \rho_{i,j}(t) & : \ds\int_{x_i}^{x_{i+1}} F_2(x,y_j) dy < 0 
    \end{array}
  \right.
\end{align}
\end{subequations}
Here $\alpha_{i,j}$ describes whether or not a given cell has a top or right edge that remains within the interior of the simplex (describing if $C_{i,j}$ is a square or triangular volume), $\beta_{i,j}$ weighs the contribution of integrals over square or triangular volumes. We use the quantities $U(x_i,y_j)$ and $V(x_i,y_j)$ to implement the upwinding convention to specify that the change of mass through boundaries should be determined by the value of the discretization of $\rho(t,x,y)$ on the volume from which the density is flowing. 

Now that we have established the general approach for performing finite volume simulations for our trimorphic PDE model of multilevel selection, we look to formulate expressions for the flux across boundaries and the group-level replication rates for our specific model of multilevel selection with individual-level and group-level replication rates arising from payoffs in our model for altruistic punishment.

\subsubsection{Calculation of Fluxes Through Boundaries for Within-Group Terms}

To describe the flux across volume boundaries for our model of within-group altruistic punishment, we look to find expressions for the functions $F_1(x,y)$ and $F_2(x,y)$. Plugging the expressions for the payoffs $\pi_C(x,y)$, $\pi_D(x,y)$, and $\pi_P(x,y)$ from Equation \eqref{eq:trimorphicpayoffs} into Equation \eqref{eq:F1F2}, we can see that $F_1(x,y)$ and $F_2(x,y)$ take the form
\begin{subequations}
\begin{align}
F_1(x,y) &= x \left[(1-x)  \left(q + k y\right) - y \left(c + q - p + px + (p+k) y \right)\right] \\
F_2(x,y) &= y \left[ (1-y) \left(c + q - p + px + (p+k) y \right) - x \left( q + k y \right) \right].
\end{align}
\end{subequations}
We can then use these expressions to calculate the integrals of $F_1(x,y)$ and $F_2(x,y)$ along the edges of grid volumes. We can compute that the fluxes across vertical volume boundaries are given by
\begin{equation}
\begin{aligned}
\int_{y_j}^{y_{j+1}} F_1(x_i,y) dy &= q x_i \left( 1 - x_i \right) \left( y_{j+1} - y_j \right) +  \frac{1}{2} \left[ \left( k + p - c - q\right) x_i + \left( k + p \right) x_{i}^2 \right]\left(y_{j+1}^2 - y_j^2 \right)\\  &- \frac{1}{3} \left( k + p\right) x_i \left( y_{j+1}^3 - y_j^3 \right) \\
\int_{y_j}^{y_{j+1}} F_1(x_{i+1},y) dy &= q x_{i+1} \left( 1 - x_{i+1} \right) \left( y_{j+1} - y_j \right) +  \frac{1}{2} \left[ \left( k + p - c - q\right) x_{i+1}  + (k+p)x_{i+1}^2 \right] \left(y_{j+1}^2 - y_j^2 \right) \\ &- \frac{1}{3} \left( k + p\right) x_{i+1} \left( y_{j+1}^3 - y_j^3 \right),
\end{aligned}
\end{equation}
while the fluxes across horizontal volumes boundaries are given by
\begin{equation}
\begin{aligned}
\int_{x_i}^{x_{i+1}} F_2(x,y_j) dy &= \left[(c + q -p)y_{j} + \left( k - c + 2p - q\right) y_{j}^2 - (p+k) y_{j}^3\right] \left( x_{i+1} - x_i \right)\\
&+ \frac{1}{2} \left[ \left(p - q \right) y_{j} - (k+p) y_{j}^2 \right]  \left( x_{i+1}^2 - x_i^2 \right) \\
\int_{x_i}^{x_{i+1}} F_2(x,y_{j+1}) dy &= \left[(c + q -p)y_{j+1} + \left( k - c + 2p - q\right) y_{j+1}^2 - (p+k) y_{j+1}^3\right] \left( x_{i+1} - x_i \right) \\
&+ \frac{1}{2} \left[ \left( p - q \right) y_{j+1} - (k+p) y_{j+1}^2 \right]  \left( x_{i+1}^2 - x_i^2 \right).
\end{aligned}
\end{equation}

\subsubsection{Calculation of Average Group-Level Reproduction Rates on Volumes}

Now, we look to calculate the average net group-level reproduction rates on each volume $\mc{C}_{i,j}$. Because we consider numerical solutions for the group-level replication rate $\mc{G}(x,y) = 1 -y$ and $\mc{G}(x,y) = \left( \frac{1}{G^* - G_*} \right) G(x,y)$ that are both polynomials of at most second-order, we can calculate the average reproduction rates $\mc{G}_{i,j}$ by integrating all two-variables polynomials up to quadratic order in $x$ and $y$. These integrals have already been calculated in previous work on finite volume simulations of multilevel selection in protocells, which we summarize in Table \ref{tab:polynomialintegrals}.

\renewcommand{\arraystretch}{2}
\begin{table}[H] 
\caption{Integrals of polynomials over grid cells $\mc{C}_{i,j}$ up to quadratic order in the variables $x$ and $y$, as calculated in previous work on finite volume simulations of PDE models for multilevel selection in protocell evolution \citep{cooney2022pde}.} 
\label{tab:polynomialintegrals}
\centering 
\begin{tabular}{|c|c|c|} 
\hline
Integral  & 
\makecell{Rectangular Cell \\ $i + j \leq N - 2$} &  
\makecell{Triangular Cell \\ $i + j = N-1$} \\
\hline
$\iint_{\mc{C}_{i,j}} 1 dy dx$ & $\ds\frac{1}{N^2} $ & $\ds\frac{1}{2N^2}$  \\
\hline
$\iint_{\mc{C}_{i,j}} x dy dx$ & $\ds\frac{1}{N^2} \left[ \frac{i}{N} + \frac{1}{2 N^2} \right]$ & $\ds\frac{1}{N^2} \left[ \frac{i}{2 N} + \frac{1}{2 N^2} \right]$  \\
\hline
$\iint_{\mc{C}_{i,j}} y dy dx$ & $\ds\frac{1}{N^2} \left[ \frac{j}{N} + \frac{1}{2 N^2} \right]$ & $\ds\frac{1}{N^2} \left[ \frac{j}{2 N} + \frac{1}{6 N^2} \right]$  \\
\hline
$\iint_{\mc{C}_{i,j}} x^2 dy dx$ & $\ds\frac{1}{N^2} \left[ \frac{i^2}{N^2} + \frac{i}{N^2} + \frac{1}{3 N^2}  \right]$ & $\ds\frac{1}{N^2} \left[ \frac{i^2}{2N^2} + \frac{i}{3N^2} + \frac{1}{12 N^2}  \right]$  \\
\hline
$\iint_{\mc{C}_{i,j}} y^2 dy dx$ & $\ds\frac{1}{N^2} \left[ \frac{j^2}{N^2} + \frac{j}{N^2} + \frac{1}{3 N^2}  \right]$ & $\ds\frac{1}{N^2} \left[ \frac{(N-i)^2}{2N^2} - \frac{2(N-i)}{3N^2} + \frac{1}{4 N^2}  \right]$  \\
\hline
$\iint_{\mc{C}_{i,j}} xy dy dx$ & $\ds\frac{1}{N^2} \left[ \frac{ij}{N^2} + \frac{i}{2N^2} + \frac{j}{2 N^2} + \frac{1}{4 N^2}  \right]$ & $\ds\frac{1}{N^2} \left[ \frac{i}{2N} - \frac{i^2}{2 N^2} - \frac{i}{2N^2} + \frac{1}{6N} - \frac{1}{8 N^2} \right]$  \\
\hline
\end{tabular}
\end{table}
\renewcommand{\arraystretch}{1}

For the net group-level reproduction rate $\mc{G}(x,y) = 1 - y$, we have that, for $i + j < N-1$, the average reproduction rates $\mc{G}_{i,j}$ on the volume $\mc{C}_{i,j}$ are given by
\begin{subequations}
\begin{equation}
\mc{G}_{i,j} = N^2 \iint_{\mc{C}_{i,j}} \left[ 1 - y \right] dy dx = \int_{x_{i}}^{x_{i+1}} \int_{y_{j}}^{y_{j+1}} \left( 1 - y \right) dy dx = 1 - \frac{j}{n} - \frac{1}{2N^2},
\end{equation}
while, for $i + j = N - 1$, the volume average is given by
\begin{equation}
\mc{G}_{i,j} = 2 N^2 \iint_{\mc{C}_{i,j}} \left[ 1 - y \right] dy dx = \int_{x_i}^{x_{i+1}} \int_{y_j}^{1-x} \left[ 1 - y\right] dy dx = 1 - \frac{j}{N} - \frac{1}{3 N^2}.   
\end{equation}
\end{subequations}
For the group-level reproduction function $\mc{G}(x,y) = \left( \frac{1}{G^* - G_{*}} \right) G(x,y)$, we can use the expression for average group payoffs to see that, for $i + j < N - 1$, the average group-level reproduction rate is given by
\begin{equation}
\begin{aligned}
\mc{G}_{i,j} &= \frac{N^2}{G^* - G_{*}} \iint_{\mc{C}_{i,j}} G(x,y) dy dx \\ &= \frac{N^2}{G^* - G_{*}} \int_{x_{i}}^{x_{i+1}} \int_{y_{j}}^{y_{j+1}}  \left[  b - c - q + q x - \left( b - c + p + q \right) y + \left( p + k \right) x y + \left( p + k \right) y^2 \right] dy dx \\
&= \left( \frac{1}{G^* - G_*} \right) \left[  b - c  - q + q \left( \frac{i}{N} + \frac{1}{2N^2} \right) - \left( b - c + p + q \right) \left( \frac{j}{N} + \frac{1}{2 N^2} \right) \right. \\
&\left. \hspace{35mm}  + (p+k) \left( \frac{ij}{N^2} + \frac{i}{2N^2} + \frac{j}{2N^2} + \frac{1}{4 N^2} \right) + (p+k) \left( \frac{j^2}{N^2} + \frac{j}{N^2} + \frac{1}{3 N^2} \right)\right].
\end{aligned}
\end{equation}
In addition, we can see that, for $i + j = N-1$, the average group-level reproduction rate is given by
\begin{equation}
\begin{aligned}
\mc{G}_{i,j} &= \frac{2 N^2}{G^* - G_{*}} \iint_{\mc{C}_{i,j}} G(x,y) dy dx  \\
&= \frac{2 N^2}{G^* - G_{*}}  \int_{x_{i}}^{x_{i+1}} \int_{y_j}^{1-x} \left[  b - c - q + q x - \left( b - c + p + q \right) y + \left( p + k \right) x y + \left( p + k \right) y^2 \right] dy dx \\
&= \left( \frac{1}{G^* - G_*} \right)  \left[ b - c - q + q \left(\frac{i}{N} + \frac{1}{2N^2} \right) - \left( b - c + p + q \right) \left( \frac{j}{N} + \frac{1}{3 N^2} \right) \right. \\
&\left. \hspace{35mm} + \left( p + k \right) \left(\frac{i}{N} - \frac{i^2}{N^2} + \frac{(N-i)^2}{N^2}  - \frac{i}{N^2} - \frac{(N-i)}{3N^2}+ \frac{1}{3N} + \frac{1}{4 N^2} \right) \right]
\end{aligned}
\end{equation}

\end{document}